\documentclass[11pt,a4paper,twoside,final]{article}

\usepackage{amssymb,amsmath,amsthm,graphicx,float}
\usepackage[a4paper,rmargin=3cm,lmargin=3cm,twoside]{geometry}
\usepackage{bbm}
\usepackage{fancyhdr}
\usepackage[english]{babel}
\usepackage[T1]{fontenc}
\usepackage[latin1]{inputenc}
\usepackage{listings}
\usepackage[all]{xy} 
\usepackage{graphicx} 
\usepackage{psfrag} 
\usepackage{dsfont} 

\newtheorem{thm}{Theorem}[section]

\newtheorem{prop}[thm]{Proposition}
\newtheorem{cor}[thm]{Corollary}
\newtheorem{lem}[thm]{Lemma}
\newtheorem{Theorem}[thm]{Theorem}
\newtheorem{Proposition}[thm]{Proposition}
\newtheorem{Lemma}[thm]{Lemma}

\newtheorem{Remark}[thm]{Remark}
\newtheorem{rem}[thm]{Remark}

\newtheorem{ex}{Example}[section]
\newcommand{\bco}{\begin{cor}}
\newcommand{\eco}{\end{cor}}

\newcommand{\CC}{{\mathbb{C}}}

\newcommand{\HH}{{\mathbb{H}}}
\newcommand{\II}{{\mathbbm{1}}}

\newcommand{\KK}{{\mathbb{K}}}
\newcommand{\LL}{{\mathbb{L}}}

\newcommand{\RR}{{\mathbb{R}}}

\newcommand{\ZZ}{{\mathbb{Z}}}

\newcommand{\bbm}{\begin{Remark}}
\newcommand{\ebm}{\end{Remark}}

\newcommand{\ben}{\begin{enumerate}}
\newcommand{\een}{\end{enumerate}}

\newcommand{\ble}{\begin{Lemma}}
\newcommand{\ele}{\end{Lemma}}

\newcommand{\bpp}{\begin{Proposition}}
\newcommand{\epp}{\end{Proposition}}

\newcommand{\bpr}{\begin{Proposition}}
\newcommand{\epr}{\end{Proposition}}

\newcommand{\bpf}{\begin{proof}}
\newcommand{\epf}{\end{proof}}

\newcommand{\btm}{\begin{Theorem}}
\newcommand{\etm}{\end{Theorem}}

\newcommand{\bre}{\begin{Remark}\rm}
\newcommand{\ere}{\end{Remark}}

\newcommand{\bex}{\begin{ex}\rm}
\newcommand{\eex}{\end{ex}}

\newcommand{\beq}{\begin{equation}}
\newcommand{\eeq}{\end{equation}}

\newcommand{\beqa}{\begin{eqnarray}}
\newcommand{\eeqa}{\end{eqnarray}}

\newcommand{\beqast}{\begin{eqnarray*}}
\newcommand{\eeqast}{\end{eqnarray*}}

\DeclareMathOperator{\I}{I} 
\DeclareMathOperator{\Sp}{Sp} \DeclareMathOperator{\SO}{SO}
\DeclareMathOperator{\OO}{O} \DeclareMathOperator{\SU}{SU}
\DeclareMathOperator{\UU}{U} \DeclareMathOperator{\Sq}{Sq}
\DeclareMathOperator{\BB}{B} \DeclareMathOperator{\sgn}{sgn}
\DeclareMathOperator{\EE}{E}
\DeclareMathOperator{\Hom}{Hom}

\newcommand{\IF}[3]{\I_{#1\,#2}^{\phantom{#1}\,#3}}
\newcommand{\OOF}[2]{\OO_{#1}^{#2}}
\newcommand{\SOF}[2]{\SO_{#1}^{#2}}
\newcommand{\UUF}[2]{\UU_{#1}^{#2}}
\newcommand{\SpF}[2]{\Sp_{#1}^{#2}}

\newcommand{\typeSHp}{[\mr SH^+]}
\newcommand{\typeSHm}{[\mr SH^-]}
\newcommand{\typeSHpm}{[\mr SH^\pm]}

\newcommand{\GL}{\mathrm{GL}}

\newcommand{\CPt}{\CC\mr P^2}
\newcommand{\StS}{\mr S^2\times\mr S^2}
\newcommand{\caseCPt}{M=\CC\mr P^2}
\newcommand{\caseStS}{M=\mr S^2\times\mr S^2}

\newcommand{\mf}[1]{\mathfrak{#1}}
\newcommand{\mr}[1]{\mathrm{#1}}

\newcommand{\comment}[1]{}
\newcommand{\ve}{\varepsilon}

\newcommand{\vp}{\varphi}

\newcommand{\todo}[1]{}

\newcommand{\ra}{\rightarrow}
\newcommand{\into}{\hookrightarrow}
\newcommand{\Z}{\mathbb{Z}}
\newcommand{\R}{\mathbb{R}}
\newcommand{\C}{\mathbb{C}}
\newcommand{\bewe}{\hfill $\Box$}
\newcommand{\im}{\mathrm{im}}
\newcommand{\id}{\mathrm{id}}
\newcommand{\lra}{\longrightarrow}
\newcommand{\A}{\mathcal{A}}
\newcommand{\M}{\mathcal{M}}
\newcommand{\G}{\mathcal{G}}

\newcommand{\rref}[1]{{\rm \ref{#1}}}
\newcommand{\Rmap}{\frak R}   
\newcommand{\ol}{\overline}
\newcommand{\stimes}{}

\newcommand{\punkt}[1]{\put(#1){\circle*{0.06}}}
\newcommand{\marke}[3]{\put(#1){\put(0.05,0.1){\makebox(-0.1,-0.2)[#2]{\tiny
   $#3$}}}}

\newcommand{\linie}[3]{\put(#1){\line(#2){#3}}}

\newcommand{\whole}[3]{\put(#1){\punkt{0,0}\put(0.05,0.1){\makebox(-0.1,
   -0.2)[#2]{\tiny $#3$}}}}

\newcommand{\lri}[3]{\put(#1){\punkt{0,0}\linie{0.1,0}{1,0}{0.8}
\put(0.05,0.1){\makebox(-0.1,-0.2)[#2]{\tiny $#3$}}}}
\newcommand{\lrii}[3]{\put(#1){\punkt{0,0}\linie{0.1,0}{1,0}{1.8}
\put(0.05,0.1){\makebox(-0.1,-0.2)[#2]{\tiny $#3$}}}}

\newcommand{\lori}[3]{\put(#1){\punkt{0,0}\linie{0.0894,0.0447}{2,1}{0.821}
\put(0.05,0.1){\makebox(-0.1,-0.2)[#2]{\tiny $#3$}}}}
\newcommand{\lorii}[3]{\put(#1){\punkt{0,0}\linie{0.0894,0.0447}{2,1}{1.821}
\put(0.05,0.1){\makebox(-0.1,-0.2)[#2]{\tiny $#3$}}}}

\newcommand{\luri}[3]{\put(#1){\punkt{0,0}\linie{0.0894,-0.0447}{2,-1}{
0.821}\put(0.05,0.1){\makebox(-0.1,-0.2)[#2]{\tiny $#3$}}}}
\newcommand{\lurii}[3]{\put(#1){\punkt{0,0}\linie{0.0894,-0.0447}{2,-1}{
1.821}\put(0.05,0.1){\makebox(-0.1,-0.2)[#2]{\tiny $#3$}}}}

\newcommand{\lorri}[3]{\put(#1){\punkt{0,0}\linie{0.0968,0.0242}{4,1}{1.8064}
\put(0.05,0.1){\makebox(-0.1,-0.2)[#2]{\tiny $#3$}}}}

\newcommand{\loori}[3]{\put(#1){\punkt{0,0}\linie{0.0707,0.0707}{1,1}{0.8586}
\put(0.05,0.1){\makebox(-0.1,-0.2)[#2]{\tiny $#3$}}}}

\newcommand{\luuri}[3]{\put(#1){\punkt{0,0}\linie{0.0707,-0.0707}{1,-1}{
0.8586}\put(0.05,0.1){\makebox(-0.1,-0.2)[#2]{\tiny $#3$}}}}

\newcommand{\lurrri}[3]{\put(#1){\punkt{0,0}\linie{0.0986,-0.0164}{6,-1}{
2.8028}\put(0.05,0.1){\makebox(-0.1,-0.2)[#2]{\tiny $#3$}}}}

\newcommand{\looori}[3]{\put(#1){\punkt{0,0}\linie{0.0554,0.0832}{2,3}{0.8892}
\put(0.05,0.1){\makebox(-0.1,-0.2)[#2]{\tiny $#3$}}}}

\newcommand{\luuuri}[3]{\put(#1){\punkt{0,0}\linie{0.0554,-0.0832}{2,-3}{
0.8892}\put(0.05,0.1){\makebox(-0.1,-0.2)[#2]{\tiny $#3$}}}}

\newcommand{\loorrri}[3]{\put(#1){\punkt{0,0}\linie{0.0949,0.0316}{3,1}{2.8102}
\put(0.05,0.1){\makebox(-0.1,-0.2)[#2]{\tiny $#3$}}}}

\newcommand{\luurrri}[3]{\put(#1){\punkt{0,0}\linie{0.0949,-0.0316}{3,-1}{
2.8102}\put(0.05,0.1){\makebox(-0.1,-0.2)[#2]{\tiny $#3$}}}}

\begin{document}

\title{\bf Gauge Orbit Types for Theories with \\ Classical Compact Gauge
Group}

\author{
    A.\ Hertsch, G.\ Rudolph and M.\ Schmidt\\
    Institut f\"ur Theoretische Physik, Universit\"at Leipzig\\
    Augustusplatz 10/11, 04109 Leipzig, Germany\\
    }

\maketitle

\comment{
\vspace{2cm}

{\bf Keywords:}~
\\

{\bf MSC:}~ 70G65, 70S15
}
\vspace{2cm}

\begin{abstract}

\noindent
We determine the orbit types of the action of the group of local gauge
transformations on the space of connections in a principal bundle with structure
group $\OO(n)$, $\SO(n)$ or $\Sp(n)$ over a closed, simply connected manifold of
dimension $4$. Complemented with earlier results on $\UU(n)$ and $\SU(n)$ this
completes the classification of the orbit types for all classical compact gauge
groups over such space-time manifolds. On the way we derive the classification
of principal bundles with  structure group $\SO(n)$ over these manifolds and the
Howe subgroups of $\SO(n)$.

\end{abstract}

\newpage

\tableofcontents


\section{Introduction}


The principle of local gauge invariance plays a fundamental role in modern
theoretical physics. Its application to the theory of particle interactions
gave rise to the standard model, which proved to be successful from both
theoretical and phenomenological points of view. There is a lot of important
results obtained within perturbation theory, which works well for high energy
processes. On the other hand, the low energy hadron physics, in particular, the
quark confinement, turns out to be dominated by nonperturbative effects, for
which there is no rigorous theoretical explanation yet. To study nonperturbative
aspects, a variety of different concepts and mathematical methods has been
developed.

One of the main reasons, which makes a nonabelian gauge theory so different from other field theories is
the rich mathematical structure of its classical configuration space (the gauge orbit space) and the
corresponding phase space. The gauge orbit space is obtained by factorizing the infite-dimensional affine space of gauge potentials by the action of the group of local gauge transformations. Since this action is not free, non-generic orbit types (singularities) occur endowing the gauge orbit space with a stratified structure.
Let us discuss some aspects indicating its physical relevance.

First, the geometry and topology of the generic (principal) stratum has been
clarified a long time ago \cite{NaRa,Singer:Gribov}.
In particular, one gets an intrinsic topological interpretation of the
Gribov-ambiguity \cite{Gribov}. We stress that the problem of
finding all Gribov copies has been discussed within specific models, see e.g.\
\cite{Langmann}. For a detailed analysis in the case of $2$-dimensional
cylindrical space time (including the Hamiltonian path integral) we refer to
\cite{Shabanov1}. Investigating the topology of the determinant line bundle over
the generic stratum, one gets an understanding of anomalies in terms of the
family index theorem \cite{Alvarez, Atiyah/Singer}, see also \cite{Carey} for
the Hamiltonian approach. In particular, one gets anomalies of purely
topological type \cite{Witten} which cannot be seen by perturbative quantum
field theory.

There are partial results and conjectures concerning the relevance of nongeneric
strata. First, generally speaking, nongeneric gauge orbits affect the
classical motion on the orbit space due to boundary conditions
and, in this way, they may produce nontrivial contributions to the path
integral. They may lead to localization of certain quantum
states, as it was suggested by finite-dimensional examples
\cite{EmmrichRoemer}, see also \cite{HRS} for a finite-dimensional gauge model.
Further, the gauge field configurations
belonging to nongeneric orbits can possess a magnetic charge, i.e.
they can be considered as a kind of magnetic monopole
configurations. Following t'Hooft \cite{tHooft}, these could be responsible for
quark confinement. The role of these configurations was investigated within the
framework of Schr\"odinger quantum mechanics on the gauge orbit space of
topological Chern-Simons theory in \cite{Asorey:Nodes}, see also
\cite{Asorey:99} for an approach to 4-dimensional Yang-Mills theories with
$\theta$-term. Within t'Hooft's concept, the idea of abelian projection is of
special importance and has been
discussed by many authors. This concept was studied within the setting of
quantum field theory at finite temperature on the $4$-torus in \cite{Tok1,Tok2}.
There, a hierarchy of defects, which should be related to the gauge orbit space
structure, was discovered. Finally, let us also mention that in
\cite{Heil:Anom} the existence of additional anomalies corresponding to
non-generic strata was suggested.

Most of the problems mentioned here are still awaiting a
systematic investigation. For that purpose, a deeper insight into the structure
of the gauge orbit space is necessary. The stratified structure of the
full gauge orbit space was investigated in detail in \cite{KoRo}.
Based on that, in a series of papers \cite{RSV:clfot,RSV:poot,RS:otG}, see also
\cite{RSV:review} for a review, we have given a complete solution to the problem
of determining the strata (including their partial ordering) for gauge theories
with structure group $\SU(n)$ on closed connected manifolds of dimension
$d=2,3,4$. Our analysis was based in particular on the observation, made e.g.\
in \cite{KoRo}, that  orbit types are in bijective correspondence with a certain
type of bundle reductions of the principal bundle underlying the gauge theory
under consideration. We call bundle reductions of this type holonomy-induced
Howe subbundles. This
observation yields a method for solving the classification problem for orbit
types, because it turns out that holonomy-induced Howe subbundles can be
classified by methods of algebraic topology. In the present paper we extend the
above results to the case of structure group $\OO(n)$, $\SO(n)$ or $\Sp(n)$.
We apply the same general method, but the classification problem for the
subbundles turns out to be much harder than in the case of $\SU(n)$.

Let us put this paper into a broader perspective. In a next step, we will
investigate the structure of the phase space (infinite-dimensional Hamiltonian
system with symmetry) by applying the method of singular
Marsden-Weistein reduction. This will yield the reduced phase
space as a Hamiltonian system with singularities, which can be
taken as a starting point for constructing the quantum theory.
To separate the geometric problems arising from the reduction process from the
analytic problems due to infinite dimensions, it is reasonable to study
finite-dimensional approximations (provided by lattice gauge theory) of the
above situation. We refer to \cite{CKRS,CRS,FRS,HRS} for first results in this
direction. In particular, in \cite{HRS} we have studied the role of
non-generic strata on the quantum level for one of these finite-dimensional
models.

The paper is organized as follows: In Section \ref{method} we
present the method and formulate the program for solving the
classification problem. In Section \ref{S-HSG} we determine the
Howe subgroups of $\OO(n)$, $\SO(n)$ and $\Sp(n)$. In Section \ref{S-HFB} we
classify principal bundles whose structure group is a Howe
subgroup under some assumptions concerning the space time manifold.
This chapter is the heart of the paper. In Section
\ref{S-HSB} we determine the Howe subbundles of a given principal
bundle and in Section \ref{S-holind} we specify the Howe
subbundles which are holonomy-induced. Section \ref{S-factoriz}
contains the factorization with respect to the action of the
structure group on bundle reductions and in Section
\ref{S-summary} we summarize the classification result. As an illustration, in
Section \ref{S-examples} we apply the results to gauge theories with structure
group $\OO(4)$, $\SO(4)$ or $\Sp(2)$, defined over a space-time manifold
diffeomorphic to $\CPt$ or $\StS$.
For the convenience of the reader, the paper is supplemented by 5 Appendices,
where we collect relevant material from algebraic topology referred to in the
text.

Finally, we note that we restrict attention to closed (compact, without
boundary) and simply connected space time manifolds of dimension $4$.
However, the topological results derived in Section \ref{S-HFB} are obtained for
$CW$-complexes, some of them hold true without the above assumptions.


\section{Orbit types and holonomy-induced Howe subbundles}
\label{method}


Let $M$ be a closed connected orientable manifold, let $G$ be a
compact connected Lie-group and let $P$ be a principal bundle over
$M$ with structure group $G$. We denote the affine Hilbert-space
of connection forms on $P$ of an appropriate Sobolev class $k$ by
$\A$ the Hilbert-Lie group of vertical automorphisms of $P$ of
Sobolev class $(k+1)$ by $\G$. For the analytic framework, see
\cite{NaRa,Singer:Gribov}. The results we are going to derive are
independent of the choice of $k$ provided $k$ is large enough.

Our aim is to investigate the structure of the configuration space of a gauge
theory on $P$,
\[
\M=\A/\G \, ,
\]
where the action of $\G$ on $\A$ is given by
\[
(g,A)\mapsto A^{(g)}= g^{-1}Ag +g^{-1}\mathrm{d}g.
\]
This space is known as the gauge orbit space. Since, in general, the
action of $\G$ is not free, $\M$ is not a smooth manifold. However, the action
is proper and admits slices \cite{KoRo}. Therefore, $\M$ is a stratified
space, where the stratification is induced by the orbit types of the action of
$\G$ on $\A$. Let us briefly
recall the construction. The stabiliser (or isotropy group) of a connection
$A\in \A$ is given by $\G_A=\{g\in \G\ |\ A^{(g)}=A\}$. It transforms under
the action of $\G$ like $\G_{A^{(g)}}=g^{-1}\G_A g$. This allows to define
the type of an orbit $[A]$ to be the conjugacy class of the subgroup $\G_A$ in
$\G$. For given orbit type $\tau$ let $\M_\tau$ denote the subset of $\M$ of
orbits of type $\tau$. This defines a disjoint decomposition
 \beq\label{G-otdeco}
\M = \bigcup_{\text{ orbit types}} \M_\tau\,.
 \eeq
The following was shown in \cite{KoRo}:

--~  each $\M_\tau$ carries the structure of a Hilbert manifold,

--~ the orbit type deomposition \eqref{G-otdeco} is locally
finite,

--~ the frontier condition holds: if $\M_\tau\cap\ol{\M_\sigma} \neq\emptyset$
then $\M_\tau\subset\ol{\M_\sigma}$.

In \cite{KoRo:Strat,KoRo}, the decomposition \eqref{G-otdeco} is therefore
called a stratification and the orbit type subsets $\M_\tau$ are referred to as
the strata of $\M$. We adopt this
terminology\footnote{There exist several stronger notions of
stratification.}. It was also shown in \cite{KoRo} that the
natural partial ordering of orbit types, which is induced from the
inclusion relation between subgroups of $\G$, satisfies
$$
\tau\leq\sigma
 ~~~\Leftrightarrow~~~
\M_\sigma\subseteq\ol{\M_\tau}\,.
$$
Thus, the the partially ordered set of orbit types contains
information about which strata occur in the gauge theory on $P$ and
how they are glued together. In the present paper we determine the
orbit types for structure groups $G=\OO(n)$, $\SO(n)$ and $\Sp(n)$
and for $M$ being a closed simply connected manifold of dimension
$4$; the partial ordering will be studied in a separate paper. For
that purpose, we use a relation between orbit types and certain
types of bundle reductions of $P$ which for the first time was
described in detail in \cite{KoRo}. Let us explain this relation. A
subgroup $H\subseteq G$ with the property $C^2_G(H)=H$ is called
\emph{Howe subgroup}. For any subgroup $H$ of $G$ there exists a
smallest Howe subgroup $\tilde H$ of $G$ such that
$H\subseteq\tilde H$. We say that $\tilde H$ is the Howe subgroup
generated by $H$. $\tilde H$ can be obtained by taking the double
centralizer, $\tilde H = \mr C^2_G(H)$. A reduction of $P$ to a
Howe subgroup is called \emph{Howe subbundle}. For any reduction
$Q$ of $P$ to a subgroup $H$ there exists a smallest Howe
subbundle $\tilde Q$ such that $Q\subseteq \tilde Q$. We say that
$\tilde Q$ is the Howe subbundle generated by $Q$. $\tilde Q$ can
be obtained from $Q$ by extending the structure group to the Howe
subgroup generated by $H$, i.e., $\tilde Q = Q \cdot \mr C_G(H)$.
A Howe subbundle is called \emph{holonomy-induced} if it is
generated by the holonomy subbundle of some connection in $P$. Due
to existence results for holonomy subbundles \cite[Ch.\ II, Thm.\
8.2]{nomizu} in dimension $d\geq 2$ this is equivalent to
requiring that the Howe subbundle be generated by a connected
reduction of $P$.

 \btm[Kondracki and Rogulski \cite{KoRo}]\label{T-orbit}

If $\dim M > 1$, there is an (order-preserving) bijection from the
set of orbit types onto the set of isomorphism classes of
holonomy-induced Howe subbundles, factorized by the action of the
structure group $G$ on bundle reductions.

 \etm

For a detailed proof, see \cite{RSV:clfot}. The idea of the proof
is based on the observation that when gauge transformations are
viewed as equivariant maps from $P$ to $G$, the stabilizer of a
connection consists of the gauge transformations which are
constant on any holonomy subbundle of that connection. It follows
that the stabilizer is the same for any two connections whose
holonomy subbundles (based at the same point) generate the same
Howe subbundle. Therefore, stabilizers correspond to
holonomy-induced Howe subbundles. Passing to classes on the level
of the subbundles then corresponds to passing to conjugacy classes
on the level of the stabilizers.
 \medskip

According to Theorem \ref{T-orbit}, in order to determine the
orbit types of the action of $\G$ on $\A$ we have to do the
following.
 \medskip

1.~ Determine the Howe subgroups of $G$.

2.~ Determine the Howe subbundles of $P$ up to isomorphy.

3.~ Specify the Howe subbundles which are holonomy-induced.

4.~ Factorize by the action of the structure group $G$ on bundle
reductions.
 \medskip

To work out the full program for $G=\OO(n)$, $\SO(n)$ or $\Sp(n)$
we have to assume that $M$ is simply connected and of dimension
$4$. Step 1 will be treated in Section \rref{S-HSG}. Step 2 is the
hardest one, it will be worked out in Sections \rref{S-HFB} and
\rref{S-HSB}. Steps 3 and 4 are treated in Sections
\rref{S-holind} and \rref{S-factoriz}, respectively. The results
are summarized in Section \rref{S-summary}.

As mentioned above, for structure group $G=\SU(n)$ the orbit types
and their partial ordering have already been determined in
\cite{RSV:clfot,RSV:poot}, see also \cite{RS:otG}. Here the
assumption that $M$ be simply connected can be dropped. For the sake of
completeness, we include these results into the summary in Section
\rref{S-summary}.

\bre\label{Bem-dimension}

The assumption that $\dim(M)$ be exactly $4$ is just made for simplicity.
Our results on the classification of Howe subbundles hold also in dimension
$d\leq 4$. In view of Theorem \rref{T-orbit}, we thus derive the orbit types
for $\dim (M) = 2,3,4$.

\ere


\section{Howe subgroups}
\label{howe1} \label{S-HSG}


Let $\KK = \RR,\CC,\HH$ denote the real numbers, complex numbers
and quaternions, respectively. Let $\I_\KK(n)$ denote the isometry
groups of the standard scalar products on $\KK^n$. The standard notation is
obtained by replacing $\I_\RR = \OO$, $\I_\CC = \UU$ and $\I_\HH = \Sp$. For a
group $G$ and a subgroup $H\subseteq G$, let $\mr
C_G(H)$ and $\mr N_G(H)$ denote the centralizer and the
normalizer, respectively, of $H$ in $G$.

The study of the Howe subgroups of $\OO(n)$ and $\Sp(n)$ amounts
to an application of the theory of finite-dimensional real and
quaternionic von Neumann algebras, respectively. We cite the results from
\cite{schmidt} 
where the more general case of the isometry group of a Hermitian
form over $\KK$ was treated. The Howe subgroups of $\SO(n)$ can be
derived from those of $\OO(n)$. Since to our knowledge this is not
documented in the literature, we give full proofs here. For this
discussion, and also for the discussion of the property of a
bundle reduction to be holonomy-induced in Section
\rref{S-holind}, we need information about the inclusion relations
between Howe subgroups of $\OO(n)$ and $\Sp(n)$. Therefore, for
these groups, from \cite{schmidt} we also cite the operations
producing the direct predecessors (up to conjugacy) of a given
Howe subgroup w.r.t.\ the natural partial ordering defined by the
inclusion relation.

Let $G = \I_\KK(n)$ be given. A Howe subgroup $H$ of $G$ is called
irreducible if the representation of $H\times\mr C_G(H)$ on
$\KK^n$ is irreducible. Orthogonally decomposing $\KK^n =
\KK^{n_1}\oplus\cdots\oplus \KK^{n_r}$ into $H\times\mr
C_G(H)$-irreducible subspaces one obtains a decomposition of $H$
into a direct product of irreducible Howe subgroups of
$\I_\KK(n_i)$. The centralizer of $H$ is then the direct product
of the centralizers of the irreducible factors of $H$ in the
groups $\I(n_i)$.


\subsection{Howe subgroups of $\OO(n)$}
\label{S-HSG-O}


For any positive integer $m$, field restriction defines homomorphisms
$$
\vp_{\CC,\RR} : \mr M_\CC(m) \to \mr M_\RR(2m)
 \,,~~~~~~
\vp_{\HH,\CC} : \mr M_\HH(m) \to \mr M_\CC(2m)
 \,,~~~~~~
\vp_{\HH,\RR} : \mr M_\HH(m) \to \mr M_\RR(4m)
 \,.
$$
For convenience, we set $\vp_{\RR,\RR} = \id_{\mr M_\RR(m)}$.
Explicitly, we choose
 \begin{align}
\vp_{\CC,\RR}(a_1 + \mr i a_2)
 & =
\left[ \begin{array}{cc}
 a_1 & a_2 \\ -a_2 & a_1
\end{array} \right]
 \,,~~~~~~ &
a_k\in\mr M_\RR(m)\,,
\\
\vp_{\HH,\CC}(a_1 + a_2\mr j)
 & =
\left[ \begin{array}{cc} \ol{a_1} & -\ol{a_2} \\ a_2 & a_1
\end{array} \right]
 \,,~~~~~~ &
a_k\in\mr M_\CC(m)\,,
 \end{align}
and $\vp_{\HH,\RR} := \vp_{\CC,\RR} \circ \vp_{\HH,\CC}$. I.e.,
 \beq\label{G-embHR}
\vp_{\HH,\RR}(a_1 + \mr i a_2 + \mr j a_3 + \mr k a_4)
 =
\left[ \begin{array}{cccc}
 a_1 & -a_2 & -a_3 & -a_4
 \\
 a_2 & a_1 & a_4 & -a_3
 \\
 a_3 & -a_4 & a_1 & a_2
 \\
 a_4 & a_3 & -a_2 & a_1
\end{array} \right]
 \,,~~~~~~
a_k\in\mr M_\RR(m)\,.
 \eeq
These mappings commute with taking adjoints w.r.t.\ the respective scalar
product. Therefore, they induce Lie group embeddings (denoted by the same
symbol)
$$
\vp_{\CC,\RR} : \UU(m)\to\OO(2m)
 \,,~~~~~~
\vp_{\HH,\CC} : \Sp(m)\to\UU(2m)
 \,,~~~~~~
\vp_{\HH,\RR} : \Sp(m)\to\OO(4m)\,.
$$
Since $\UU(m)$ is connected, $\vp_{\CC,\RR}(\UU(m))\subseteq\SO(2m)$ and hence
$\vp_{\HH,\RR}(\Sp(n))\subseteq\SO(4m)$. Moreover,
$\vp_{\HH,\CC}(\Sp(n))\subseteq\SU(2m)$.

The conjugacy classes of irreducible Howe subgroups of $\OO(n_i)$
are in bijective correspondence with the solutions of the equation
$\delta_i m_i k_i = n_i$, where $(m_i,k_i)$ is an ordered pair of
positive integers and $\delta_i = 1,2,4$. For given
$\delta_i,m_i,k_i$, a representative for the corresponding class
of Howe subgroups is given by $H_i = \IF{\KK_i}{m_i}{k_i}$ where
 \beq\label{G-irrHSG-O-1}
\IF{\KK_i}{m_i}{k_i}
 =
\left\{\vp_{\KK,\RR}(a) \otimes_\RR \II_{k_i} ~:~ a\in\I_{\KK_i}(m_i)\right\}
 \eeq
w.r.t.\ the decomposition $\RR^{n_i} = \RR^{\delta_i m_i} \otimes_\RR \RR^{k_i}$
or, alternatively,
 \beq\label{G-irrHSG-O-2}
\IF{\KK_i}{m_i}{k_i} =
 \left\{
\vp_{\KK,\RR}(a) \oplus\stackrel{k_i}{\cdots}\oplus \vp_{\KK,\RR}(a)
 ~:~
a\in\I_{\KK_i}(m_i)
 \right\}
 \eeq
w.r.t.\ the decomposition
 $
\RR^{n_i}
 =
\RR^{\delta_i m_i} \oplus\stackrel{k_i}{\cdots}\oplus \RR^{\delta_i m_i}
$.
Here $\KK_i = \RR$ for $\delta_i = 1$, $\KK_i = \CC$ for $\delta_i = 2$
and $\KK_i = \HH$ for $\delta_i = 4$. We call $\KK_i$ the base field, $m_i$ the
rank and $k_i$ the multiplicity of $H_i$. Moreover, we denote $\OOF{m_i}{k_i} =
\IF{\RR}{m_i}{k_i}$, $\UUF {m_i}{k_i} = \IF{\CC}{m_i}{k_i}$ and $\SpF {m_i}{k_i}
= \IF{\HH}{m_i}{k_i}$. A representative for the conjugacy
class of the centralizer of $H_i$ is given by $\IF{\KK_i}{k_i}{m_i}$ (i.e., rank
and multiplicity are interchanged).
Thus, up to conjugacy, the Howe subgroups of
$\OO(n)$ are given by the direct products
 \beq\label{G-HSG-O}
H = \IF{\KK_1}{m_1}{k_1} \times\cdots\times \IF{\KK_r}{m_r}{k_r}
 \,,~~~~~~
\sum\nolimits_{i=1}^r \dim_\RR\KK_i \, m_i k_i = n
 \,,~~~~~~
r=1,2,3,\dots\,.
 \eeq
where the factors are embedded by virtue of the homomorphisms
$\vp_{\KK_i,\RR}$. The factors are referred to as
$\OO$-factors, $\UU$-factors and $\Sp$-factors according to the base field
being $\RR$, $\CC$ or $\HH$, respectively.
A representative for the conjugacy class of the centralizer of $H$ is
obtained from $H$ by interchanging ranks and multiplicities.
The subgroup $\IF\RR 1n\equiv \OOF1n$ coincides
with the center of $\OO(n)$ and the subgroup $\IF\RR n1\equiv\OOF
n1$ coincides with $\OO(n)$ itself.

 \comment{

Das ist nur noetig, wenn man den Zentralisator direkt (unkonjugiert)
aufschreiben will:

In addition, in case of $\mr I=\Sp$, $\vp_\Sp$ has to be replaced by the map
$\psi_\Sp \colon \Sp(k) \to \OO(4k)$, defined as
$$
\psi_{\Sp}(a_1 + \mr i a_2 + \mr j a_3 + \mr k a_4)
 =
\left[ \begin{array}{cccc}
 a_1 & a_2 & a_3 & -a_4
 \\
 -a_2 & a_1 & -a_4 & -a_3
 \\
 -a_3 & a_4 & a_1 & a_2
 \\
 a_4 & a_3 & -a_2 & a_1
\end{array} \right]\,.
$$
Notice that $\vp_\Sp$ is constructed from the action of
$\GL(s,\HH)$ on $\HH^s$ by left multiplication while $\psi_\Sp$ is
constructed from the action by right multiplication by the adjoint
matrix.
 }%

The operations which produce the direct predecessors of a given
Howe subgroup of $\OO(n)$ (up to conjugacy) are

--~ {\it Merging:}~ replacing a double factor $\IF\KK m{k_1} \times
\IF\KK m{k_2}$ by a single factor $\IF\KK m{k_1+k_2}$; $\I_\KK(m)$
is diagonally embedded into $\I_\KK(m)\times \I_\KK(m)$;

--~ {\it Splitting:}~ replacing a factor $\IF\KK mk$ with $m>1$
by the double factor $\IF\KK{m_1}k\times\IF\KK{m_2}k$, where $m_1 + m_2 = m$;
$\I_\KK(m_1) \times \I_\KK(m_2)$ is embedded into $\I_\KK(m)$ in the obvious
way;

--~ {\it Inverse field restriction:}~ replacing a factor $\OOF{2m}k$
or $\UUF{2m}k$ by a factor $\UUF mk$ or $\SpF mk$, embedded via $\vp_{\CC,\RR}$
or $\vp_{\HH,\CC}$, respectively;

--~ {\it Inverse field extension:}~ replacing a factor $\UUF mk$
or $\SpF mk$ by a factor $\OOF m{2k}$ or $\UUF m{2k}$, embedded in the obvious
way.

 \comment{ Ohne die Vfh in der Bezeichnung:

--~ {\it Merging:}~ replacing a double factor $\I_{\KK}(m) \times
\I_{\KK}(m)$ with multiplicities $k_1$ and $k_2$ by a single
factor $\I_\KK(m)$ with multiplicity $k = k_1 + k_2$; $\I_\KK(m)$
is diagonally embedded into $\I_\KK(m)\times \I_\KK(m)$;

--~ {\it Splitting:}~ replacing a factor $\I_\KK(m)$ with $m>1$
and multiplicity $k$ by the double factor
$\I_\KK(m_1)\times\I_\KK(m_2)$ with multiplicities $k_1 = k_2 =
k$, where $m_1 + m_2 = m$; $\I_\KK(m_1) \times \I_\KK(m_2)$ is
embedded into $\I_\KK(m)$ in the obvious way;

--~ {\it Inverse field restriction:}~ replacing a factor $\OO(2m)$
or $\mr U(2m)$ by a factor $\mr U(m)$ or $\Sp(m)$ of the same
multiplicity, embedded via $\vp_{\CC,\RR}$ or $\vp_{\HH,\CC}$,
respectively;

--~ {\it Inverse field extension:}~ replacing a factor $\mr U(m)$
or $\mr Sp(m)$ of multiplicity $k$ by a factor $\mr O(m)$ or $\mr
U(m)$ of multiplicity $2k$, respectively, embedded in the obvious
way.
 }

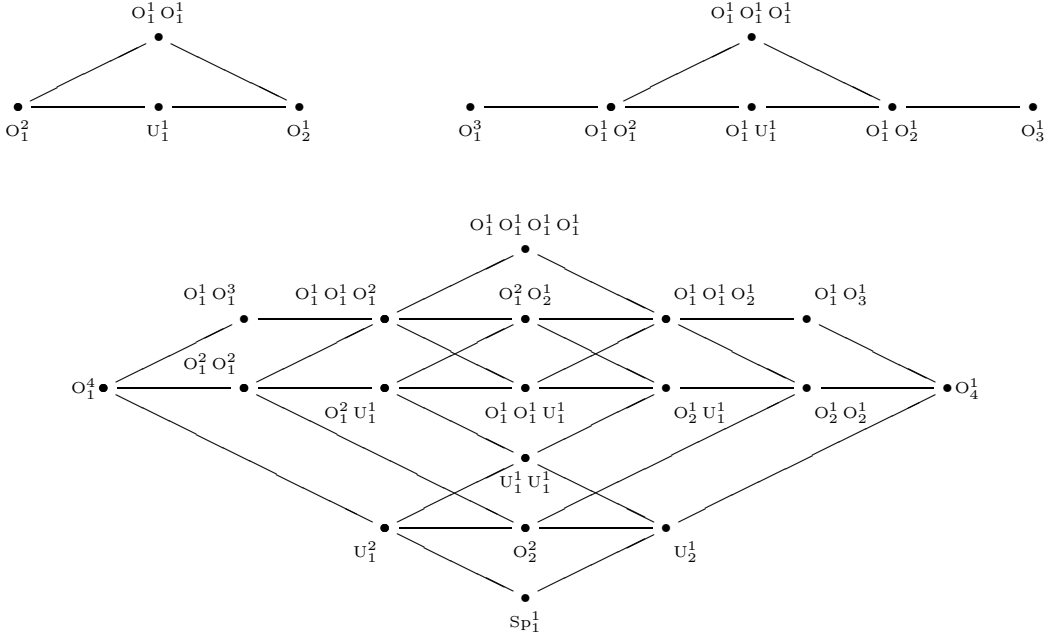
\begin{figure}

\begin{center}

\unitlength1.85cm

\begin{picture}(2,1.5)
\put(0,0.5){
 \lori{0,0}{tc}{\OOF12}
 \lri{0,0}{tc}{}
 \lri{1,0}{tc}{\UUF11}
 \luri{1,0.5}{bc}{\OOF11\stimes\OOF11}
 \whole{2,0}{tc}{\OOF21}
 }
\end{picture}
 \hspace{2cm}
\begin{picture}(4,1.5)
\put(0,0.5){
 \lri{0,0}{tc}{\OOF13}
 \lori{1,0}{tc}{\OOF11\stimes\OOF12}
 \lri{1,0}{tc}{}
 \lri{2,0}{tc}{\OOF11\stimes\UUF11}
 \luri{2,0.5}{bc}{\OOF11\stimes\OOF11\stimes\OOF11}
 \lri{3,0}{tc}{\OOF11\stimes\OOF21}
 \whole{4,0}{tc}{\OOF31}
 }
\end{picture}

\begin{picture}(6,3.5)
\put(0,2){
 \lori{0,0}{cr}{\OOF14}
 \lri{0,0}{}{}
 \lurii{0,0}{}{}

 \lri{1,0.5}{br}{\OOF11\stimes\OOF13}
 \lori{1,0}{br}{\OOF12\stimes\OOF12}
 \lri{1,0}{}{}
 \lurii{1,0}{}{}

 \lori{2,0.5}{br}{\OOF11\stimes\OOF11\stimes\OOF12}
 \lri{2,0.5}{}{}
 \luri{2,0.5}{}{}
 \lori{2,0}{tr}{\OOF12\stimes\UUF11}
 \lri{2,0}{}{}
 \luri{2,0}{}{}
 \lori{2,-1}{tr}{\UUF12}
 \lri{2,-1}{}{}
 \luri{2,-1}{}{}

 \luri{3,1}{bc}{\OOF11\stimes\OOF11\stimes\OOF11\stimes\OOF11}
 \lri{3,0.5}{bc}{\OOF12\stimes\OOF21}
 \luri{3,0.5}{}{}
 \lori{3,0}{tc}{\OOF11\stimes\OOF11\stimes\UUF11}
 \lri{3,0}{}{}
 \lori{3,-0.5}{tc}{\UUF11\stimes\UUF11}
 \luri{3,-0.5}{}{}
 \lorii{3,-1}{tc}{\OOF22}
 \lri{3,-1}{}{}
 \lori{3,-1.5}{tc}{\SpF11}

 \lri{4,0.5}{bl}{\OOF11\stimes\OOF11\stimes\OOF21}
 \luri{4,0.5}{}{}
 \lri{4,0}{tl}{\OOF21\stimes\UUF11}
 \lorii{4,-1}{tl}{\UUF21}

 \luri{5,0.5}{bl}{\OOF11\stimes\OOF31}
 \lri{5,0}{tl}{\OOF21\stimes\OOF21}

 \whole{6,0}{cl}{\OOF41}
 }
\end{picture}

\end{center}

\caption{\label{F-HSG-O} Hasse diagrams of the sets of conjugacy
classes of Howe subgroups of $\OO(2)$, $\OO(3)$, $\OO(4)$. For
brevity, direct product signs are omitted, i.e., $\OOF11\OOF11$
stands for $\OOF11\times\OOF11$ etc.}

\end{figure}

As an example, the Hasse diagrams of the sets of conjugacy classes
of Howe subgroups of $\OO(2)$, $\OO(3)$ and $\OO(4)$ are displayed
in Figure \rref{F-HSG-O}.

We will also have to study the identity connected component $H_0$ of a Howe
subgroup $H$. One has
$$
H_0
 =
(\IF{\KK_1}{m_1}{k_1})_0
 \times\cdots\times
(\IF{\KK_r}{m_r}{k_r})_0\,,
$$
where $(\IF{\KK_i}{m_i}{k_i})_0$ denotes the identity connected component of
the factor $\IF{\KK_i}{m_i}{k_i}\,$. For $\KK_i=\RR$, this is
$(\IF{\RR}{m_i}{k_i})_0 \equiv \SOF{m_i}{k_i}$, whereas for $\KK_i = \CC$, $\HH$,
it coincides with $\IF{\KK_i}{m_i}{k_i}$. Thus, $H_0$ consists of
$\SO$-factors, $\UU$-factors and $\Sp$-factors.


\subsection{Howe subgroups of $\Sp(n)$}
\label{S-HSG-Sp}


We use the convention that on $\HH^n$, scalars act by
multiplication from the right and endomorphisms by matrix
multiplication from the left. The conjugacy classes of irreducible
Howe subgroups of $\Sp(n_i)$ are in bijective correspondence with
ordered pairs of positive integers $(m_i,k_i)$ such that $m_i k_i
= n_i$, together with a choice of $\KK_i=\RR,\CC,\HH$. For given
$(m_i,k_i)$ and $\KK_i$, a representative for the corresponding
class of Howe subgroups is given by $H_i=\IF{\KK_i}{m_i}{k_i}$,
where
 \beq\label{G-irrHSG-Sp-1}
\IF{\KK_i}{m_i}{k_i} = \left\{a \otimes_\HH \II_{k_i}
 ~:~
a\in\I_{\KK_i}(m_i)\right\}
 \eeq
w.r.t.\ the decomposition $\HH^{n_i} = \HH^{m_i} \otimes_\HH \HH^{k_i}$ or,
alternatively,
 \beq\label{G-irrHSG-Sp-2}
\IF{\KK_i}{m_i}{k_i} =
 \left\{
a \oplus\stackrel{k_i}{\cdots}\oplus a
 ~:~
a\in\I_{\KK_i}(m_i)
 \right\}
 \eeq
w.r.t.\ the decomposition
 $
\HH^{n_i}
 =
\HH^{m_i} \oplus\stackrel{k_i}{\cdots}\oplus \HH^{m_i}
 $.
Recall that the tensor product over $\HH$ is defined w.r.t.\ the
actions of $\HH$ by right multiplication on the first factor and
by left multiplication on the second factor. Thus, operators of
the type $a\otimes_\HH b$ can be defined provided the entries of
$b$ are central, i.e., real.
 \comment{

insbesondere ist der Ztr in der TePr-Dst schwer zu erkennen, ist
zumindest nicht $\II_m\otimes a$, da das nur fuer reelle $a$
erklaert ist. Lasse evtl. TePr-Dst bei Sp(n) weg und stelle
folgenden Text auf die Summen-Darstellung um.

 }%
We will use the same terminology as in the case of $\OO(n)$, i.e.,
$\KK_i$ will be referred to as the base field of $H_i$, $m_i$ as
the rank of $H_i$ and $k_i$ as the multiplicity of $H_i$ and we will speak of
$\OO$-, $\UU$- and $\Sp$-factors. A
representative for the conjugacy class of the centralizer of
$\IF{\KK_i}{m_i}{k_i}$ is given by $\IF{\LL_i}{k_i}{m_i}$, where $\LL_i = \HH$
if $\KK_i=\RR$, $\LL_i=\RR$ if $\KK_i=\HH$ and $\LL_i=\CC$ otherwise.

Thus, up to conjugacy, the Howe subgroups of $\Sp(n)$ are given by
direct products
 \beq\label{G-HSG-Sp}
H = \IF{\KK_1}{m_1}{k_1} \times \cdots\times \IF{\KK_r}{m_r}{k_r}
 \,,~~~~~~
n = \sum\nolimits_{i=1}^r m_i k_i\,,
 \,,~~~~~~
r=1,2,3,\dots\,,
 \eeq
and a representative for the conjugacy class of the centralizer of $H$ is
obtained from $H$ by interchanging ranks and multiplicities as well as base
fields $\RR$ and $\HH$. The subgroup $\IF\RR 1n \equiv\OOF1n$
coincides with the center of $\Sp(n)$ and the subgroup $\IF\HH n1
\equiv \SpF n1$ coincides with $\Sp(n)$ itself.

The operations which produce the direct predecessors of a given
Howe subgroup of $\Sp(n)$ (up to conjugacy) are

--~ {\it Merging} and {\it splitting} similar to the case of
$\OO(n)$

--~ {\it Inverse field restriction:}~ replacing a factor $\OOF{2m}k$
or $\UUF{2m}k$ by a factor $\UUF m{2k}$ or $\SpF m{2k}$, embedded via
$\vp_{\CC,\RR}$ or $\vp_{\HH,\CC}$, respectively;

--~ {\it Inverse field extension:}~ replacing a factor $\UUF mk$
or $\SpF mk$ by a factor $\OOF mk$ or $\UUF mk$, respectively, embedded in the
obvious way.

 \comment{ ohne Bezeichnung der Vfh:

--~ {\it Merging} and {\it splitting} similar to the case of
$\OO(n)$

--~ {\it Inverse field restriction:}~ replacing a factor $\OO(2m)$
or $\mr U(2m)$ of multiplicity $k$ by a factor $\mr U(m)$ or
$\Sp(m)$ of multiplicity $2k$, embedded via $\vp_{\CC,\RR}$ or
$\vp_{\HH,\CC}$, respectively;

--~ {\it Inverse field extension:}~ replacing a factor $\mr U(m)$
or $\mr Sp(m)$ by a factor $\mr O(m)$ or $\mr U(m)$ of the same
multiplicity, respectively, embedded in the obvious way.

 }

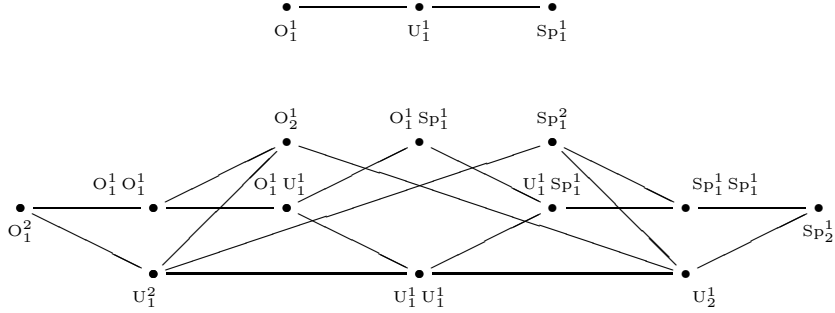
\begin{figure}

\begin{center}

\unitlength1.75cm

\begin{picture}(2,1)
\put(0,0.5){
 \lri{0,0}{tc}{\OOF11}
 \lri{1,0}{tc}{\UUF11}
 \whole{2,0}{tc}{\SpF11}
 }
\end{picture}

\begin{picture}(6,2)
\put(0,1){
 \lri{0,0}{tc}{\OOF12}
 \luri{0,0}{}{}

 \lori{1,0}{br}{\OOF11\stimes\OOF11}
 \lri{1,0}{}{}
 \loori{1,-0.5}{tr}{\UUF12\!\!\!}
 \loorrri{1,-0.5}{}{}
 \lrii{1,-0.5}{}{}

 \luurrri{2,0.5}{bc}{\OOF21}
 \lori{2,0}{br}{\OOF11\stimes\UUF11\!\!\!\!\!\!\!\!}
 \luri{2,0}{}{}

 \luri{3,0.5}{bc}{\OOF11\stimes\SpF11}
 \lori{3,-0.5}{tc}{\UUF11\stimes\UUF11}
 \lrii{3,-0.5}{}{}

 \luri{4,0.5}{bc}{\SpF12}
 \luuri{4,0.5}{}{}
 \lri{4,0}{bc}{\UUF11\stimes\SpF11}

 \lri{5,0}{bl}{\SpF11\stimes\SpF11}
 \lori{5,-0.5}{tl}{\UUF21}

 \whole{6,0}{tc}{\SpF21}
 }
\end{picture}

\end{center}

\caption{\label{F-HSG-Sp} Hasse diagrams of the sets of conjugacy
classes of Howe subgroups for $\Sp(1)$ and $\Sp(2)$. For the
notation, see Figure \rref{F-HSG-O}. For $\Sp(1)$ one has the
identifications $\SpF11 = \Sp(1) \equiv \SU(2)$, $\UUF11 =$
maximal toral subgroup of $\SU(2)$ and $\OOF11\equiv\ZZ_2 =$
center of $\SU(2)$.}

\end{figure}


\subsection{Howe subgroups of $\SO(n)$}
\label{S-HSG-SO}

First, we treat the case of odd $n$. Here
 \beq\label{G-deco-odd}
\OO(n) = Z\cdot\SO(n)\,,
 \eeq
where $Z = \{\II,-\II\}$ is the center of $\OO(n)$.

\bpp\label{P-HSG-odd}

For $n$ odd, intersection with $\SO(n)$ defines a bijection from
the set of Howe subgroups of $\OO(n)$ onto the set of Howe
subgroups of $\SO(n)$. The bijection preserves the equivalence
relation of conjugacy. Its inverse is given by multiplication by
the center $Z$.

\epp

{\it Proof.}~ We start with deriving some formulae. As an
immediate consequence of \eqref{G-deco-odd}, for any subgroup $H$
of $\OO(n)$,
 \beq\label{G-HSG-odd-1}
\mr C_{\OO(n)} (H) = \mr C_{\OO(n)} (Z\cdot H)\,.
 \eeq
Furthermore, the chain of inclusions $\SO(n)\cap H \subseteq H \subseteq Z\cdot
(\SO(n)\cap H)$ implies
$$
\mr C_{\OO(n)}(\SO(n)\cap H)
 \supseteq
\mr C_{\OO(n)}(H)
 \supseteq
\mr C_{\OO(n)}\big(Z\cdot(\SO(n)\cap H)\big)\,.
$$
Then \eqref{G-HSG-odd-1}, applied to the subgroup $\SO(n)\cap H$, yields
 \beq\label{G-HSG-odd-2}
\mr C_{\OO(n)} (H) = \mr C_{\OO(n)} (\SO(n)\cap H)\,.
 \eeq
Finally, by writing the centralizer in $\SO(n)$ as $\mr C_{\SO(n)} = \SO(n) \cap
\mr C_{\OO(n)}$ one sees that \eqref{G-deco-odd} implies that, for any subgroup
$\tilde H$ of $\SO(n)$,
 \beq\label{G-HSG-odd-3}
Z\cdot \mr C_{\SO(n)} (\tilde H) = \mr C_{\OO(n)} (\tilde H)\,.
 \eeq
Now let $H$ be a Howe subgroup of $\OO(n)$. Then $H = \mr C_{\OO(n)}(H')$ for
some subgroup $H'$ of $\OO(n)$. According to \eqref{G-HSG-odd-2}, then $H = \mr
C_{\OO(n)}\big(\SO(n)\cap H'\big)$. Intersection with $\SO(n)$ yields
$$
\SO(n)\cap H = \mr C_{\SO(n)}\big(\SO(n)\cap H'\big)\,.
$$
Hence, $\SO(n)\cap H$ is a Howe subgroup of $\SO(n)$. Conversely,
let $\tilde H$ be a Howe subgroup of $\SO(n)$. Consider $H =
Z\cdot \tilde H$. By construction, $\tilde H = \SO(n) \cap H$.
Moreover, since $\tilde H = \mr C_{\SO(n)}(\tilde H')$ for some
subgroup $\tilde H'$ of $\SO(n)$, \eqref{G-HSG-odd-3} implies $H =
\mr C_{\OO(n)}(\tilde H')$. Hence, $H$ is a Howe subgroup of
$\OO(n)$.

Finally, in view of \eqref{G-deco-odd} it is obvious that
subgroups of $\OO(n)$ that are conjugate under $\OO(n)$ are also
conjugate under $\SO(n)$.
 $\qed$
 \relax\bigskip

When $n$ is even, there exist Howe subgroups of $\OO(n)$ whose
intersection with $\SO(n)$ is not a Howe subgroup of $\SO(n)$. The
simplest example of such a situation is provided by the center of
$\OO(2)$ which is contained in $\SO(2)$ but is not a Howe subgroup
there, because $\SO(2)$ is Abelian. Moreover, distinct Howe
subgroups of $\OO(n)$ may have the same intersection with
$\SO(n)$. Again, the simplest example is provided by $\OO(2)$, and
the Howe subgroups $\SO(2)$ and $\OO(2)$.

To begin with, let us introduce some notation. For a subgroup $H$
of $\mr O(n)$ let
$$
\mr CH := \mr C_{\mr O(n)}(H)
 \,,~~~~~~
\mr SH := \SO(n) \cap H
 \,,~~~~~~
\mr MH := \mr C^2\mr SH\,.
$$
By construction, $\mr MH$ is the Howe subgroup of $\OO(n)$ generated by $\mr
SH$, i.e., the smallest Howe subgroup of $\OO(n)$ containing $\mr SH$. For
convenience, $\mr C$, $\mr S$ and $\mr M$ will be viewed as maps on the set of
subgroups of $\mr O(n)$. The basic properties of these maps are monotony,
$$
H\subseteq K
 ~~~\Rightarrow~~~
\mr CH \supseteq \mr CK
 \,,~~~
\mr C^2H \subseteq \mr C^2K
 \,,~~~
\mr SH \subseteq \mr SK
 \,,~~~
\mr MH \subseteq \mr MK
$$
and periodicity resp.\ idempotence,
$$
\mr C^3 = \mr C
 \,,~~~~~~
\mr S^2 = \mr S
 \,,~~~~~~
\mr M^2 = \mr M\,.
$$
For $\mr C$, these properties are well known. For $\mr S$, they
are obvious. For $\mr M$, monotony follows from that of $\mr C^2$
and $\mr S$. Idempotence can be seen as follows. $\mr MH$ is a
Howe subgroup of $\OO(n)$ containing $\mr S\mr MH$. Since $\mr
M^2H$ is the smallest such subgroup, $\mr M^2H \subseteq \mr MH$.
Conversely, we have both $\mr SH\subset \mr MH$ and $\mr S\mr
MH\subset \mr M^2H$. Applying $\mr S$ to the first inclusion and
composing with the second one we obtain $\mr SH \subseteq \mr S\mr
MH \subseteq\mr M^2H$. I.e., $\mr M^2H$ is a Howe subgroup
containing $\mr SH$. As $\mr MH$ is the smallest such subgroup,
$\mr MH\subseteq\mr M^2H$.

A Howe subgroup $H$ of $\OO(n)$ will be called $\mr S$-admissible
if $\mr SH$ is a Howe subgroup of $\SO(n)$ and there is no smaller
Howe subgroup of $\OO(n)$ containing $\mr SH$. Using $\mr M$, this
can be reformulated as follows. $H$ is $\mr S$-admissible if and
only if $\mr SH$ is a Howe subgroup of $\SO(n)$ and $\mr MH=H$.

\bpp\label{P-S-adm}

$\mr S$ induces a bijection from the set of $\mr S$-admissible
Howe subgroups of $\OO(n)$ onto the set of Howe subgroups of $\mr
SO(n)$. The inverse is given by $\mr M$.
 \comment{
Let $\tilde H$ be a Howe subgroup of $\SO(n)$. Then $\tilde H =
\mr S\mr M \tilde H$.
 }

\epp

{\it Proof.}~ By definition of $\mr S$-admissibility, $\mr S$
defines a map from the set of $\mr S$-admissible Howe subgroups of
$\OO(n)$ to the set of Howe subgroups of $\SO(n)$. Since the
intersection of arbitrarily many Howe subgroups is a Howe
subgroup, this map is injective. To see that it is also
surjective, let $\tilde H$ be a Howe subgroup of $\SO(n)$. We show
that $\tilde H = \mr S\mr M \tilde H$ (then evidently $\mr M\tilde
H$ is $\mr S$-admissible and $\mr M$ is inverse to $\mr S$). Since
$\mr S\tilde H = \tilde H$, $\tilde H\subseteq\mr M\tilde H$.
Application of $\mr S$ yields $\tilde H \subseteq \mr S\mr M\tilde
H$. Conversely, by $\mr S\mr C\tilde H \subseteq \mr C\tilde H$
and monotony of $\mr C$ and $\mr S$,
$$
\mr S\mr M\tilde H
 =
\mr S\mr C\big(\mr C\tilde H\big)
 \subseteq
\mr S\mr C \big(\mr S\mr C\tilde H\big)\,.
$$
The rhs.\ equals $\mr C_{\SO(n)}^2(\tilde H)$. Since $\tilde H$ is
a Howe subgroup of $\SO(n)$, $\mr C_{\SO(n)}^2(\tilde H) = \tilde
H$ and hence $\mr S\mr M\tilde H \subseteq \tilde H$.
 \comment{
{\it Proof.}~ Since $\mr S\tilde H = \tilde H$, $\tilde
H\subseteq\mr M\tilde H$. Application of $\mr S$ yields $\tilde H
\subseteq \mr S\mr M\tilde H$. Conversely, by $\mr S\mr C\tilde H
\subseteq \mr C\tilde H$ and monotony of $\mr C$ and $\mr S$,
$$
\mr S\mr M\tilde H
 =
\mr S\mr C\big(\mr C\tilde H\big)
 \subseteq
\mr S\mr C \big(\mr S\mr C\tilde H\big)\,.
$$
The rhs.\ equals $\mr C_{\SO(n)}^2(\tilde H)$. Since $\tilde H$ is a Howe
subgroup of $\SO(n)$, $\mr C_{\SO(n)}^2(\tilde H) = \tilde H$ and hence $\mr S\mr
M\tilde H \subseteq \tilde H$.
 }
 $\qed$
\relax\bigskip

Note that $\mr S$-admissibility is not a necessary condition for a Howe
subgroup of $\OO(n)$ to yield a Howe subgroup of $\SO(n)$ by intersection. In
fact, for Howe subgroups $H$ that are not $\mr S$-admissible, $\mr SH$ may or
may not be a Howe subgroup of $\SO(n)$, as is shown by the Howe subgroups
$H=\OO(2)$ and $H=$ center of $\OO(2)$.
\bigskip

Next, we determine the $\mr S$-admissible Howe subgroups of
$\OO(n)$. As noted above, any such subgroup is stable under $\mr
M$.

\ble\label{L-Hcond}

Let $H$ be an $\mr M$-stable Howe subgroup of $\OO(n)$. Then $\mr
SH$ is a Howe subgroup of $\SO(n)$ if and only if $H = (\mr M\mr
C)^2 H$.

\ele

{\it Proof.}~ First, assume that $H = (\mr M\mr C)^2 H$ holds.
Application of $\mr S$ yields $\mr SH = \mr S(\mr M\mr C)^2 H$. We
observe: if $K$ is a Howe subgroup of $\OO(n)$ then $\mr
SK\subseteq\mr MK \subseteq K$. Application of $\mr S$ yields $\mr
S\mr MK = \mr SK$. In particular, this implies $\mr S\mr M\mr C =
\mr S\mr C$. Using in addition the obvious relation $\mr C\mr M =
\mr C\mr S$, the expression $\mr S(\mr M\mr C)^2 H$ can be
rewritten as $\mr S \mr C (\mr S \mr C H)$, i.e., as the
centralizer of the subgroup $\mr S\mr CH$ of $\mr SO(n)$, taken in
$\SO(n)$. It follows that $\mr SH$ is a Howe subgroup of $\SO(n)$.

Conversely, assume that $\mr SH$ is a Howe subgroup of $\SO(n)$. We rewrite
$(\mr M\mr C)^2 H$ as $\mr C^2(\mr S\mr C)^2 H$. Applying $\mr C$ to
$H = \mr MH$ and using $\mr C\mr M = \mr C\mr S$ again we obtain $\mr CH =
\mr C\mr SH$. Hence, in the expression $\mr C^2(\mr S\mr C)^2 H$ we can replace
$H$ by $\mr SH$, thus obtaining $\mr C^2(\mr S\mr C)^2 \mr S H$. Now
$(\mr S\mr C)^2\mr S H$ is the double centralizer of $\mr SH$ in $\SO(n)$ and
hence equals $\mr SH$. Upon using $\mr CH = \mr C\mr SH$ once more, we arrive at
$(\mr M\mr C)^2H = H$, as asserted.
 $\qed$
 \relax\bigskip

As a result, in order to determine the Howe subgroups of $\SO(n)$,
we may first determine the Howe subgroups that are stable under
$\mr M$ and then, among these, the Howe subgroups that are stable
under $(\mr M\mr C)^2$.

\ble\label{L-Mstable}

A Howe subgroup $H$ of $\OO(n)$ is stable under $\mr M$ except for
the following cases.

{\em (A)}~ $H$ has a factor $\OOF 2k$ with $k$ odd and no
other $\OO$-factor of odd multiplicity. Here, $\mr MH$ arises from
$H$ by inverse field restriction, i.e., by replacing $\OOF 2k$ by
$\UUF1k$.

{\em (B)}~ $H$ has a double factor $\OOF1k\times\OOF1l$ with
$k$, $l$ odd and no other $\OO$-factor of odd multiplicity. Here,
$\mr MH$ arises from $H$ by merging this double factor to $\OOF1{k+l}$.

\ele

{\it Proof.}~ First, we determine $\mr MH$ in cases (A) and (B).
In case (A), let $K$ be the Howe subgroup obtained from $H$ by
inverse field restriction of the factor $\OOF2k$ under consideration.
Thus, this factor is replaced by the factor $\UUF1k$. Since the original
factor $\OOF2k$
is the only $\OO$-factor of odd multiplicity of $H$, an element of $H$ has
negative determinant if and only if its entry in this factor has so. Hence, $\mr
SH = K$. Since $K$ is Howe, then $\mr MH = K$.
In case (B), let $K$ be the Howe subgroup obtained from $H$ by
merging the double factor $\OOF1k\times\OOF1l$ to
$\OOF1{k+l}$. Since $H$ has no other
$\OO$-factors of odd multiplicity, an element of $H$ has negative
determinant if and only if its entries in the two factors $\OOF1k$ and
$\OOF1l$ are distinct. Hence, $\mr SH = K$ and so $\mr M H = K$.
Thus, in cases (A) and (B), $H$ is not stable under $\mr M$.

Conversely, assume that $H$ is not stable under $\mr M$. Since
$\mr SH$ has the same dimension as $H$, $\mr MH$ has the same
dimension as $H$. It follows that $H$ has a direct predecessor
$K$, wrt.\ the natural partial ordering of Howe subgroups modulo
conjugacy, of the same dimension. By comparing the dimension of
the factors that are replaced by one another through the
operations of merging, splitting, inverse field restriction and
inverse field extension, one finds that there are only two
situations where the dimension does not change. These are

--~ inverse field restriction of a factor $\OOF2k$, which yields
the factor $\UUF1k$ instead,

--~ merging of a double factor $\OOF1k\times\OOF1l$, which
yields the factor $\OOF1{k+l}$.

If, in the first situation, $k$ is even or if $H$ contains further
$\OO$-factors of odd
multiplicity, $\mr SH$ contains elements whose entry in the
factor $\OOF2k$ under consideration has negative determinant, hence
$\mr SK \neq \mr SH$. Similarly, in the second situation, if $k$ or
$l$ is even or if $H$ contains further $\OO$-factors of odd multiplicity,
$\mr SH$ contains elements whose entries in the factors $\OOF1k$ and
$\OOF1l$ are distinct. Thus, we are either in case (A) or case (B).
 $\qed$
 \relax\bigskip

\ble\label{L-MC2stable}

Let $H$ be a Howe subgroup of $\OO(n)$ which is stable under $\mr M$. Then $H$
is stable under $(\mr M\mr C)^2$ if and only if $\mr CH$ is stable under $M$.

\ele

{\it Proof.}~ If $\mr CH$ is $M$-stable then $(\mr M\mr C)^2
H = \mr M\mr C\mr M\mr CH = \mr M\mr C^2H = \mr MH = H$. If $\mr CH$ is not
$M$-stable then by Lemma \rref{L-Mstable} $\mr CH$ and $\mr M\mr CH$ are of
the form
$$
\mr CH = (\mr CH)^{(0)}\times\OOF2k
 \,,~~~~~~
\mr M\mr CH = (\mr CH)^{(0)}\times\UUF1k
$$
or
$$
\mr CH = (\mr CH)^{(0)}\times\OOF1k\times\OOF1l
 \,,~~~~~~
\mr M\mr CH = (\mr CH)^{(0)}\times\OOF1{k+l}\,,
$$
where $k$ and $l$ are odd and $(\mr CH)^{(0)}$ does not contain an
$\OO$-factor of odd multiplicity. Then $H$ and $\mr C\mr M\mr CH$ are of the
form
$$
H = H^{(0)}\times\OOF k2
 \,,~~~~~~
\mr C\mr M\mr CH = H^{(0)}\times\UUF k1
$$
or
$$
H = H^{(0)}\times\OO(k)_1\times\OOF l1
 \,,~~~~~~
\mr C\mr M\mr CH = H^{(0)}\times\OOF{k+l}1\,,
$$
respectively, where $H^{(0)}$ corresponds to $(\mr CH)^{(0)}$
under taking the centralizer. In both cases, $H$ and $\mr C\mr
M\mr CH$ have different dimension. On the other hand, $(\mr M\mr
C)^2H$ arises from $\mr C\mr M\mr CH$ by application of  $\mr M$
and hence has the same dimension as $\mr C\mr M\mr CH$. Therefore,
$H \neq (\mr M\mr C)^2H$.
 \qed
 \bigskip

Lemmas \rref{L-Hcond}--\rref{L-MC2stable} imply

\bpr\label{P-HSG-SO}

A Howe subgroup of $\OO(n)$, $n$ even, is $\mr S$-admissible if
and only if neither itself nor its centralizer in $\OO(n)$ belong
to cases {\rm (A)} or {\rm (B)} of Lemma \rref{L-Mstable}.
 $\qed$

\epr

For convenience, we reformulate the conditions for $\mr CH$ to
belong to cases (A) or (B) of Lemma \rref{L-Mstable} as conditions
on $H$. $\mr CH$ belongs to case (A) of Lemma \rref{L-Mstable} iff
$H$ has an $\OO$-factor of odd rank and multiplicity $2$ and no
further $\OO$-factor of odd rank. $\mr CH$ belongs to case (B) of
Lemma \rref{L-Mstable} iff $H$ has two $\OO$-factors of odd rank
and multiplicity $1$ and no further $\OO$-factor of odd rank.
\bigskip

\bre

Propositions \rref{P-S-adm} and \rref{P-HSG-SO} apply trivially to
the case of odd $n$.

\ere

\bex\label{E-HSG-SO}

We determine the $\mr S$-admissible Howe subgroups of $\OO(2)$ and
$\OO(4)$, see Figure \rref{F-HSG-O}. Cases (A) and (B) refer to
Lemma \rref{L-Mstable}.
 \bigskip

$\OO(2)$:~ $\OOF21$ belongs to case (A), $\OOF11\times\OOF11$
belongs to case (B). Since $\OOF21$ is the centralizer of $\OOF12$, the only
$\mr S$-admissible Howe subgroup of $\OO(2)$ is therefore $\UUF11$ which amounts
to $\SO(2)$ itself. This is consistent with the fact that $\SO(2)$ is abelian.
 \bigskip

$\OO(4)$:~ $\OOF21 \times \OOF12$ and $\OOF21\times\UUF11$ belong
to case (A). $\OOF11\times\OOF11\times\UUF11$,
$\OOF11\times\OOF11\times\OOF12$ and
$\OOF11\times\OOF11\times\OOF13$ belong to case (B). By cancelling
these subgroups and their centralizers we arrive at the following
list of $\mr S$-admissible Howe subgroups of $\OO(4)$:~ $\OOF14$,
$\OOF12\times\OOF12$, $\UUF12$,
$\OOF11\times\OOF11\times\OOF11\times\OOF11$, $\OOF22$,
$\UUF11\times\UUF11$, $\SpF11$, $\UUF21$, $\OOF21\times\OOF21$,
$\OOF41$.

\eex

It remains to discuss the passage to conjugacy classes in $\SO(n)$, $n$ even. We
will show

\bpr\label{P-cjg}

Let $H$ be an $\mr S$-admissible Howe subgroup $H$ of $\OO(n)$,
$n$ even. On intersection with $\SO(n)$, the $\OO(n)$-conjugacy
class of $H$ passes to a single conjugacy class of Howe subgroups
of $\SO(n)$ exactly in the following cases:

--~ $H$ contains an $\OO$-factor of odd rank or odd multiplicity.

--~ $H$ contains a $\mr U$-factor of odd rank and odd multiplicity.

Otherwise, the $\OO(n)$-conjugacy class of $H$ gives rise to two
distinct $\SO(n)$-conjugacy classes, generated by $\mr SH$ and
$\mr S(aHa^{-1})$ for some $a\in\OO(n)$ with negative determinant.

\epr

We start with some preliminary observations. Since $\SO(n)$ is
normal in $\OO(n)$, intersection with $\SO(n)$ commutes with
conjugation under either $\OO(n)$ and $\SO(n)$. It is therefore
sufficient to show that the conjugacy classes of $H$ under
$\OO(n)$ and $\SO(n)$ coincide if and only if $H$ belongs to one
of the two cases listed in the proposition. If the two conjugacy classes do not coincide, it is clear that the
$\OO(n)$-class splits into two distinct $\SO(n)$-classes given by
$H$ and $aHa^{-1}$, where $a$ is an arbitrary element of $\OO(n)$
with $\det (a) = -1$.
Since the conjugacy classes of $H$ under $\OO(n)$ and $\SO(n)$ coincide iff
$\mr N_{\OO(n)}(H)$ contains an element with negative determinant, we have to
determine the normalizer of $H$. This will be done in two lemmas. We include the
case of the identity connected component $H_0$ because this will be needed in
Section \rref{S-factoriz} and we get it for granted here.

\ble\label{L-nmr-1}

The normalizer of a Howe subgroup $H = \IF{\KK_1}{m_i}{k_i}
\times\cdots\times \IF{\KK_r}{m_r}{k_r}$ of $\OO(n)$ consists of
all elements of $\OO(n)$ which can be written in the form $a b$
where
$$
a
 =
 \big(
a_1\oplus\stackrel{k_1}{\cdots}\oplus a_1
 \big)
 \oplus\cdots\oplus
 \big(
a_r\oplus\stackrel{k_r}{\cdots}\oplus a_r
 \big)
$$
with $a_i\in\mr N_{\OO(\delta_i
m_i)}\big(\IF{\KK_i}{m_i}1\big)$ and $b\in\mr C_{\OO(n)}(H)$. The
assertion still holds if $H$ is replaced by $H_0$ and $\IF{\KK_i}{m_i}{k_i}$
by $\big(\IF{\KK_i}{m_i}{k_i}\big)_0$. \todo{Bezeichnung einfuehren}

\ele

{\it Proof.}~ It is evident that any element of $\OO(n)$ of the form given in
the lemma belongs to the normalizer of $H$ in $\OO(n)$. Conversely, let
$c\in\mr N_{\OO(n)}(H)$. Consider the standard decomposition
$$
\RR^n
 =
 \big(
\RR^{\delta_1 m_1}\oplus\stackrel{k_1}{\cdots}\oplus\RR^{\delta_1 m_1}
 \big)
 \oplus\cdots\oplus
 \big(
\RR^{\delta_rm_r}\oplus\stackrel{k_r}{\cdots}\oplus\RR^{\delta_rm_r}
 \big)
$$
into $H$-irreducible subspaces, used in the definition of $H$ according to
\eqref{G-irrHSG-O-2}. The linear
transformation $c$ maps this decomposition to another orthogonal decomposition
of $\RR^n$ into $H$-irreducible subspaces. As abstract direct sums of orthogonal
representations of $H$, the two decompositions are isometrically isomorphic.
Hence, there exists $b_1\in\OO(n)$ commuting with $H$ such that $cb_1$
leaves invariant each $H$-irreducible subspace $\RR^{\delta_i m_i}$ in the
standard decomposition above separately. Then
$$
cb_1
 =
 \left(
c_{11}\oplus\cdots\oplus c_{1k_1}
 \right)
 \oplus\cdots\oplus
 \left(
c_{r1}\oplus\cdots\oplus c_{rk_r}
 \right)
$$
with $c_{ij}\in\mr N_{\OO(\delta_im_i)}(\IF{\KK_i}{m_i} 1)$. Since
the action of conjugation by $c_{ij}$ on the operators of $\IF{\KK_i}{m_i}
1$ is independent of $j$, $c_{ij} = a_i b_{ij}$, where $a_i\in\mr
N_{\OO(\delta_im_i)}(\IF{\KK_i}{m_i} 1)$ and $b_{ij}\in\mr
C_{\OO(\delta_im_i)}(\IF{\KK_i}{m_i} 1)$. Setting
 $
b
 =
 \left(
b_{11}\oplus\cdots\oplus b_{1k_1}
 \right)
 \oplus\cdots\oplus
 \left(
b_{r1}\oplus\cdots\oplus b_{rk_r}
 \right)
b_1^{-1}
 $
we arrive at the assertion. The argument for $H_0$ is completely analogous.
 \qed
 \bigskip

\ble\label{L-nmr-2}

The normalizers $\mr N_{\OO(\delta_im_i)}\big(\IF{\KK_i}{m_i}1\big)$ are as
follows.

$\KK_i=\RR$:~ $\mr N_{\OO(m_i)}\big(\OOF{m_i}1\big) = \mr
N_{\OO(m_i)}\big(\SOF{m_i}1\big) = \OO(m_i)$\,. \todo{Bezeichnung $\SOF{}{}$
einfuehren}

$\KK_i=\CC$:~ $\mr N_{\OO(2m_i)}\big(\UUF{m_i}1\big)$ is generated by
$\UUF{m_i}1$, its centralizer and
 $
\II_{m_i,m_i} =
 $
 {\footnotesize
 $
 \begin{bmatrix} \II_{m_i} & 0 \\ 0 & -\II_{m_i} \end{bmatrix}
 $
 }.

$\KK_i=\HH$:~ $\mr N_{\OO(4m_i)}\big(\SpF{m_i}1\big)$ is generated
by $\SpF{m_i}1$ and its centralizer.

With the exception of the subgroups $\OOF{m_i}1$ and $\UUF21$, in any case
the induced homomorphism from the normalizer to the automorphism group of
$\I_{\KK_i}(m_i)$ or $\I_{\KK_i}(m_i)_0$ is surjective. In the case of the
subgroup $\UUF21$, the image of this homomorphism is generated by inner
automorphisms and the outer automorphism of complex conjugation of matrices.

\ele

{\it Proof.}~ For $\KK_i = \RR$, the normalizers are obvious. As for the
automorphism groups it is known that for odd $m_i$ all automorphisms of
$\OO(m_i)$ are inner whereas for even $m_i$ they are generated
by inner automorphisms and the outer automorphism $D:a\mapsto \det(a) a$. The
automorphisms of $\SO(m_i)$ are restrictions of automorphisms of $\OO(n)$.
Since in case $m_i$ is even, $D$ acts trivially on $\SO(m_i)$, they are
restrictions of inner automorphisms of $\OO(n)$, as asserted.

For $\KK=\CC$, let $b\in\mr N_{\OO(2m_i)}(\UUF {m_i} 1)$.
Conjugation by $b$ defines an automorphism of $\UU(m_i)$. In case $m_i\neq 2$,
the automorphism group of $\UU(m_i)$ is generated by inner automorphisms and by
the outer automorphism of complex conjugation of matrices. In case of $\UU(2)$
there is one further generator, given by the outer automorphism $D(a) =
\ol{\det a} \cdot a$. When $\UU(m_i)$ is embedded into $\OO(2m_i)$ via
$\vp_{\CC,\RR}$, complex conjugation of matrices can be represented by
conjugation by $\II_{m_i,m_i}$, but $D$ can not. Thus, $b = Tb_1b_2$ where $T =
\II_{2m_i}$ or $\II_{m_i,m_i}$, $b_1\in\UUF{m_i}1$ and $b_2\in\mr
C_{\OO(2m_i)}\big(\UUF{m_i}1\big)$. Conversely, any element of $\OO(2m_i)$ of
this form normalizes $\UUF{m_i}1$.

For $\KK_i=\HH$ the argument is similar. Here all automorphisms
are inner.
 \qed
 \relax\bigskip

{\it Proof of Proposition \rref{P-cjg}.}~ According to the above considerations,
we have to show that $\mr N_{\OO(n)}(H)$ contains an element with negative
determinant if and only if $H$ belongs to one of the two cases given in the
proposition. In the first case, $H$ itself or its centralizer
contains an element of negative determinant. In the second case,
by Lemmas \rref{L-nmr-1} and \rref{L-nmr-2}, $\mr N_{\OO(n)}(H)$ contains an
element which in the notation of Lemma \rref{L-nmr-1} is given by $ab$ with
$b= \II_n$, $a_i = \II_{m_i,m_i}$ for the relevant $\UU$-factor and
$a_i=\II_{\delta_im_i}$ for the other factors. The determinant is $\det(ab) =
(-1)^{m_ik_i} = -1$.

Conversely, let an element of $\mr N_{\OO(n)}(H)$ with negative
determinant be given. We can write it in the form $a b$ where
$b\in\mr C_{\OO(n)}(H)$ and $a$ is given in Lemma \rref{L-nmr-1}.
Then
$$
\det(ab)
 =
(\det a_1)^{k_1} \cdots (\det a_r)^{k_r} \det b
 =
-1\,.
$$
If $\det b = -1$, $\mr C_{\OO(n)}(H)$ contains an $\OO$-factor of
odd multiplicity, hence $H$ contains an $\OO$-factor of odd rank
and we are in the first case of the proposition. If $\det b = 1$,
one of the $a_i$ must have negative determinant and the
corresponding multiplicity $k_i$ must be odd. By Lemma
\rref{L-nmr-2}, if $a_i$ has negative determinant then
$\KK_i=\RR$, where we are in the first case of the proposition, or
$\KK_i=\CC$ and $m_i$ odd, where we are in the second case of the
proposition.
 \qed
 \relax\bigskip

To define a standard form for Howe subgroups of $\SO(n)$, we fix
$a_{(n)}\in\OO(n)$ with $\det(a_{(n)}) = -1$, e.g.\ $a_{(n)} =
${\footnotesize$\begin{bmatrix} -1 & 0 \\ 0 & \II_{n-1}
\end{bmatrix}$}. For a Howe subgroup $H$ of $\OO(n)$ let
 \beq\label{G-def-SHpm}
\mr SH^+ := \mr SH
 \,,~~~~~~
\mr SH^- := a_{(n)}^{-1} \mr SH a_{(n)}\,.
 \eeq
By a subgroup of $\SO(n)$ of type $\typeSHpm$ we mean a
subgroup of the form $\mr SH^\pm$ where $H$ is a Howe subgroup of
$\OO(n)$ in standard form. For simplicity, when treating $\SO(n)$
in the following, all results will be derived for both these types
of subgroups, irrespective of that some of them may not be Howe or
that $\mr SH^+$ and $\mr SH^-$ may define the same conjugacy class
in $\SO(n)$. If the latter is true we may just forget about $\mr
SH^-$ and consider $\mr SH^+$ alone. This applies in particular
when $n$ is odd.

\bex\label{E-cjg}

\begin{figure}

\begin{center}

\unitlength2cm

\begin{picture}(4,3.1)
\put(0,1.3){
 \lori{0,0}{cr}{\OOF14}
 \lri{0,0}{}{}
 \luri{0,0}{}{}

 \lori{1,0.5}{br}{\UUF12{}^+\!\!\!\!\!\!}
 \lri{1,0.5}{}{}
 \luri{1,0.5}{}{}
 \looori{1,0}{br}{\OOF12\stimes\OOF12\!\!\!\!}
 \lori{1,0}{}{}
 \luri{1,0}{}{}
 \lori{1,-0.5}{tr}{\UUF12{}^-\!\!\!\!}
 \lri{1,-0.5}{}{}
 \luri{1,-0.5}{}{}

 \luuuri{2,1.5}{bc}{\OOF11\stimes\OOF11\stimes\OOF11\stimes\OOF11}
 \luri{2,1}{tc}{\Sp11{}^+}
 \lri{2,0.5}{tc}{\OOF22{}^+}
 \luri{2,0.5}{}{}
 \lori{2,0}{tc}{\UUF11\stimes\UUF11\,}
 \luri{2,0}{}{}
 \lori{2,-0.5}{tc}{\OOF22{}^-}
 \lri{2,-0.5}{}{}
 \lori{2,-1}{tc}{\SpF11{}^-}

 \luri{3,0.5}{bl}{\!\!\!\UUF21{}^+}
 \lri{3,0}{bl}{\!\!\!\!\OOF21\stimes\OOF21}
 \lori{3,-0.5}{tl}{\!\!\UUF21{}^-}

 \whole{4,0}{cl}{\OOF41}
 }
\end{picture}

\end{center}

\caption{\label{F-HSG-SO} Hasse diagram of the set of conjugacy
classes of Howe subgroups of $\SO(4)$. For the notation, see
Figure \rref{F-HSG-O}.}

\end{figure}
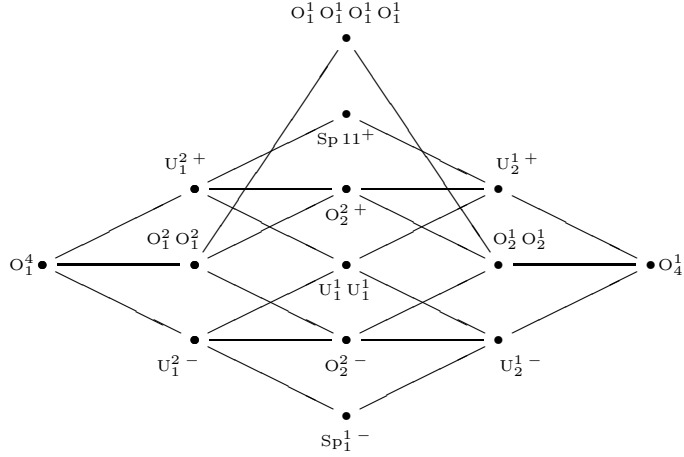

Consider the $\mr S$-admissible Howe subgroups of $\OO(2)$ and
$\OO(4)$ derived in Example \rref{E-HSG-SO}. We determine their
conjugacy classes in $\SO(2)$ and $\SO(4)$, respectively. For
$\OO(2)$, the situation is trivial. Let us check consistency,
anyhow. The only $\mr S$-admissible Howe subgroup of $\OO(2)$ is
$\UUF11$; it possesses a $\UU$-factor of odd rank and odd
multiplicity and hence defines a single conjugacy class in
$\SO(2)$ by Proposition \rref{P-cjg} (consisting of $\SO(2)$
itself). For $\OO(4)$, the $\mr S$-admissible Howe subgroups which
give rise to two distinct conjugacy classes in $\SO(4)$ by
Proposition \rref{P-cjg} are $\UUF12$, $\UUF21$, $\OOF22$ and
$\SpF11$. Thus, there are altogether 14 conjugacy classes of Howe
subgroups in $\SO(4)$. The corresponding Hasse diagram is shown in
Figure \rref{F-HSG-SO}. It is interesting to note that one of the
two classes of type $\SpF 1 1$ corresponds to the subgroup of left
isoclinic rotations in $\SO(4)$ and the other one to the subgroup
of right isoclinic rotations. These two subgroups are known to be
conjugate in $\OO(4)$ but not in $\SO(4)$, indeed.

\eex

\bre

$\OO(n)$ contains Howe subgroups which give rise to two distinct conjugacy
classes in $\SO(n)$ if and only if $n$ is a multiple of $4$. Indeed, any
$\OO$-factor of such a Howe subgroup must have even rank and even multiplicity
and any $\UU$-factor must have even rank or even multiplicity. Hence,
any irreducible factor contributes to the dimension by a multiple of $4$.
Conversely, if $n= 4l$ then e.g.\ the Howe subgroup $\OOF{2l}2$ gives rise to
two distinct conjugacy classes in $\SO(n)$ by Proposition \rref{P-cjg}.

\ere


\section{Principal bundles with a Howe subgroup as structure group}
\label{S-HFB}


The aim of this section is to classify principal bundles whose
structure group is a Howe subgroup of $\OO(n)$, $\SO(n)$ or
$\Sp(n)$ and whose base space is a simply connected CW-complex of
dimension $4$ without $2$-torsion in $H^4(X,\Z)$. We will use
classifying spaces and classifying maps to construct
characteristic classes which distinguish the bundles, see
Appendices \ref{universal bundles} and \rref{characteristic
classes} for a brief overview. We start with introducing some
notation. For a Lie group $G$, let $\BB G$ denote the classifying
space and let $[X,\BB G]$ denote the set of homotopy classes of
maps from $X$ to $\BB G$ or, equivalently, the set of isomorphism
classes of principal $G$-bundles over $X$.

Let $H$ be a Howe subgroup of $G=\OO(n)$ or $G=\Sp(n)$. Due to
\eqref{G-HSG-O} and \eqref{G-HSG-Sp}, as a Lie group,
$$
H \cong \I_{\KK_1}(m_1) \times\cdots\times \I_{\KK_r}(m_r)\,.
$$
Accordingly, the classifying space of $H$ decomposes as
$$
\BB H = \BB\I_{\KK_1}(m_1) \times\cdots\times \BB\I_{\KK_r}(m_r)
$$
and any principal $H$-bundle $Q$ over $X$ decomposes as
$$
Q = Q_1 \oplus\cdots\oplus Q_r
$$
(fiber product), where the $Q_i$ are $\I_{\KK_i}(m_i)$-bundles
over $X$. We refer to them as the factors of $Q$. More precisely,
extending the terminology used for the Howe subgroups, we speak of
$\OO$-, $\UU$- and $\Sp$-factors, depending on whether $\KK_i =
\RR$, $\CC$ or $\HH$, respectively. We will also have to consider
the identity connected component $H_0$ of $H$. As a Lie group,
$$
H_0 \cong \I_{\KK_1}(m_1)_0 \times\cdots\times \I_{\KK_r}(m_r)_0\,,
$$
where $\I_{\KK_i}(m_i)_0 = \SO(m_i)$ for $\KK_i = \RR$ and
$\I_{\KK_i}(m_i)_0 = \I_{\KK_i}(m_i)$ for $\KK_i=\CC$, $\HH$. The
classifying space is
$$
\BB H_0 = \BB\I_{\KK_1}(m_1)_0 \times\cdots\times
\BB\I_{\KK_r}(m_r)_0
$$
and principal $H_0$-bundles over $X$ decompose as
$$
Q_0 = Q_{01} \times_X\cdots\times_X Q_{0r}\,,
$$
where the $Q_{0i}$ are $\I_{\KK_i}(m_i)_0$-bundles over $X$,
referred to as the factors of $Q_0$. Similarly to the Howe
subbundles we speak of $\SO$-, $\UU$- or $\Sp$-factors here. Since
$X$ is simply connected, any $H$-bundle or $\mr SH^+$-bundle (in
the case $G=\OO(n)$) can be reduced to $H_0$, where the reductions
are given by the connected components of the given $H$-bundle. As
a consequence, the classification of $H$-bundles and $\mr
SH^+$-bundles can be derived from the classification of
$H_0$-bundles. To classify $H_0$-bundles it suffices to classify
bundles with the factors of $H_0$ as structure group, i.e.,
$\SO(m)$, $\UU(m)$ or $\Sp(m)$. The cases of $\UU(m)$ and $\Sp(n)$
are standard; the necessary results will be cited below.
Thus, the
problem is reduced to the classification of principal
$\SO(n)$-bundles. Since the solution is obvious for $n= 1,2$, it
remains to solve the problem for $n \geq 3$. We use methods
introduced by Woodward \cite{woodward} and improved by {\v{C}}adek
and Van{\v{z}}ura \cite{cadek2, cadek1, cadek3}. Since $X$ has
dimension $4$, we have stability beginning with $n=5$ i.e.
$[X,\BB\SO(5)]=[X,\BB\SO(6)]=\dots =[X,\BB\SO]$. Since $X$ is
simply connected, $[X,\BB\SO] = [X,\BB\OO]$.

\bre\label{Bem-simplyconnected}

The reason for the assumption that $X$ be simply connected
is that it allows to reduce the study of $[X,\BB\OO(4)]$ to the
study of $[X,\BB\SO(4)]$. As we have to make it anyhow, we can use
it further to make simplify arguments by, e.g., treating $H$-bundles and $\mr
SH$-bundles simultaneously through $H_0$-bundles or switching
freely between $\OO(m)$-bundles and $\SO(m)$-bundles. The assumption of simple
connectedness is however not as general as possible; in fact it is sufficient
that $H^1(X,\ZZ_2) = 0$.

\ere


\subsection{Principal bundles with structure group $\OO(n)$ or $\SO(n)$}


Let $\theta_*:H^*(X,\Z_2)\ra H^*(X,\Z_4)$ be the map induced by
the inclusion homomorphism $\theta : \Z_2\ra\Z_4$ and let $\rho$
and $\rho_n$ denote the maps induced by reduction $\Z_4\ra\Z_2$
and $\Z\ra\Z_n$ repectively. Let $\frak P$ and $\Sq^1$ denote the
cohomology operations of Pontryagin square and Steenrod square,
repectively, see Appendix \ref{cohomology operations}. Define a
mapping
$$
\Rmap
 :
H^1(X,\Z_2)\oplus H^2(X,\Z_2)\oplus H^4(X,\Z_2)\oplus H^4(X,\Z)
 \to
H^4(X,\Z_4)
$$
by
$$
\Rmap(u_1,u_2,u_4,v_4) := \rho_4 v_4 -\frak{P}u_2 +
\theta_*(u_1\mathrm{Sq}^1u_2+u_4) \, ,
$$
for all spaces $X\,.$
Let $w_k \in H^k(\BB\OO,\Z_2)$ be the $k$-th Stiefel-Whitney class, $p_1\in
H^4(\BB\OO,\Z)$ the first Pontryagin class and denote $\alpha =
(w_1,w_2,w_4,p_1)\, .$
It can be shown, see Theorem \ref{wu-formel-thm}, that
\begin{align}
\label{rel1}
\Rmap(\alpha)=0
\end{align}
holds for characteristic classes of real vector bundles of
arbitrary dimension $n\, .$ For any CW-complex $X$ the $4$-tuple
$\alpha$ of cohomology elements defines a mapping
$$
\alpha_*:[X,\BB\OO]\ra H^1(X,\Z_2)\oplus H^2(X,\Z_2)\oplus H^4(X,\Z_2)\oplus H^4(X,\Z)
$$
by
\beq
\label{Def-alpha}
\alpha_*(\xi): = \xi^*(\alpha) \, , \quad \xi \in [X,\BB\OO] \, .
\eeq
Obviously, we have $ \alpha_*(\xi) = \big(w_1(\xi),w_2(\xi),w_4(\xi),p_1(\xi)\big) \, .$

\begin{thm}
\label{chs1}

Let $X$ be a  CW-complex of dimension four. The mapping $\alpha_*$ defined by \eqref{Def-alpha}
is injective if $H^4(X,\Z)$ has no element of order 4.

\end{thm}

\begin{rem}
\label{rem-BO} Let $X$ be a $4$-dimensional CW-complex without
$4$-torsion in $H^4(X,\Z)$. Then, according to the above theorem
the sets $[X,\BB\OO(n)],\ n\geq 5 \, ,$ are classified by tuples
$(w_1,w_2,w_4,p_1)$. Moreover, via the natural projection
$\BB\SO(n)\ra \BB\OO(n)$ we obtain a classification of
$[X,\BB\SO(n)]\, , $ $ n\geq 5\, ,$ given by $(w_2,w_4,p_1)$.
\end{rem}

\emph{Proof of Theorem \ref{chs1}}:
The strategy will be to replace $\BB\OO$ by a $5$-equivalent space.
For that purpose, consider the following product of Eilenberg-MacLane spaces \footnote{See Appendix \ref{e-m}.}
\[
K=K(\Z_2,1)\times K(\Z_2,2)\times K(\Z_2,4)\times K(\Z,4) \, .
\]
By virtue of the natural isomorphism $H^n(\BB\OO,\cdot)=[\BB\OO,K(\cdot,n)]\, ,$
see Appendix \ref{A-em},  the $4$-tuple $\alpha$ of cohomology
elements induces a mapping
$\BB\OO \to K\,, $ which will be also denoted by the letter $\alpha\, .$\\

Let
$$
(\kappa_1,\kappa_2,\kappa_4,\chi_4) \in H^1(K(\Z_2,1),\Z_2)\oplus H^2(K(\Z_2,2),\Z_2)\oplus H^4(K(\Z_2,4),\Z_2)\oplus
H^4(K(\Z,4),\Z)
$$
denote the fundamental classes and let $(u_1,u_2,u_4,v_4)\in H^1(X,\Z_2)\oplus H^2(X,\Z_2)\oplus H^4(X,\Z_2)\oplus
H^4(X,\Z)$. Then via the isomorphism $H^n(X,\cdot)\ra [X,K(\cdot,n)]$ we get a map
$f=\varphi_1\times \varphi_2\times \varphi_4\times \phi_4\in [X,K]$ such that
$$
(u_1,u_2,u_4,v_4)=f^*(\kappa_1\otimes 1\otimes 1\otimes 1,1\otimes \kappa_2\otimes 1\otimes 1,
1\otimes 1\otimes \kappa_4\otimes 1, 1\otimes 1\otimes 1\otimes \chi_4).
$$
In what follows we leave out the 1's in the formulae.
Since $\Rmap$ is built of cohomology operations we have:
$$
\Rmap\circ f^*(\kappa_1,\kappa_2,\kappa_4,\chi_4)=
f^*\circ\Rmap(\kappa_1,\kappa_2,\kappa_4,\chi_4).
$$
That means $\Rmap$ is uniquely determined by its value on the fundamental classes. Moreover, since
$\Rmap(\kappa_1,\kappa_2,\kappa_4,\chi_4)\in H^4(K,\Z_4)$, it defines a unique
mapping $k : K \to  K(\Z_4,4) $ by
 \beq\label{G-Rmap}
k^* (\iota_4)  = \Rmap(\kappa_1, \kappa_2, \kappa_4, \chi_4) \, ,
 \eeq
where $\iota_4\in H^4(K(\Z_4,4),\Z_4)$ denotes the fundamental class.\\

By passing to a homotopy equivalent $K$ we can turn $k$ into a fibration \cite{at}.
The homotopy fiber is given by $F = k^{-1}(\ast)\, ,$ where $\ast$ denotes the base point of $K(\Z_4,4)\, .$
We show that equation \eqref{rel1} implies that $k\circ \alpha$ is nullhomotopic:
\begin{align*}
(k \circ \alpha)^* (\iota_4) & = \alpha^* \circ k^* (\iota_4)\\
& = \alpha^* \circ \Rmap(\kappa_1, \kappa_2, \kappa_4, \chi_4)\\
& = \Rmap \circ \alpha^*(\kappa_1, \kappa_2,\kappa_4, \chi_4)\\
& = \Rmap(\alpha^*\kappa_1,\alpha^*\kappa_2,\alpha^*\kappa_4 ,\alpha^*\chi_4)\\
& = \Rmap(\alpha) = 0\, ,
\end{align*}
where we have used again the isomorphism $H^n(X,\cdot)=[X,K(\cdot,n)]\, .$
By lifting the homotopy $k\circ\alpha \sim 0$ to $K$ we can achieve that $\alpha(\BB\OO)\subseteq F$.
Thus, there exists a lift $\tilde \alpha: \BB\OO \to F$ and we have the following
commutative diagram
\begin{align}
\label{rel2}
\xymatrix{
&F\ar[r]^j & K\ar[r]^-{k} & K(\Z_4,4)\\
& &\BB\OO\ar@{-->}[lu]^{\tilde \alpha}\ar[u]^{\alpha}\\
}
\end{align}

\begin{Lemma}
\label{5-equival-alpha} The mapping $\tilde \alpha$ is a
$5$-equivalence.
\end{Lemma}

\emph{Proof of the Lemma}:
We will use that $j_\ast\circ {\tilde \alpha}_\ast = \alpha_\ast$ and determine
${\tilde \alpha}_\ast$ from $j_\ast$ and $\alpha_\ast$. The low-dimensional
homotopy groups of $\BB\OO$ are
$$
\pi_1(\BB\OO)=\pi_2(\BB\OO)=\Z_2
 \,,~~~~~~
\pi_3(\BB\OO)=0
 \,,~~~~~~
\pi_4(\BB\OO) = \Z\,.
$$
The homotopy groups of $K$ are
$$
\pi_1(K)=\pi_2(K)=\Z_2
 \,,~~~~~~
\pi_4(K) = \Z_2\oplus\Z
 \,,~~~~~~
\pi_i(K) = 0 \,,~~~ i=3,i\geq 5\,.
$$
The exact homotopy sequence of the fibration \eqref{rel2} decomposes into the
portions
$$
0 \to \pi_i(F) \stackrel{j_\ast}{\to} \pi_i(K) \stackrel{k_\ast}{\to} 0
 \,,~~~~~~
i\neq3,4\,,
$$
and
 \beq\label{G-sqcK}
0 \to \pi_4(F)
 \stackrel{j_\ast}{\to}
\pi_4(K)
 \stackrel{k_\ast}{\to}
\pi_4(K(\Z_4,4))
 \to
\pi_3(F)
 \stackrel{j_\ast}{\to}
0\,.
 \eeq
It follows that $j_\ast$ is an isomorphism for all $i \neq 3,4$ and injective
for $i=4$.
First, this implies $\pi_5(F) = \pi_5(K) = 0$, hence ${\tilde \alpha}_*$ is onto for $i= 5$.
Second, we conclude that for $i=1,2$, ${\tilde \alpha}_\ast$ is an isomorphism iff so is
$\alpha_\ast$.

We determine ${\tilde \alpha}_*$ in dimension $i=3$ using homology. Let $\nu$ be an element of the subgroup
$H_4(K(\Z,4)) \subset H_4(K)$ . Evaluation of
\eqref{G-Rmap} on $1 \times 1 \times 1 \times \nu$ yields for the l.h.s.
$$
k^\ast(\iota_4) (1 \times 1 \times 1 \times \nu)
 =
\iota_4(k_\ast (1 \times 1 \times 1 \times \nu))
$$
and for the r.h.s.\
$$
\Rmap(\kappa_1,\kappa_2,\kappa_4,\chi_4) (1 \times 1 \times 1 \times \nu)
 =
\rho_4\circ\chi_4(\nu) \,.
$$
Thus,
$$
k_\ast (1 \times 1 \times 1 \times \nu)
 =
{\iota_4}^{-1}\circ\rho_4\circ\chi_4(\nu)
\,.
$$
Since $\chi_4$ and $\iota_4$ are isomorphisms and $\rho_4$ is surjective,
$k_\ast$ is surjective. Hence, $\pi_3(F) = 0$. Since also
$\pi_3(\BB\OO) = \pi_2(\mr O) = 0$, ${\tilde \alpha}_\ast$ is trivially an isomorphism in
dimension $i=3$.

Thus, it remains to show that $\alpha_\ast$ is an isomorphism in dimensions $i=1,2$
and ... in dimension $i=4$. We have
$\alpha_* =  w_{1*} \times w_{2*} \times w_{4*} \times p_{1*} \, .$ Thus, in dimension $i = 1,2$ we get
$$
\alpha_* = w_{i_*} : \pi_i(\BB\OO) \to \pi_i(K(\Z_2,i))
$$ and in dimension 4 we obtain
$$
\alpha_* =   w_{4*} \times p_{1*} : \pi_4(\BB\OO) \to \pi_4(K(\Z_2,4)) \times \pi_4(K(\Z,4))\, .
$$
We calculate $w_{i_*}$ for $i = 1,2,4$ : Let $[\xi] \in \pi_i(\BB\OO) = [S^i, \BB\OO]\, .$ Then
$w_{i_*}\left([\xi]\right) = [w_i \circ \xi] \, .$ By the isomorphism $[S^i,K(\Z_2,i)] \cong H^i(S^i,\Z_2)$ the element
$[w_j \circ \xi]$ corresponds to
$$
(w_i \circ \xi)^*(\kappa_i) = \xi^* \circ w_i^* (\kappa_i) = \xi^* w_i = w_i(\xi)\, ,
$$
where $w_i(\xi)$ is the $i$-th Stiefel-Whitney class of the real vector bundle $\xi\, .$
Since $\pi_i(K(\Z_2,i)) = \Z_2$ and since there exist real vector bundles with nonvanishing $i$-th Stiefel-Whitney class in all cases (the real, complex and quaternionic Hopf bundles), $ w_{i_*}$ is surjective. Since $\pi_1(\BB\OO) = \Z_2 = \pi_2(\BB\OO)$ and $\pi_4(\BB\OO) = \Z$ we conclude that $w_{i_*}$  is an isomorphism for $i = 1,2$ and $w_{4_*}$ is reduction modulo $2\, .$ This finishes the proof for  $i = 1,2 \, .$

Next, we calculate $p_{1*}$: As above, for $[\xi] \in \pi_4(\BB\OO)\, ,$ we get
$$
p_{1_*}\left([\xi]\right) = p_1(\xi)\, .
$$
Using the standard relation $\rho_2 p_1 = w_2^2 \, ,$ see
Proposition \ref{P-realrelCC}, together with $H^2(S^4,Z_2) = 0$,
we conclude $\rho_2p_1(\xi)=0 \, .$ Thus, $p_{1_*}$ is
multiplication by an even number. Consider the quaternionic line
bundle associated with the quaternionic Hopf bundle $S^3\ra S^7\ra
S^4\, ,$ which can be viewed either as a real bundle $\xi$ or as a
complex bundle $\eta\, .$ Then, $\xi^{\C} = \eta \oplus \eta$ and
the third relation in Proposition \ref{P-relBvp} together with the
Whitney sum formula for the Chern classes yields
$$
p_1(\xi)= -c_2(\xi^{\C})= -2c_2(\eta)-c_1(\eta)^2 = -2c_2(\eta)\, .
$$
Since $c_2(\eta)$ is a generator of $H^4(S^4,\Z)$ we conclude that $(p_1)_*$ is multiplication by two. Hence, we get
$$
\alpha_*=(w_4)_*\times(p_1)_* :\pi_4(\BB\OO)=\Z\ra\Z_2\oplus \Z=\pi_4(K) \, , \quad n\mapsto (n\mod 2, 2n) \, .
$$
In particular, $\alpha_*$ is injective. Since $j_*$ is also injective and $\alpha_*$ and $j_*$ have the same image,
${\tilde \alpha}_*$ is an isomorphism.
\bewe\\

We continue with the proof of the Theorem.
Since $\tilde \alpha$ is a $5$-equivalence the sets $[X,\BB\OO]$ and $[X,F]$
are isomorphic. By iterating the process of forming homotopy fibers one obtains a sequence of fibrations associated with the fibration \eqref{rel2}:
$$
\ldots \lra \Omega F \lra \Omega K \lra \Omega B \lra F \lra K \lra B\, ,
$$
with $B = K(\Z_4,4) \, .$
Consider the following exact sequence related to this fibration (obtained by applying the functor $[X,-]$):
\[
\cdots\lra [X,\Omega K]\stackrel{\Omega k_*}{\lra}[X,K(\Z_4,3)] \stackrel{i_*}{\lra} [X,F]
\stackrel{j_*}{\lra} [X,K]\stackrel{k_*}{\lra} [X,K(\Z_4,4)] \, ,
\]
where $\Omega B = K(\Z_4,3)$ has been used.
The aim is to find a condition which ensures the injectivity of the map $j_*$ in the sequence above.
Obviously, $j_*$ is injective if and only if $i_*$ is the zero map. The latter is true if and only if
$\Omega k_*$ is surjective. Thus, we have to calculate $\Omega k \, .$
We know that $\Omega k\in [\Omega K,K(\Z_4,3)]=H^3(\Omega K, \Z_4)$. Under the isomorphism
$ H^3(K(\Z_4,3),\Z_4) = [K(\Z_4,3),K(\Z_4,3)]\, ,$ the element $\iota_3\in H^3(K(\Z_4,3),\Z_4)$
obviously corresponds to the identity $[\id]\, .$ Thus,
$(\Omega k)^*(\iota_3)$ corresponds to $[\id \circ \Omega k] = [\Omega k]$ and
computing
$\Omega k$ is equivalent to computing $(\Omega k)^*(\iota_3)\, .$
For this calculation consider the commutative diagram built from the path loop fibrations of
$K$ and $K(\Z_4,4)$ respectively:
\begin{align*}
\xymatrix{
&\Omega K\ar[r]^-{\Omega k}\ar[d] &  K(\Z_4,3)\ar[d]\\
&PK\ar[r]\ar[d] & PK(\Z_4,4)\ar[d]\\
& K\ar[r]^-{k} & K(\Z_4,4)\\
}
\end{align*}
In what follows all suspension homomorphisms\footnote{see Appendix
\ref{A-em}} occuring in the text will be denoted by $\sigma\,.$
Using \eqref{sigma-EMcL} together with $(\Omega k)^* \circ \sigma
= \sigma \circ k^*\,, $ see \cite[Chapter 6.2]{mccleary}, and
\eqref{G-Rmap} we get:
\begin{align*}
(\Omega k)^* (\iota_3) &= (\Omega k)^*\big(\sigma (\iota_4)\big)\\
& = \sigma \big( k^*(\iota_4)\big)\\
& = \sigma \big( \Rmap(\kappa_1,\kappa_2,\kappa_4,\chi_4) \big)\\
& = \sigma \big(\rho_4 \chi_4 -\frak{P}\kappa_2 + \theta_*(\kappa_1\mathrm{Sq}^1 \kappa_2+ \kappa_4)\big)\, .
\end{align*}
Obviously, $\Omega k_*$ is surjective if so is its restriction to $K(\ZZ,3)$,
which is the map induced by the first summand in the last line above. We have, see
\cite[Chapter 6.2]{mccleary},
$$
\sigma \circ \rho_4(\chi_4)=\rho_4 \circ \sigma (\chi_4)
= \rho_4(\chi_3) \, ,
$$
where $\chi_3 \in H^3(K(\Z,3),\Z)$ is the fundamental class and $K(\Z,3) = \Omega K(\Z,4) \subset \Omega K\,.$
Now let $f\in [K(\Z,3),K(\Z_4,3)]$ denote the restriction of $\Omega k$ induced by $\rho_4(\chi_3)\, ,$ i.e.
\[
f^*(\iota_3)=\rho_4(\chi_3)\, .
\]
On the level of cohomology equivalence classes that means \beq
\label{omega-k-rho4} [\iota_3\circ\Omega
k]_{\mathrm{co}}=[\rho_4\circ\chi_3]_{\mathrm{co}}\, . \eeq We
claim that the following diagram is commutative:
\begin{align*}
\xymatrix{
[X,K(\Z,3)]\ar[d]\ar[rr]^-{f_*} & & [X,K(\Z_4,3)]\ar[d]\\
H^3(X;\Z)\ar[rr]^-{\rho_4} & & H^3(X;\Z_4)
}
\end{align*}
Here the unnamed arrows are the natural isomorphisms, see Appendix \ref{A-em}.
Let $g\in [X,K(\Z,3)]\, .$ Using \eqref{omega-k-rho4} we get
\[
\rho_4g^*(\chi_3) = [\rho_4\circ\chi_3\circ g]_{\mathrm{co}} =
[\iota_3\circ f\circ g]_{\mathrm{co}} = \big(f\circ g\big)^*(\iota_3)\, .
\]
Thus, the above diagram is commutative, indeed.
That means $j_*$ is injective if the map
\[
H^3(X,\Z)\ra H^3(X,\Z_4):\ x\mapsto \rho_4x
\]
is surjective. For the final step consider the following portion of the Bockstein sequence associated to
$\Z\stackrel{4}{\ra}\Z\stackrel{\rho_4}{\ra}\Z_4$:
\begin{align*}
\xymatrix{
H^3(X,\Z)\ar[r]^-{\rho_4} & H^3(X,\Z_4)\ar[r]^-{\beta} & H^4(X,\Z)\ar[r]^-{4} &  H^4(X,\Z)
}
\end{align*}
Thus, $\rho_4$ is surjective iff $\beta=0 \, ,$ that means iff multiplication by $4$ is injective, that means iff $H^4(X,\Z)$ has no elements of order $4$. This finishes the proof.

\bewe

\begin{cor}
\label{chs02} The image of the mapping $\alpha_*$ defined by
\eqref{Def-alpha} coincides with the set of $4$-tuples
$$
\beta = (u_1,u_2,u_4,v_4) \in H^1(X,\Z_2)\oplus H^2(X,\Z_2)\oplus
H^4(X,\Z_2)\oplus H^4(X,\Z)
$$
fulfilling the relation
\begin{equation}
\label{Rmap-beta} \Rmap (u_1,u_2,u_4,v_4) = 0 \, .
\end{equation}
\end{cor}

\emph{Proof}: Let $\beta $ be a $4$-tuple fulfilling relation
\eqref{Rmap-beta} and let us denote the corresponding mappping $X
\ra K$ also by $\beta\, .$ By the same arguments as above, there
exists a lift $\tilde \beta$ of $\beta\, , $ i.e. the following
diagram commutes:
\begin{align}
\label{diag-beta}
\xymatrix{
&F\ar[r]^j & K\ar[r]^-{k} & K(\Z_4,4)\\
& &X\ar@{-->}[lu]^{\tilde \beta}\ar[u]^{\beta}\\
}
\end{align}
Putting together the diagrams \eqref{rel2} and \eqref{diag-beta} we obtain
\begin{align}
\label{diag-alpha-beta} \xymatrix{
& \BB\OO\ar[ld]_-{\tilde \alpha}\ar[d]^-{\alpha}\\
F\ar[r]^-{j} & K\ar[r]^-{k} & K(\Z_4,4)\\
& X\ar[lu]^-{\tilde \beta}\ar[u]^-{\beta}
}
\end{align}
By Lemma \ref{5-equival-alpha}, $\tilde \alpha$ induces an
isomorphism
$$
\tilde \alpha_* \colon [X,F] \ra [X,\BB\OO] \, .
$$
Thus, using the inverse isomorphism $(\tilde \alpha_*)^{-1}$ we get a mapping class
$\hat \beta := (\tilde \alpha_*)^{-1} (\tilde \beta) \in [X,\BB\OO]\, .$
We compute
$$
\alpha \circ \hat \beta = j \circ {\tilde \alpha} \circ \hat \beta
= j \circ {\tilde \alpha_*} \big((\tilde \alpha_*)^{-1} (\tilde \beta) \big) = j \circ \tilde \beta = \beta\, ,
$$
or on the level of cohomology classes $ \alpha_* (\hat \beta) = \beta \, .$
\bewe\\ \\

Next, in analogy to the situation in Theorem \ref{chs1} we
consider the $3$-tuple $\alpha'= (w_1,w_2,p_1)$ of cohomology
elements with $w_k \in H^k(\BB\OO(3),\Z_2)$ and $p_1\in
H^4(\BB\OO(3),\Z)$ being the Stiefel-Whitney classes and the
Pontryagin class respectively.  As above, $\alpha'$ defines the
mapping
\begin{equation}
\label{Def-alpha'}
\alpha'_*:[X,\BB\OO(3)]\ra H^1(X,\Z_2)\oplus H^2(X,\Z_2)\oplus H^4(X,\Z)\, , \quad
\alpha'_*(\xi): = \xi^*(\alpha') \, ,
\end{equation}
for any CW-complex $X \, .$ This means $ \alpha'_*(\xi) = \big(w_1(\xi),w_2(\xi),p_1(\xi)\big) \, .$

\begin{thm}
\label{chs01}

Let $X$ be a complex of dimension four. Then, the mapping $\alpha'_*$ defined by \eqref{Def-alpha'}
is injective if $H^4(X,\Z)$ has no element of order $4$.
\end{thm}
\begin{rem}
\label{rem-BO3} Let $X$ be a $4$-dimensional CW-complex without
$4$-torsion in $H^4(X,\Z)$. Then, according to the above theorem
the sets $[X,\BB\OO(3)]$ are classified by tuples $(w_1,w_2,p_1)$.
Moreover, via the natural projection $\BB\SO(3)\ra \BB\OO(3)$ we
gain a classification of $[X,\BB\SO(3)]$ given by $(w_2,p_1)$.
\end{rem}
The proof of Theorem \ref{chs01} is completely analogous to the proof of Theorem \ref{chs1} and will be omitted.
We also have the analogous
\begin{cor}
\label{Cor-BO(3)} The image of the mapping $\alpha'_*$ defined by
\eqref{Def-alpha'} coincides with the set of $3$-tuples
$$
(u_1,u_2,v_4) \in H^1(X,\Z_2)\oplus H^2(X,\Z_2)\oplus H^4(X,\Z)
$$
fulfilling the relation
\begin{equation}
\label{relation-BO(3)} \rho_4 v_4 -\frak{P}u_2
-\theta_*(u_1\mathrm{Sq}^1 u_2)=0 \, .
\end{equation}
\end{cor}

The next result generalizes a classical one (see \cite{thomas2}).

\begin{thm}\label{chs3}

Let $X$ be a complex of even dimension $n$. Then the number of different isomorphism classes of
$n$-dimensional orientable vector bundles which have the same Euler class and which are stably equivalent
is bounded from above by the number of  $2$-torsion elements in $H^n(X,\Z)$.

\end{thm}

\begin{cor}\label{chs4}

If $H^n(X,\Z)$ has no $2$-torsion then two vector bundles $\xi$ and $\eta$ over $X$ are isomorphic iff
they are stably equivalent and have the same Euler class.

\end{cor}

\bre \label{stableEq-4}

For the case $n=4$ and under the
assumption that $H^4(X,\Z)$ has no $2$-torsion, it follows from
Theorem \ref{chs1} that two principal $\SO(4)$-bundles are stably
equivalent iff their second and fourth Stiefel-Whitney classes and
their first Pontryagin classes coincide. Thus, principal
$\SO(4)$-bundles $P$ over $X$ are classified by tuples
$(w_2(P),w_4(P),p_1(P),e(P))\, .$ This follows from the fact that
$(w_2(P),w_4(P),p_1(P)) = f_P^* \circ \mu^*(w_2,w_4,p_1)\, ,$
where $\mu : \BB \SO(4) \to \BB \SO$ is the natural projection,
$f_P$ is the classifying map of $P$ and $(w_2,w_4,p_1)$ are the
Stiefel-Whitney classes and the first Pontryagin class in the
cohomology of $\BB \SO \, .$

\ere

\emph{Proof of Theorem \ref{chs3}}:
We formulate the statement in terms of a lifting property: Let there be given $2$ elements
$f_i \in [X,\BB\SO(n)]\, ,$ $i = 1,2\, ,$ which define stably equivalent bundles. Stable equivalence means
$\pi \circ f_1 = \pi \circ f_2 \,,$ with $\pi$ being the projection of the canonical fibration
$S^n \into \BB\SO(n) \ra \BB\SO(n+1) \, .$ Thus, the question is how many lifts (defining the same Euler class) of a given map
$f:X \ra \BB\SO(n+1)$ do exist. This problem will be solved by using the
Moore-Postnikov tower for the fibration $\BB\SO(n)\ra \BB\SO(n+1)$ (see
\cite{thomas1} for the details of this construction). It turns out that it is sufficient to consider
its first nontrivial stage:
\begin{align*}
\xymatrix{
& & S^n\ar[d]^-{\iota'} & K(\Z,n)\ar[d]^{\iota}\\
& & \BB\SO(n)\ar[d]^-{\pi}\ar[r]^-{\varphi} & E \ar[d]^-{p}\\
&X\ar[r]\ar@<0.5ex>[ur]^{f_1}\ar@<-0.5ex>[ur]_{f_2}& \BB\SO(n+1)\ar@{=}[r]& \BB\SO(n+1)\ar[r]^-{k}&K(\Z,n+1)\\
}
\end{align*}

For the convenience of the reader we recall that this diagram is commutative and has the following properties:
 \smallskip

1.~ The map $\varphi$ is an $(n+1)$-equivalence.
 \smallskip

2.~ The fibration $p : E \ra \BB\SO(n+1)$ is the pull-back of the path loop fibration of $K(\Z,n+1)$ under the mapping $k\,.$
 \smallskip

3.~ The map $k : \BB\SO(n+1) \ra K(\Z,n+1)$ is given by
\begin{equation}
\label{k-inv0} k^*(\iota_{n+1}) = \tau' (o_n)\, ,
\end{equation}
where $\tau'$ is the transgression homomorphisms, see Appendix \ref{A-em},
associated with the fibration $\pi : \BB\SO(n)\ra \BB\SO(n+1)\, ,$ $\iota_{n+1}$
is the fundamental class of $K(\Z,n+1)$ and $o_n \in H^n(S^n, \Z)$ is the $n$-th
cohomology generator.

\begin{lem}
\label{lem-k-inv}

Let $w_n$ be the $n$-th Stiefel-Whitney class of $\BB \SO(n+1)\,
.$ We have
\begin{equation}
\label{k-inv} \tau' (o_n) = \beta(w_n) \, ,
\end{equation}
where $\beta$ is the Bockstein homomorphism associated to $0 \to \Z \to \Z \to \Z_2 \to 0 \,.$

\end{lem}

\emph{Proof of the Lemma}:~
Consider the following portion of the Serre exact sequence associated to the
fibration $\pi : \BB\SO(n)\ra \BB\SO(n+1)$:
\[
 \cdots\lra H^{n}(S^n,\Z)\stackrel{\tau'}{\lra} H^{n+1}(\BB\SO(n+1),\Z)\stackrel{\pi^*}{\lra} H^{n+1}(\BB\SO(n),\Z).
\]
We have $\im\, (\tau') = \ker(\pi^*)$. Recall that $n$ is even.
According to \eqref{G-ESW} in Appendix \rref{S-cc}, the Euler class of
$\BB\SO(n+1)$ satisfies $\rho_2 e = w_{n+1}$. On the other hand, we have
$$
\rho_2 \circ \beta (w_{n}) = \Sq^1 (w_n) = w_{n+1} \, .
$$
Since $\rho_2 $ is injective on the torsion elements of
$H^*(\BB\SO(n+1),\Z)$, we conclude that $e = \beta(w_n) \,
.$ Now consider the Gysin sequence of the fibration $S^n\into
\BB\SO(n)\stackrel{\pi}{\ra}\BB\SO(n+1)$:
$$
\cdots\lra H^0(\BB\SO(n+1),\Z)\stackrel{\beta(w_n)}{\lra} H^{n+1}(\BB\SO(n+1),\Z) \stackrel{\pi^*}{\lra}
H^{n+1}(\BB\SO(n),\Z)\lra\cdots.
$$
Obviously, $\ker\pi^*$ is generated by $\beta(w_n)$ which proves
the lemma.
 \bewe
 \bigskip

By point 1 of the above list, the induced mapping
$$
\varphi_* : [X,\BB\SO(n)] \ra [X,E]
$$
is an isomorhism. Thus, instead of $f_i$ we consider their images under this
isomorphism, which we denote by the same letter. By point 2 as well as the above Lemma we have
$$
E= \{(b,\gamma)\in \BB\SO(n+1)\times PK(\Z,n+1)\ |\ \beta (w_n) (b) = \gamma (1)\} \, .
$$
Since $K(\Z,n)=\Omega K(\Z,n+1)$  we can define a multiplication $m:E\times K(\Z,n)\ra E$ by
$$
m\big((b,\gamma),\alpha\big) = (b,\gamma\circ \alpha) \, .
$$
In every point $x\in X$ the two lifts $f_1$ and $f_2$ differ only by an element
$\alpha\in \Omega K(\Z,n+1)=K(\Z,n)$. Thus, there is a map $d: X\ra K(\Z,n)$ such that
\begin{align}
\label{m}
f_2=m(f_1, d) \, ,
\end{align}
where $m$ is viewed as point-wise product.
\begin{lem}
\label{lem-m*} The induced mapping $m^*: H^n(E,\Z)\ra H^n(E\times
\Omega K(\Z,n+1))$ is given by \beq \label{induced-m*} m^*(x_n) =
x_n\otimes 1 + 1\otimes \iota^*(x_n) \, , \eeq where $\iota$ is
the inclusion of the fiber into the fibration $\Omega
K(\Z,n+1)\into E\ra \BB\SO(n+1)$.
\end{lem}

\emph{Proof of the Lemma}:
Using the Künneth Theorem we obtain
\begin{align*}
H^n(E\times \Omega K(\Z,n+1),\Z)& = H^n(E,\Z)\otimes H^0(\Omega K(\Z,n+1),\Z)\\
& \oplus H^0(E,\Z)\otimes H^n( \Omega K(\Z,n+1),\Z).
\end{align*}
Since $H^0(E,\Z)=\Z= H^0(\Omega K(\Z,n+1),\Z)$ we get
\begin{align*}
m^*(x_n)&=(\sum_k (y_n)_k\otimes \mu_k)+(\sum_r \nu_r\otimes (z_n)_r)\\
&= (\sum_k \mu_k(y_n)_k\otimes 1)+(1\otimes\sum_r \nu_r(z_n)_r)\\
&= (y_n\otimes 1)+(1\otimes z_n) \,  .
\end{align*}
For the computation of $y_n$ and $z_n$ let
the canonical inclusions and projections of $E\times \Omega K(\Z,n+1)$ to its factors be
denoted according to the following diagram:
\begin{align*}
\xymatrix{
&  E\times\Omega K(\Z,n+1) \ar@<0.35ex>[ld]^-{pr_1}\ar@<-0.35ex>[rd]_-{pr_2}& &\\
E\ar@<0.35ex>[ru]^-{i_1} & & \Omega K(\Z,n+1)\ar@<-0.35ex>[ul]_-{i_2}
}
\end{align*}
Obviously, we have $m\circ i_1=\id$ and hence:
\[
x_n = i_1^* \, m^*(x_n) = i_1^*(y_n \otimes 1+ 1\otimes z_n) = y_n
\]
where the last equality simply follows from the fact that $i_1^* \, pr_1^*(y_n)= y_n$ and $i_1^* \, pr_2^*(y_n)=0$.
We show that $z_n=\iota^*(x_n)$. For that purpose note that $ m \circ i_2 = \iota$ and hence:
\[
\iota^*(x_n)=i_2^* \, m^*(x_n) = i_2^*(x_n \otimes 1+ 1\otimes z_n) = z_n \, .
\]
\bewe
\\

>From Proposition \ref{Lemma-SerreS} applied to the map
$\varphi:\BB\SO(n)\ra E$ we know that
$$
\varphi^*:H^n(E,\Z)\ra H^n(\BB\SO(n),\Z)
$$
is an isomorphism. In what follows we
identify the Euler class $e\in H^n(\BB\SO(n),\Z)$ with its image
$(\varphi^*)^{-1}(e)$ in $H^n(E,\Z)\,.$

\begin{lem}
\label{lem-iota-en}
Let $\iota_n\in H^n(K(\Z,n),\Z)$ be the fundamental class. Then we have
$$
\iota^*(e)=(\pm)2\iota_n \, .
$$
\end{lem}

\emph{Proof of the Lemma}:
Let us
consider the following portion of the Serre sequence for the fibration
$K(\Z,n) \stackrel{\iota}{\into} E\stackrel{p}{\ra} \BB\SO(n+1)$ as well as for the
fibration $S^n\stackrel{\iota'}{\into} \BB\SO(n)\stackrel{\pi}{\ra} \BB\SO(n+1)$:
\begin{align*}
\xymatrix{
H^n(\BB\SO(n+1),\Z)\ar[r]^-{p^*}\ar@{=}[d] & H^n(E,\Z)\ar[r]^-{\iota^*}\ar[d]^-{\varphi^*} &
H^n(K(\Z,n),\Z)\ar[d]^-{(\varphi_{\upharpoonright_{S^n}})^*}\\
H^n(\BB\SO(n+1),\Z)\ar[r]^-{\pi^*} & H^n(\BB\SO(n),\Z)\ar[r]^-{(\iota')^*} & H^n(S^n,\Z)\\
H^n(K(\Z,n),\Z)\ar[d]^-{(\varphi_{\upharpoonright_{S_n}})^*}\ar[r]^-{\tau} & H^{n+1}(\BB\SO(n+1),\Z)\ar[r]^-{p^*}\ar@{=}[d] &
H^{n+1}(E,\Z)\ar[d]^-{\varphi^*}\\
H^n(S^n,\Z)\ar[r]^-{\tau'} & H^{n+1}(\BB\SO(n+1),\Z)\ar[r]^-{\pi^*} &
H^{n+1}(\BB\SO(n),\Z)
}
\end{align*}
Since $p\circ\varphi = \pi$ we have $\pi^*=\varphi^*\circ p^*$ and
hence $\ker \pi^* = \ker \varphi^*\circ p^*$. Moreover, applying
Proposition \ref{Lemma-SerreS} again to the mapping
$\varphi:\BB\SO(n)\ra E$ we see that $\varphi^*:H^{n+1}(E,\Z)\ra
H^{n+1}(\BB\SO(n),\Z)$ is injective and hence $\ker \pi^* = \ker
p^*$. Since the Serre sequence is exact this implies $\im\, \tau'
= \im\, \tau$. Since both, $H^n(S^n,\Z)$ and $H^n(K(\Z,n),\Z)$ are
generated by a single element, namely $o_n$ and $\iota_n\, ,$
Lemma \ref{lem-k-inv} implies
$$
\tau'(o_n) = \beta(w_n) = \tau(\iota_n) \, .
$$
Since $\beta(w_n)$ is a $2$-torsion element the kernel of $\tau$
is given by the multiplication map by $2$. Since the Sequence is
exact this kernel equals $\im\, (\iota^*)$, i.e. the image of
$\iota^*$ is generated by $2 \iota_n\, .$ It is well-known that
$H^n(\BB\SO(n),\Z)/\im\,\pi^*$ is generated by the Euler class
$e$. Since $\pi^*=\varphi^*\circ p^*$ we see that
$H^n(E,\Z)/\im\, p^*$ is generated by $e$, too. (Recall that
we have identified $e$ with its image under $(\varphi^*)^{-1}$.)
Thus, we obtain the result $\iota^*(e)= \pm 2\iota_n\, .$
\bewe\\

For the final step
of the proof we write down  \eqref{m} more precisely as the composition:
\begin{align*}
\xymatrix{
&X\ar@/_1.5pc/[rrr]_-{f_2}\ar[r]^-{\Delta}& X\times X \ar[r]^-{f_1\times d}&
E\times K(\Z,n)\ar[r]^-{m}& E \\
}
\end{align*}
and calculate:
\begin{align*}
f_2^*(e)&=\Delta^*(f_1\times d)^*m^*(e)\\
& =\Delta^*(f_1\times d)^*\big(e\otimes 1 +1\otimes \iota^*(e)\big)\\
& =\Delta^*(f_1\times d)^*(e\otimes 1  \pm 1\otimes 2\iota_n)\\
& = \Delta^*\big(f_1^*(e)\otimes 1  \pm 1\otimes 2d^*(\iota_n)\big)\\
& = f_1^*(e) \pm 2d^*(\iota_n) \, .
\end{align*}
Assume that the two stably equivalent bundles $f_1$ and $f_2$ have the same
Euler class. Then, the last equation
implies that they can only differ by a map $d$ with the property $2d = 0\, .$
Realizing that the homotopy classes of such maps are in one-to-one
correspondence with $2$-torsion elements of $H^n(X,\Z)$ it completes the proof
of the theorem.
\bewe
 \bigskip

\bre \label{Rem-w3}

The Stiefel Whitney class $w_3$ does not enter in any of the above
classification results. This suggests that there should be a
relation expressing $w_3$ in terms of the other characteristic
classes. Indeed, \eqref{wu-formel} implies \beq
\label{rel-STW3} w_3(P) = w_1 w_2 + \Sq^1 w_2(P)\, . \eeq In
particular, for $X$ being simply connected or for principal
$SO(n)$-bundles $P$ we have $w_1(P) = 0 $ and, therefore,
 \beq\label{Rel-STW3}
w_3(P) = \Sq^1 w_2(P)\, .
 \eeq

\ere

For further use, we will now formulate the classification of
$\SO(n)$- and $\OO(n)$-bundles over $4$-dimensional CW-complexes
$X$ in a universal way. For convenience, in both cases we assume
that $X$ is simply connected, although this assumption is
necessary only for the $\OO(n)$-bundles\footnote{See however
Remark \rref{Bem-simplyconnected}}.

\btm\label{T-SObun}

Let $X$ be a simply connected CW-complex of dimension $4$ without
$2$-torsion in $H^4(X,\Z)$. $\SO(n)$-bundles over $X$ are
classified by the characteristic classes $w$, $p$, and $e$. More
precisely, the map
 \begin{eqnarray}\nonumber
[X,\BB\SO(n)]
 & \to &
H^2(X,\ZZ_2) \times H^4(X,\ZZ_2)
 \times H^4(X,\ZZ) \times H^n(X,\ZZ)
\\ \label{thm-classmap}
P & \mapsto & \big(w_2(P),w_4(P),p_1(P),e(P)\big)
 \end{eqnarray}
is a bijection onto the subset defined by the fundamental relation
\beq
\label{fundRel}
\Rmap(0,w_2(P),w_4(P),p_1(P)) \equiv \rho_4 \, p_1(P) -\frak{P}w_2(P) + \theta_* w_4(P) = 0
\eeq
and the following special relations:
 \smallskip

$n=1$:~ $w_2(P) = w_4(P) = p_1(P) = e(P) = 0$.
 \smallskip

$n=2$:~ $w_2(P) = \rho_2 e(P)$, $w_4(P) = 0$, $p_1(P) =
e(P)^2$.
 \smallskip

$n=3$:~ $w_4(P) = 0$, $\rho_2 e(P) = \Sq^1 w_2(P)$.
 \smallskip

$n=4$:~ $w_4(P) = \rho_2 e(P)$.

\etm

\bre

In case $n = 3$, the relation for $e$ determines $e(P)$ uniquely
because $e(P)$ is a $2$-torsion element here, see \cite[Prop.\
3.13]{vb}. Hence, the Euler class is a relevant parameter for
$n=2$ and $n=4$ only.

\ere

{\it Proof.}~We have $w_1(P) = 0 $ and $w_3(P) = \Sq^1 w_2(P)\, ,
$ by Remark \ref{Rem-w3}. For the stable case $n \geq 5\, ,$ the
theorem follows from Theorem \ref{chs1} and Corollary \ref{chs02}.
As mentioned earlier, see \eqref{rel1}, the fundamental relation
is fulfilled in all cases. For $n=1$, the assertion is obvious.
For $n=2$ we have the isomorphism $\vp_{\CC,\RR} : \UU(1) \to
\SO(2)$. According to Proposition \rref{P-relBvp},
$\big(\BB\vp_{\CC,\RR}\big)^\ast e = c_1$. Since $c_1$ defines a
bijection from $[X,\BB\UU(1)]$ onto $H^2(X,\ZZ)$, the same holds
for $e$ and $[X,\BB\SO(2)]$. It remains to check the relations.
The first one follows from the relation $\rho_2 e = w_2$, see
\eqref{G-ESW}. The second one is obvious. The third one follows
from Theorem \ref{App-thm2}.

For $n=3$, $w_4=0$. The relation for $e$ is due to
\eqref{rel-STW3} and \eqref{G-ESW}. Hence, the assertion follows
from Theorem \rref{chs01} and Corollary \rref{Cor-BO(3)}.

For $n=4 \, ,$ the classification is given by tuples
$(w_2(P),w_4(P),p_1(P),e(P))$ of characteristic classes according
to Theorem \rref{chs1} and Corollary \rref{chs4}. It remains to
show that the image of the mapping \eqref{thm-classmap} is defined
by the above relations: For that purpose, observe that for every
$\SO(4)$-bundle with classifying map $f$ there exists a stable
bundle $\mu \circ f$ induced via the natural projection $\mu:
\BB\SO(4) \to \BB\SO\,.$ For a given tuple $(w_2,w_4,p_1)$,
fulfilling the fundamental relation, there is always a stable
bundle and hence an $\SO(4)$-bundle $P$ with classifying map $f_P$
and classes $(w_2(P),w_4(P),p_1(P)) = f_P^* \circ
\mu^*(w_2,w_4,p_1)\, ,$ which also fulfil the fundamental
relation. This follows from Corollary \rref{chs02}, see also
Remark \ref{stableEq-4}. For any class $e \in H^4(X,\Z)$
fulfilling the relation $w_4(P) = \rho_2 e$ we take $c_2 = \BB
\varphi_{\C,\R} (e)\, .$ This is the second Chern class of a
principal $\UU(2)$-bundle with classifying map $g\, .$ Then, $\BB
\varphi_{\C,\R} \circ g$ defines a principal $\SO(4)$-bundle with
Euler class $e\, .$ The relation $w_4(P) = \rho_2 e$ follows from
\eqref{G-ESW}.
 \qed
 \bigskip

Since $X$ is simply connected, from the classification of
$[X,\BB\SO(n)]$ we obtain a classification of $[X,\BB\OO(n)]$ by
identifying $\SO(n)$-bundles which are reductions of the same
$\OO(n)$-bundle, i.e., which differ at most in the sign of the
Euler class. To formalize that, for $\alpha\in H^k(X,\ZZ)$ let
$[\alpha]$ denote the class of $\alpha$ under the equivalence
relation $\alpha\sim\beta \Leftrightarrow \alpha + \beta = 0$ and
let $\mr PH^k(X,\ZZ)$ denote the set of equivalence classes. On
$\mr PH^\ast(X,\ZZ)$, the following operations can be defined:
 \smallskip

--~ the cup product by $[\alpha][\beta] = [\alpha\beta]$,
 \smallskip

--~ the square as a map $\mr PH^k(X,\ZZ) \to H^{2k}(X,\ZZ)$ by
$[\alpha]^2 = \alpha^2$,
 \smallskip

--~ reduction mod $2$ as a map $\mr PH^\ast(X,\ZZ) \to
H^\ast(X,\ZZ_2)$ by $\rho_2([\alpha]) = \rho_2 \alpha$.
 \smallskip

Define a map $[e]:[X,\BB\OO(n)] \to \mr PH^n(X,\ZZ)$ by $[e](P) :=
[e(P_0)]$, where $P_0$ is some reduction of $P$ to $\SO(n)$. We
will loosely speak of $[e]$ as a characteristic class for
$\OO(n)$-bundles, although it is not a cohomology element of $X$
and it is not defined by a cohomology element of $\BB\OO(n)$. Passing
to equivalence classes of $e(P)$ in Theorem \rref{T-SObun} we
obtain

\bco\label{C-Obun}

Theorem \rref{T-SObun} remains true when $\SO(n)$ is replaced by
$\OO(n)$, $H^n(X,\ZZ)$ by $\mr PH^n(X,\ZZ)$ and $e(P)$ by
$[e](P)$.
 \qed

\eco


\subsection{Principal bundles with structure groups $H_0$, $H$ and $\mr
SH^\pm$}
 \label{SS-HFB-H}


We recall the classification of $\UU(n)$ and $\Sp(n)$-bundles over
a CW-complex $X$ of dimension $4$:
 \smallskip

--~ The Chern class $c$ of $\BB\UU(n)$ defines a bijection from
$[X,\BB\UU(n)]$ onto $H^2(X,\ZZ)$ for $n=1$ and onto
$H^2(X,\ZZ)\times H^4(X,\ZZ)$ for $n\geq 2$.
 \smallskip

--~ The Chern class $c$ of $\BB\Sp(n)$, defined by
$c\big(\Sp(n)\big) = \big(\BB\vp_{\HH,\CC}\big)^\ast
c\big(\UU(2n)\big)$, defines a bijection from $[X,\BB\Sp(n)]$ onto
$H^4(X,\ZZ)$. This is a consequence of the isomorphy of $\Sp(1)$
and $\SU(2)$ and the fact that in $4$ dimensions, $\Sp(1)$ is
already the stable case.
 \smallskip

Putting all of the above facts together we obtain the following
classification result for Lie groups, which appear as Howe
subgroups of $\OO(n)$ and $\Sp(n)$.

\btm \label{thm-classif-HoweB}\label{T-Hbun}

Let $X$ be a simply connected CW-complex of dimension $4$ without
$2$-torsion in $H^4(X,\Z)$ and let
 $
H = \I_{\KK_1}(m_1) \times\cdots\times \I_{\KK_r}(m_r)\,,
 $
where $\KK_i = \RR$, $\CC$, $\HH$. Principal bundles $Q$ over $X$
with structure group $H_0$ or $H$ are classified by the collection
of characteristic classes of their factors $Q_i$, i.e.,
 \smallskip

--~ in case of $H_0$ by $w(Q_i)$, $p(Q_i)$ and $e(Q_i)$ for
$\KK_i=\RR$ and $c(Q_i)$ for $\KK_i=\CC,\HH$,
 \smallskip

--~ in case of $H$ by $w(Q_i)$, $p(Q_i)$ and $[e](Q_i)$ for
$\KK_i=\RR$ and $c(Q_i)$ for $\KK_i=\CC,\HH$.
 \smallskip

For $\KK_i = \R\, ,$ the characteristic classes are subject to the relations
given in Theorem \rref{T-SObun}, Corollary \rref{C-Obun} and Formula
\eqref{rel-STW3}.
\qed

\etm

More precisely, the statement of the theorem reads as follows. Let
$K=H_0$ or $H$ and let $J_i$ denote the image of the
characteristic classes of the factor $Q_i$ in the cohomology of
$X$ (where it is understood that in case $K=H$ the image of the
Euler class is a subset of $\mr PH^{m_i}(M,\ZZ)$). Then the
collection of characteristic classes of the factors $Q_i$ of $Q$
defines a bijection from $[X,\BB K]$ onto $J_1\times\cdots\times
J_r$.
 \bigskip

It remains to treat the case of principal bundles whose structure
group is a subgroup of type $\typeSHpm$ of $\SO(n)$. In the following we assume
that $H$ has an $\OO$-factor (otherwise $H\subseteq\SO(n)$ and there is nothing
to be discussed). First,
consider type $\typeSHp$. For the notation, see
\eqref{G-def-SHpm}. Let $H$ be a Howe subgroup of $\OO(n)$ in
standard form and let $Q$ be a principal $\mr SH^+$-bundle over
$X$. Since the identity connected component $H_0$ of $H$ is also the identity
connected component of $\mr SH^+$ and since
$X$ is simply connected, any connected component of $Q$ is a
principal $H_0$-bundle. One may use the classification of $H_0$-bundles to
describe $Q$ in terms of the characteristic classes of its connected components.
In general, the latter will not be
isomorphic, so that one would have to identify their characteristic classes
appropriately to obtain a unique description of $Q$. Although this can be done
easily we will pursue another approach here.

For a Lie subgroup $G\subseteq G'$ and a principal $G$-bundle $P$ over $X$, let
$P^{G'}$ denote the extension of $P$ to the structure group
$G'$. This is the unique principal $G'$-bundle which contains $P$ as a reduction
to the subgroup $G$.\footnote{$P^{G'}$ can be constructed as the fiber bundle
associated with $P$ by virtue of the action of $G$ on $G'$ by left
multiplication. The action of $G'$ on $P^{G'}$ is induced from the action of
$G'$ on itself by right multiplication.} Consider the principal $H$-bundle
$Q^H$ obtained from $Q$
by extension of the structure group to $H$. $Q$ is a reduction of
that bundle to the subgroup $\mr SH^+$. Since 
$H/\mr SH^+ = \ZZ_2$, there 
exists exactly one further such reduction $Q'$, which might be
isomorphic to $Q$ but in general is not. Thus, we may classify
$\mr SH^+$-bundles by the characteristic classes of $H$-bundles
and a certain $\ZZ_2$-valued quantity. To construct it, choose
unique representatives in $H^\ast(X,\ZZ)$ of the classes in $\mr
PH^\ast(X,\ZZ)$ and define a map $\sigma:H^\ast(X,\ZZ) \to \ZZ_2$
by
$$
\sigma(\alpha)
 =
 \begin{cases}
\phantom{-}1 & \alpha \text{ is a representative,}
\\
-1 & \alpha \text{ is not.}
 \end{cases}
$$
For a principal $H_0$-bundle $Q_0$ define
$$
\sigma(Q_0)
 :=
\prod_{\KK_i = \RR} \sigma\big(e(Q_{0i})\big)^{k_i}\,.
$$

\ble\label{L-SHbun}

Let $Q_0$ and $Q'_0$ be principal $H_0$-bundles over $X$ which have
isomorphic extensions to $H$. Then $Q_0$ and $Q'_0$ have isomorphic
extensions to $\mr SH^+$ if and only if $\sigma(Q_0) = \sigma(Q'_0)$.

\ele

{\it Proof.}~ We may identify $Q_0$ and $Q_0'$ with subsets of the
extension $Q_0^H \equiv Q_0^{\prime H} = \tilde Q$. We claim that there exists
$h\in H$ transforming $Q_0$ to $Q_0'$ inside $\tilde Q$. To see this,
consider the quotient bundle $\tilde Q/H_0$. This is a principal bundle over $X$
with structure group $H/H_0$. The natural projection $p:\tilde Q\to
\tilde Q/H_0$ together with the homomorphism $H\to H/H_0$ yields a principal
bundle morphism covering the identity on $X$. There exist sections $s,s'$ in
$\tilde Q/H_0$ such that $Q_0 = p^{-1}(s(X))$ and $Q_0' = p^{-1}(s'(X))$. Since
$\tilde Q/H_0$
has finite fibers, $s$ and $s'$ are determined by their values at
a single point. Hence, there exists an element of $H/H_0$
transforming $s$ to $s'$. Then any representative of this element
transforms $Q_0$ to $Q_0'$.

Now let $h_i$ denote the projections of $h$ to the factors of $H$. Then
 \beq\label{G-deth-1}
 \textstyle
\det(h) = \prod_{\KK_i=\RR} \det(h_i)^{k_i}\,.
 \eeq
The group element $h_i$ transforms the factor $Q_{0i}$ of $Q_0$ to the factor
$Q_{0i}'$ of $Q_0'$, where both factors are viewed as subbundles of the
corresponding factor of $\tilde Q$. Hence, for the factors with $\KK_i=\RR$
there holds
 \beq\label{G-eQi}
e(Q_{0i}') = \det(h_i) e(Q_{0i})\,.
 \eeq
In case $e(Q_{0i}) = - e(Q_{0i})$, the factors $Q_{0i}$ and
$Q_{0i}'$ are isomorphic, no matter what determinant $h_i$ has. By
passing from $Q_0'$ to an isomorphic subbundle, denoted by the same symbol,
we may assume that $h_i = \II$ in that case. Under this assumption on $h$,
\eqref{G-deth-1} and \eqref{G-eQi} yield
 \beq\label{G-deth}
\sigma(Q_0') = \det(h) \, \sigma(Q_0)\,.
 \eeq
If $\sigma(Q_0') = \sigma(Q_0)$, \eqref{G-deth} implies that $h\in\mr
SH^+$ and hence $Q_0^{\mr SH^+} = Q_0^{\prime\mr SH^+}$. Conversely,
if $Q_0^{\mr SH^+} = Q_0^{\prime\mr SH^+}$ then $Q_0$ and $Q_0'$ project
to the same section in the quotient bundle $\tilde Q/\mr SH^+$. Then
$h$ projects to the class of the identity in $H/\mr SH^+$, i.e., $h\in\mr
SH^+$. Hence, \eqref{G-deth} implies $\sigma(Q_0') = \sigma(Q_0)$.
 \qed
 \bigskip

Lemma \rref{L-SHbun} yields in particular that for an $\mr
SH^+$-bundle $Q$ we may define $\sigma(Q)$ to be $\sigma(Q_0)$ for
some connected component $Q_0$. By construction, this definition applies when 
$H$ possesses an $\OO$-factor. If it does not, we define $\sigma(Q)=1$. Then
Theorem \rref{T-Hbun} and Lemma \rref{L-SHbun} imply

\btm\label{T-SHbun}

Let $X$ be a simply connected CW-complex of dimension $4$ without
$2$-torsion in $H^4(X,\Z)$ and let
 $
H
 =
\IF{\KK_1}{m_1}{k_1} \times\cdots\times \IF{\KK_r}{m_r}{k_r}
 $
be a Howe subgroup of $\OO(n)$ of standard form. Principal $\mr
SH^+$-bundles are classified by the characteristic classes of the
factors of their extension to the structure group $H$ and the
invariant $\sigma$.
 \qed

\etm

Finally, consider type $\typeSHm$. For a principal bundle $Q$ with
structure group $\mr SH^-$ we form the extension $Q^{\OO(n)}$ and
transform $Q$ inside $Q^{\OO(n)}$ by the element $a_{(n)}$ of
$\OO(n)$ we had chosen for the definition of $\mr SH^-$. (We could
take any $a\in\OO(n)$ such that $a \mr SH^- a^{-1} = \mr SH^+$.)
Since the transformed bundle has structure group $\mr SH^+$, we
can define $\sigma(Q)$ and the characteristic classes of $Q$ to be
given by the corresponding quantities of the transformed bundle.

\bre

Since the Euler class is a $2$-torsion element in odd dimension,
only $\SO(2)$- and $\SO(4)$-factors contribute to $\sigma$. If
moreover one is interested in holonomy-induced Howe subbundles
only, $\SO(2)$-factors do not appear, hence here it is just the
Euler classes of the $\SO(4)$-factors which determine $\sigma$.

\ere

We now pass to manifolds as base spaces. It was shown by Thom
\cite{thom} and Milnor \cite{milnor3} that every closed
$n$-dimensional manifold $M$ is homeomorphic to a finite
$n$-dimensional CW-complex. Thus, all the above classification
results carry over to bundles over closed manifolds. Since in our
situation $M$ is in addition simply connected and hence
orientable, Poincar\'e duality holds. This implies:

--~ $H^n(M,\ZZ)$ is torsion-free. Hence, Theorems \rref{T-SObun},
\rref{T-Hbun} and \rref{T-SHbun} as well as Corollary
\rref{C-Obun} apply to closed and simply connected manifolds of
dimension $4$.

--~ $H^3(M,\ZZ)=0$. Hence the Euler class of any $\SO(3)$-bundle
vanishes.
 \comment{

under the weaker assumption $H^1(M,\ZZ_2)=0$, use the universal
coefficient thm to obtain that $\Hom(H_1M,\ZZ_2) = 0$ and
Poincar\'e duality to deduce that $H^3(M,\ZZ)$ does not contain
elements with $2$-torsion.

 }%

--~ $H^3(M,\ZZ_2) = 0$. Hence, the induced homomorphism
$\theta_\ast$ is injective in dimension $4$. Then $w_4(P)$ is
uniquely determined by the fundamental relation \eqref{fundRel} in
terms of $w_2(P)$ and $p_1(P)$:
 \beq\label{G-w4}
\theta_\ast w_4(P)
 =
\mf P w_2(P) - \rho_4 p_1(P)\,.
 \eeq
This means that we can ignore $w_4$ in the classification of
$\SO(n)$- and $\OO(n)$-bundles over $M$. Replacing $w_4$ by \eqref{G-w4} in the
relations listed in Theorem \rref{T-SObun}, we obtain new
relations involving only $w_2$, $p_1$ and $e$ or $[e]$,
respectively, and the fundamental relation is automatically
satisfied. In particular, for $\OO(n)$ and $\SO(n)$ with $n\geq 5$ there are no
relations any more.

\bre

If we drop the restriction on $X$ to be simply connected, the
bundles need not be orientable so that the classification of
$[X,\BB\OO(4)]$ may not be reduced to the classification of
$[X,\BB\SO(4)]$ anymore. Presumabely the classification can be
done for orientable four dimensional base complexes using similar
methods as in the proof of Theorem \ref{chs3}, but replacing the
ordinary Euler class by the Euler class with local coefficients.

\ere


\section{Howe subbundles}
\label{Howe-subbundles} \label{S-HSB}


In this section we determine the isomorphism classes of Howe
subbundles of principal bundles $P$ with structure group $G =
\OO(n)$, $\SO(n)$ or $\Sp(n)$ over closed  simply connected
manifolds of dimension $4$. Since

--~ subsequently we have to further factorize the isomorphism
classes by the action of the structure group $G$ of $P$,

--~ this action transforms the structure group of the reduction by
conjugation,

--~ any Howe subgroup is conjugate to one of standard form, see
Section \rref{S-HSG},

it suffices to discuss reductions of $P$ to Howe subgroups of
standard form. Such reductions will be referred to as Howe
subbundles of standard form.

First, we explain the idea. If a principal $H$-bundle $Q$ and a
principal $G$-bundle $P$ are given then $Q$ is a reduction of $P$
if and only if the extension $Q^G$ of $Q$ to the structure group
$G$ is isomorphic to $P$. According to the results of Section
\rref{S-HFB}, in the case where $G=\OO(n)$, $\SO(n)$ or $\Sp(n)$
and $M$ is closed  simply connected and of dimension $4$
it suffices to compare the characteristic classes. Thus, all we
have to do is to compute the characteristic classes of the
$G$-bundle $Q^G$ in terms of the characteristic classes of the
$H$-bundle $Q$. The classifying map of $Q^G$ is $\BB j\circ f$,
where $f:M\to\BB H$ is the classifying map of $Q$ and $\BB j:\BB H \to \BB G$ is
the classifying map of the extension of the universal $H$-bundle to
the structure group $G$.
 \comment{
The situation is characterized by the
following diagram
 \begin{align}
 \label{diagram}
\xymatrix{
~~Q^G\ar[d]^-{G}~~ & ~~Q\ar[d]^-{H}~~ & ~~\EE H\ar[d]^-{H} & (\EE H)^G\ar[d]^-{G} & \EE G\ar[d]^-{G}
\\
M \ar[r]^-{\id} & M\ar[r]^-{f} & \BB H\ar[r]^-{\id} & \BB H\ar[r]^-{\BB j} & \BB G \, ,
}
 \end{align}
 }%
Consequently, $\alpha(Q^G) = f^* \circ (B j)^* \alpha$. It follows that
$Q$ is a reduction of $P$ iff there holds $\alpha(P) = f^* \circ (B j)^* \alpha$
for any characteristic class $\alpha$ of $G$-bundles. This provides a set of
equations in the cohomology of $M$ whose solutions classify the reductions of
$P$ to the structure group $H$.


\subsection{Howe subbundles of principal $\OO(n)$- and $\SO(n)$-bundles}
\label{S-HSB-O}


We treat reductions of $\OO(n)$ to Howe subgroups and reductions
of $\SO(n)$ to subgroups of type $\typeSHpm$ simultaneously. Let
$H = \IF{\KK_1}{m_1}{k_1} \times \cdots \times
\IF{\KK_r}{m_r}{k_r}$ be a Howe subgroup of $\OO(n)$. We observe:
since $M$ is simply connected, a principal $H$-bundle $Q$ is a
reduction of a principal $\OO(n)$-bundle $P$ iff there exists a
common reduction of $Q$ and $P$ to $H_0$. A similar statement
holds for $H$ and $\OO(n)$ replaced by $\mr SH^+$  and $\SO(n)$.
Hence, it suffices to study the case of an $H_0$-bundle $Q$ and to
compute the characteristic classes of its extension $Q^{\SO(n)}$
to $\SO(n)$, i.e., $\alpha\big(Q^{\SO(n)}\big)$, where $\alpha =
w$, $p$, $e$. Let $f_i:M\to\BB\I_{\KK_i}(m_i)_0$ denote the
classifying map of the factor $Q_i$ of $Q$. For any set $X$ let
$\Delta_r:X\to X\times\stackrel{r}{\cdots}\times X$ denote the
diagonal map. For clarity, in the calculation to follow we label characteristic
classes by the group they belong to.

The classifying map of $Q$ is given by
 \beq\label{G-clfmap-Q0}
f =
\left(\prod\nolimits_if_i\right)\circ\Delta_r\,.
 \eeq
The classifying map of $Q^{\SO(n)}$ is $\BB j\circ f$, where $\BB
j:\BB H_0 \to \BB \SO(n)$ is the map of classifying spaces induced
by $j$.  Hence,
 \beq\label{G-cc-ext-a}
\alpha(Q^{\SO(n)})
 =
f^\ast \, (\BB j)^\ast \alpha(\SO(n))
 \,,~~~~~~
\alpha = w,p,e\,.
 \eeq
We decompose $j:H_0\to \SO(n)$ as
$$
H_0
 =
\prod\nolimits_i\I_{\KK_i}(m_i)_0
 \stackrel{\prod\nolimits_i\vp_{\KK_i,\RR}}{\longrightarrow}
\prod\nolimits_i\SO(\delta_im_i)
 \stackrel{\prod\nolimits_i\Delta_{k_i}}{\longrightarrow}
\prod\nolimits_i\SO(\delta_im_i)
 \times
\stackrel{k_i}{\cdots}
 \times
\SO(\delta_im_i)
 \stackrel{i}{\longrightarrow}
\SO(n)\,,
$$
where $\delta_i = \dim_\RR(\KK_i)$ and $i$ denotes the standard blockwise
embedding. Hence,
 \beq\label{G-Bja}
(\BB j)^\ast ~ \alpha(\SO(n))
 =
\left(\prod\nolimits_i \big(\BB\vp_{\KK_i,\RR}\big)^\ast\right)
 \circ
\left(\prod\nolimits_i \Delta_{k_i}^\ast\right)
 \circ
(\BB i)^\ast ~ \alpha\big(\SO(n)\big)\,.
 \eeq
Since for the standard embedding $\vp:\SO(l)\to\SO(l+1)$ there
holds $(\BB\vp)^\ast\alpha(\SO(l+1)) = \alpha(\SO(l))$,
 \beq\label{G-Bia}
(\BB i)^\ast\alpha(\SO(n))
 =
 \prod\nolimits_i
\alpha(\SO(\delta_im_i))
 \times\stackrel{k_i}{\cdots}\times
\alpha(\SO(\delta_im_i))\,,
 \eeq
where $\times$ and $\prod_i$ refer to the cohomology cross
product. Eqs.\ \eqref{G-clfmap-Q0}, \eqref{G-cc-ext-a},
\eqref{G-Bja} and \eqref{G-Bia} imply
 \beq\label{G-ext-SO-a}
\alpha\left(Q^{\SO(n)}\right)
 =
\Big(f_1^\ast ~ \big(\BB\vp_{\KK_1,\RR}\big)^\ast
 \,\alpha(\SO(\delta_1m_1))\Big)^{k_1}
 \cdots
\Big(f_r^\ast ~ \big(\BB\vp_{\KK_r,\RR}\big)^\ast
 \,\alpha(\SO(\delta_rm_r))\Big)^{k_r}\,,
 \eeq
where powers and products refer to the cup product on $M$. Since
$\BB\vp_{\KK_i,\RR}\circ f_i$ is the classifying map of the
extension $\hat Q_i := Q_i^{\SO(\delta_im_i)}$ of the factor $Q_i$
('realification'), we can simplify \eqref{G-ext-SO-a} to
 \beq\label{G-ext-SO-a-bun}
\alpha\left(Q^{\SO(n)}\right)
 =
\alpha\big(\hat Q_1\big)^{k_1}
 \cdots
\alpha\big(\hat Q_r\big)^{k_r}\,.
 \eeq
For $\KK_i=\RR$, we have $\delta_i=1$ and $\alpha\big(\hat
Q_i\big) = \alpha\big(Q_i\big)$. For $\KK_i=\CC,\HH\, ,$ denote
$\tilde p = \sum_k (-1)^k p_k$ and $\tilde c = \sum_k (-1)^k c_k$.
The standard relations between Chern classes and real
characteristic classes listed in Proposition \rref{P-relBvp} imply
 \beq\label{G-cc-tiQi-O}
w\big(\hat Q_i\big) = \rho_2\,c\big(Q_i\big)
 \,,~~~~~~
\tilde p\big(\hat Q_i\big)
 =
c\big(Q_i\big) \tilde c\big(Q_i\big)
 \,,~~~~~~
e\big(\hat Q_i\big) = c_{\,\frac12\delta_i m_i}\big(Q_i\big)\,.
 \eeq

\btm \label{T-HSB-H0}

Let $H = \IF{\KK_1}{m_1}{k_1} \times\cdots\times\IF{\KK_r}{m_r}{k_r}$
be a Howe subgroup of $\OO(n)$ and let $M$ be a closed
simply connected $4$-manifold. Let $P$ be a principal
$\SO(n)$-bundle over $M$ and let $Q$ be a principal $H_0$-bundle
over $M$. Then $Q$ is a reduction of $P$ if and only if
\begin{eqnarray} \label{G-redfml-w}
w(P)
 & = &
\prod\nolimits_{\KK_i=\RR} ~ w(Q_i)^{k_i}
 \cdot
\prod\nolimits_{\KK_i=\CC,\HH} ~ \rho_2\left(c(Q_i)\right)^{k_i}\,,
\\ \label{G-redfml-p}
\tilde p(P)
 & = &
\prod\nolimits_{\KK_i=\RR} ~ \tilde p(Q_i)^{k_i}
 \cdot
\prod\nolimits_{\KK_i=\CC,\HH} ~ c(Q_i)^{k_i}\,\tilde
c(Q_i)^{k_i}\,,
\\ \label{G-redfml-e}
e(P)
 & = &
\prod\nolimits_{\KK_i=\RR} ~ e(Q_i)^{k_i}
 \cdot
\prod\nolimits_{\KK_i=\CC} ~ c_{m_i}(Q_i)^{k_i}
 \cdot
\prod\nolimits_{\KK_i=\HH} ~ c_{2 m_i}(Q_i)^{k_i}\,,
 \end{eqnarray}

\etm

{\it Proof.}~As noted above, $Q$ is a reduction of $P$ iff
$Q^{\SO(n)}$ is isomorphic to $P$, hence iff $\alpha(Q^{\SO(n)}) =
\alpha(P)$ for $\alpha = w$, $p$ and $e$. Equations
\eqref{G-redfml-w}--\eqref{G-redfml-e} then follow from Equations
\eqref{G-ext-SO-a-bun} and \eqref{G-cc-tiQi-O}, where for
\eqref{G-redfml-p} one has to use that $p\mapsto\tilde p$ commutes
with the cup product.
 \qed
 \medskip

By the arguments given above, Theorem \rref{T-HSB-H0} solves the
problem of Howe subbundles of principal bundles with structure
groups $\OO(n)$ and $\SO(n)$. It remains to rewrite the reduction
equations \eqref{G-redfml-w}--\eqref{G-redfml-e} in terms of the
quantities which classify $H$-bundles and $\mr SH^\pm$-bundles,
respectively.

\bco \label{C-HSB-O}

Under the assumptions of Theorem \rref{T-HSB-H0}, let $P$ be a
principal $\OO(n)$-bundle over $M$ and let $Q$ be a principal
$H$-bundle over $M$. Then $Q$ is a reduction of $P$ if and only if
Equations \eqref{G-redfml-w} and \eqref{G-redfml-p} hold, as well
as
 \begin{eqnarray}
 \label{G-redfml-O-e}
[e](P)
 & = &
\prod\nolimits_{\KK_i=\RR} ~ [e](Q_i)^{k_i}
 \cdot
\prod\nolimits_{\KK_i=\CC} ~ \left[c_{m_i}(Q_i)\right]^{k_i}
 \cdot
\prod\nolimits_{\KK_i=\HH} ~ \left[c_{2m_i}(Q_i)\right]^{k_i}\,,
 \end{eqnarray}
where taking the class $\left[c_{m_i}(Q_i)\right]$ means that
$c_{m_i}(Q_i)$ is identified with $-c_{m_i}(Q_i)\, .$

\eco

{\it Proof.}~ As noted above, $Q$ is a reduction of $P$ iff there
exists a common reduction $Q_0$ to $H_0$. This is equivalent to
the condition that there exist reductions $Q_0$ of $Q$ to $H_0$
and $P_0$ of $P$ to $\SO(n)$ such that $Q_0$ is a reduction of
$P_0$, i.e., such that Equations
\eqref{G-redfml-w}--\eqref{G-redfml-e} hold with $P$ replaced by
$P_0$ and $Q_i$ replaced by $Q_{0i}$. By using that for
$\KK_i=\CC$ and $\HH$ one has $Q_{0i} = Q_i$, in
\eqref{G-redfml-w} and \eqref{G-redfml-p} we can then replace
$P_0$ by $P$ and $Q_{0i}$ by $Q_i$ again, with $Q$ and $P$ now
meaning the bundles given here. Equation \eqref{G-redfml-O-e}
follows from \eqref{G-redfml-e} using that the cup product
commutes with passing to equivalence classes w.r.t.\ the
equivalence relation $\alpha\sim\beta$ $\Leftrightarrow$
$\alpha+\beta = 0$.
 \qed
 \bigskip

\bco \label{C-HSB-SO}

Under the assumptions of Theorem \rref{T-HSB-H0}, let $P$ be a
principal $\SO(n)$-bundle over $M$ and let $Q$ be a principal $\mr
SH^\pm$-bundle over $M$. Then $Q$ is a reduction of $P$ if and
only if for the factors $Q_i$ of the extension of $Q$ to the
structure group $H$ there hold Eqs.\ \eqref{G-redfml-w} and
\eqref{G-redfml-p}, as well as
 \begin{eqnarray}
 \label{G-redfml-SO-e}
e(P)
 & = &
\pm\sigma(Q)
 \cdot
\prod\nolimits_{\KK_i=\RR} ~ e^\circ(Q_i)^{k_i}
 \cdot
\prod\nolimits_{\KK_i=\CC} ~ c_{m_i}(Q_i)^{k_i}
 \cdot
\prod\nolimits_{\KK_i=\HH} ~ c_{2m_i}(Q_i)^{k_i}\,,
 \end{eqnarray}
where the postitive sign applies to $\mr SH^+$ and the negative
sign to $\mr SH^-$ and where $e^\circ(Q_i)$ denotes the
representative for the equivalence class $[e](Q_i)$ chosen in the
definition of $\sigma$.

\eco

{\it Proof.}~ If $Q$ is an $\mr SH^+$-bundle, it is a reduction of
$P$ iff there exists a common reduction $Q_0$ to $H_0$, i.e., iff
there exists an $H_0$-bundle such that
\eqref{G-redfml-w}--\eqref{G-redfml-e} hold with $Q_i$ replaced by
$Q_{0i}$. By the same argument as in the proof of Corollary
\rref{C-HSB-O}, in \eqref{G-redfml-w} and \eqref{G-redfml-p} we
can replace $Q_{0i}$ by $Q_i$ again, with $Q_i$ now meaning the
factors of the extension $Q^H$. Equation \eqref{G-redfml-SO-e}
follows from \eqref{G-redfml-e} by replacing $e(Q_{0i}) =
\sigma\big(e(Q_{0i})\big) \,e^\circ(Q_i)$. If $Q$ is an $\mr
SH^-$-bundle, we pass to $Q^{\OO(n)}$ and $P^{\OO(n)}$ and
transform both $Q$ inside $Q^{\OO(n)}$ and $P$ inside $P^{\OO(n)}$
by the element $a_{(n)}$ of $\OO(n)$ chosen to define $\mr SH^-$.
Let the transforms be denoted by $\tilde Q$ and $\tilde P$,
respectively. We have $Q\subseteq P$ iff $\tilde Q\subseteq\tilde
P$, i.e., iff \eqref{G-redfml-w}, \eqref{G-redfml-p} and
\eqref{G-redfml-SO-e} hold with $Q$ and $P$ replaced by $\tilde Q$
and $\tilde P$, respectively. By definition of the classifying
data of $Q$, in these equations we can replace $\tilde Q$ by $Q$
again. Since $w(\tilde P) = w(P)$, $p(\tilde P) = p(P)$ and
$e(\tilde P) = -e(P)$, the assertion follows.
 \qed
 \bigskip

\bre\label{R-powers}

The powers of characteristic classes appearing in Corollaries
\rref{C-HSB-O} and \rref{C-HSB-SO} amount to
 \begin{align*}
 \textstyle
w(Q_i)^{k_i}
 & =
 \textstyle
1 + k_i w_2(Q_i) + k_i w_3(Q_i) + k_i w_4(Q_i) + {k_i \choose 2} w_2(Q_i)^2\,,
\\
 \textstyle
c(Q_i)^{k_i}
 & =
 \textstyle
1 + k_i c_1(Q_i) + k_i c_2(Q_i) + {k_i \choose 2} c_1(Q_i)^2\,,
\\
\tilde p(Q_i)^{k_i}
 & =
1 - k_i p_1(Q_i)\,,
\\
c(Q_i)^{k_i} \tilde c(Q_i)^{k_i}
 & =
1 + 2 k_i c_2(Q_i) - k_i c_1(Q_i)^2\,.
 \end{align*}
We observe:

--~ Equation \eqref{G-redfml-w} has to be analyzed in dimension $2$ only;

--~ Equations \eqref{G-redfml-e}, \eqref{G-redfml-O-e} and \eqref{G-redfml-SO-e}
have to be analyzed for $n\neq2,4$ only.

Indeed, due to \eqref{rel-STW3}, if \eqref{G-redfml-w} holds in
dimension $2$ then it holds in dimension $3$. Similarly, if
\eqref{G-redfml-p} is satisfied and if \eqref{G-redfml-w} holds in
dimension $2$ then it is automatically satisfied in dimension $4$,
because $w_4(P)$ is determined by $w_2(P)$ and $p_1(P)$, see
\rref{G-w4}. Finally, for $n\neq2,4$, Equations
\eqref{G-redfml-e}, \eqref{G-redfml-O-e} and \eqref{G-redfml-SO-e}
are automatically satisfied, because for $n=3$, $H$ has an
$\OO$-factor of odd rank and for $n\geq 5$ the product of Euler
classes vanishes identically because of the dimension.

\ere

\bre\label{Bem-redfml-rel}

The reduction equations \eqref{G-redfml-w}--\eqref{G-redfml-e},
\eqref{G-redfml-O-e} or \eqref{G-redfml-SO-e} are supplemented by
the relations between the characteristic classes of the $\SO$- or
$\OO$-factors, respectively, listed in Theorem \rref{T-SObun} or
Corollary \rref{C-Obun}, respectively. As observed above, in these
relations, $w_4(Q_i)$ can be replaced by $w_2(Q_i)$ and $p_1(Q_i)$
using \eqref{G-w4}.

\ere


\subsection{Howe subbundles of principal $\Sp(n)$-bundles}
\label{S-HSB-Sp}


Let a Howe subgroup $H = \IF{\KK_1}{m_1}{k_1} \times \cdots \times
\IF{\KK_r}{m_r}{k_r}$ of $\Sp(n)$ and a principal $H$-bundle $Q$
be given. We have to compute $c\big(Q^{\Sp(n)}\big)$. The
classifying map of $Q$ is given by
 $
f = \left(\prod\nolimits_i f_i\right)\circ\Delta_r\,,
 $
where $f_i:M\to\BB\I_{\KK_i}(m_i)$ are the classifying maps of the
factors $Q_i$ of $Q$. The classifying map of $Q^{\Sp(n)}$ is $\BB
j\circ f$, where $j:H\to\Sp(n)$ denotes the natural embedding.
Hence
$$
c\big(Q^{\Sp(n)}\big)
 =
f^\ast (\BB j)^\ast c\big(\Sp(n)\big)\,.
$$
We decompose $j$ as
$$
H
 =
\prod\nolimits_i\I_{\KK_i}(m_i)
 \stackrel{\prod_i j_{\KK_i,\HH}}{\longrightarrow}
\prod\nolimits_i\Sp(m_i)
 \stackrel{\prod\nolimits_i\Delta_{k_i}}{\longrightarrow}
\prod\nolimits_i\Sp(m_i)
 \times
\stackrel{k_i}{\cdots}
 \times
\Sp(m_i)
 \stackrel{i}{\longrightarrow}
\Sp(n)\,,
$$
where $j_{\KK_i,\HH} : \I_{\KK_i}(m_i) \to \Sp(m_i)$ denotes the
natural embedding as a subset. Argueing as in Subsection
\rref{S-HSB-O} we arrive at
 \beq\label{G-ext-Sp-a}
c\left(Q^{\Sp(n)}\right)
 =
\Big(f_1^\ast ~ \big(\BB j_{\KK_1,\HH}\big)^\ast
 \,c(\Sp(m_1))\Big)^{k_1}
 \cdots
\Big(f_r^\ast ~ \big(\BB j_{\KK_r,\HH}\big)^\ast
 \,c(\Sp(m_r))\Big)^{k_r}\,.
 \eeq
Since $\BB j_{\KK_i,\HH}\circ f_i$ is the classifying map of the extension
$Q_i^{\Sp(m_i)}$, \eqref{G-ext-Sp-a} can be written
as
 \beq\label{G-ext-Sp-a-bun}
c\left(Q^{\Sp(n)}\right)
 =
c\big(Q_1^{\Sp(m_1)} \big)^{k_1}
 \cdots
c\big(Q_r^{\Sp(m_r)} \big)^{k_r}\,.
 \eeq
For $\KK_i=\HH$ one has $Q_i^{\Sp(m_i)} = Q_i$ and hence
$c\big(Q_i^{\Sp(m_i)} \big) = c(Q_i)$. The standard relations
between characteristic classes listed in Proposition \ref{P-relBvp}
yield
 $$
c\big(Q_i^{\Sp(m_i)}\big)
 =
 \begin{cases}
\tilde p(Q_i)^2 & \KK_i = \RR\,,
\\
c(Q_i)\, \tilde c(Q_i) & \KK_i = \CC\,.
\end{cases}
 $$
Combining this with \eqref{G-ext-Sp-a-bun} we arrive at

\btm\label{T-red-Sp}

Let $H = \IF{\KK_1}{m_1}{k_1} \times\cdots \IF{\KK_r}{m_r}{k_r}$
be a Howe subgroup of $\Sp(n)$ and let $M$ be a closed
simply connected $4$-manifold. Let $P$ be a principal
$\Sp(n)$-bundle over $M$ and let $Q$ be a principal $H$-bundle
over $M$. Then $Q$ is a reduction of $P$ if and only if
\beq\label{G-redfml-c} c(P)
 =
\prod\nolimits_{\KK_i=\RR} ~ \tilde p(Q_i)^{2 k_i}
 \cdot
\prod\nolimits_{\KK_i=\CC} ~ c(Q_i)^{k_i} \, \tilde c(Q_i)^{k_i}
 \cdot
\prod\nolimits_{\KK_i=\HH} ~ c(Q_i)^{k_i} \, .
 \eeq
 \qed

\etm

For explicit formulae for the powers appearing in
\eqref{G-redfml-c} see Remark \rref{R-powers}. Like in the case of
$\OO(n)$-bundles and $\SO(n)$-bundles, the reduction equation is
supplemented by the relations between the characteristic classes of
$\OO$-factors, see Remark \rref{Bem-redfml-rel} for details.



\section{Holonomy-induced Howe subbundles}
\label{Holonomy-induced Subbundles}
\label{S-holind}


According to our programme, next we have to specify the Howe
subbundles which are holonomy-induced, i.e., generated by a
connected subbundle. Over a simply connected manifold, this turns
out to be a purely algebraic condition.

\ble \label{L-holind-1}

Let $P$ be a principal $G$-bundle over $M$. If $M$ is simply
connected then a reduction of $P$ to a Howe subgroup $H$ of $G$ is
holonomy-induced if and only if $H$ is generated by its identity
connected component $H_0$. That is to say, $H = \mr C^2_G(H_0)$
or, alternatively, $\mr C_G(H) = \mr C_G(H_0)$.

\ele

{\it Proof.}~ Let $H$ be given and let $Q$ be a reduction of $P$
to $H$. First, assume that $H = \mr C^2_G(H_0)$. Since $M$ is
simply connected, $Q$ can be reduced to $H_0$. By assumption, $Q$
is the Howe subbundle generated by the reduction. Since $H_0$ is
connected, the reduction is connected\footnote{The implication
under consideration holds without the assumption that $M$ be
simply connected; using this assumption simplifies the proof
though.}.
 \comment{

allgemeines Argument: if $Q$ is not connected, there exists a
connected reduction $Q_1$ with structure group $H_1$. By
assumption $\mr C_G^2(H_1) = H$, hence $Q$ is generated as a HSB
by $Q_1$.
 }%
Conversely, assume that $Q$ is holonomy-induced. Let $Q_1$ be a
connected subbundle generating $Q$ as a Howe subbundle and let
$H_1$ be the structure group of $Q_1$. Then $H = \mr C^2_G(H_1)$.
We show that $H_1$ is connected, hence $H_1 = H_0$. For that
purpose, it suffices to show that the fibers of $Q_1$ are
connected. Thus, let $q_a,q_b\in Q_1$ with common base point $m\in
M$. Since $Q_1$ is connected, there exists a path $\gamma$ from
$q_a$ to $q_b$. The path $\gamma$ projects to a closed path in $M$
based at $m$. Since $M$ is simply connected, the projected path is
homotopic to the constant path through $m$. Consequently, by the
homotopy lifting property, there exists a map $H:[0,1]\times[0,1]\to Q_1$
such that $H(0,\cdot) = \gamma$ and $H(1,\cdot)$ is a path in the
fiber of $Q_1$ over $m$. Then the three pieces $H(\cdot,0)$,
$H(1,\cdot)$ and $H(1-\cdot,1)$ establish a path from $q_a$ to
$q_b$ inside this fiber.
 $\qed$
 \bigskip

Thus, we have to determine the Howe subgroups satisfying
 \beq\label{G-holcond}
H = \mr C^2_G(H_0)
 ~~~\Leftrightarrow~~~
\mr C_G(H) = \mr C_G(H_0)\,.
 \eeq
First, we will treat the cases $G=\OO(n)$ and $G=\Sp(n)$.

\ble\label{L-holind-2}

A Howe subgroup of a compact Lie group satisfies \eqref{G-holcond}
if and only if it does not contain another Howe subgroup of the
same dimension.

\ele

{\it Proof.}~ Let $G$ be compact and let $H$ be a Howe subgroup.
The subgroup $\mr C_G^2(H_0)$ is a Howe subgroup. Since $H_0
\subseteq \mr C_G^2(H_0) \subseteq H$, it has the same dimension
as $H$. Hence, if $H$ does not contain another Howe subgroup of
the same dimension, \eqref{G-holcond} holds. Conversely, let
$K\subset H$ be another Howe subgroup of the same dimension. Since
$K_0$ and $H_0$ are closed (i.e. compact without boundary) connected manifolds of the same
dimension, invariance of domain (see \cite[p. 217, Ex. 6.5]{massey}) implies that the embedding $K_0\subseteq H_0$ is an
open map. Thus $K_0$ is an open and
closed subset of $H_0$ and hence there holds $K_0 = H_0$.
It follows $\mr C_G^2(H_0) = \mr C_G^2(K_0) \subseteq K \neq H$.
 $\qed$
 \bigskip

\ble\label{L-holind-3}

A Howe subgroup of $\OO(n)$ or $\Sp(n)$ contains another Howe
subgroup of the same dimension if and only if it contains a factor
$\OO(2)$ or a double factor $\OO(1)\times\OO(1)$.

\ele

{\it Proof.}~ We give the argument for $G=\OO(n)$; the case of
$\Sp(n)$ is completely analogous. Consider the operations
producing direct predecessors in the set of Howe subgroups of
$\OO(n)$, listed in Section \rref{S-HSG-O}. A given Howe subgroup
$H$ contains another Howe subgroup of the same dimension iff an
operation can be applied to $H$ which does not change the
dímension. As noted in the proof of Lemma \rref{L-Mstable}, the
operations which do not change the dimension are inverse field
restriction of a factor $\OO(2)$ and merging of a double factor
$\OO(1)\times\OO(1)$.
 \qed
 \bigskip

Lemmas \rref{L-holind-1}--\rref{L-holind-3} imply

\btm\label{T-holind-O-Sp}

A Howe subbundle of a principal $\OO(n)$- or $\Sp(n)$-bundle is
holonomy-induced if and only if its structure group does not
contain a factor $\OOF2k$ nor a double factor $\OOF1k\times
\OOF1l$.
 \qed

\etm

Next, we will treat the case of $G = \SO(n)$. Since in the case of
odd $n$ the argument is not simpler than in the case of even $n$
and since the description of Howe subgroups of $\SO(n)$, $n$ even,
in terms of $\mr S$-admissible Howe subgroups of $\OO(n)$
trivially applies to the case of odd $n$, too, we can treat both
cases at once, always working with $\mr S$-admissible Howe
subgroups of $\OO(n)$.

\ble\label{L-holind-4}

Let $H$ be an $\mr S$-admissible Howe subgroup of $\OO(n)$. Then
$\mr SH$ contains another Howe subgroup of $\SO(n)$ of the same
dimension if and only if $H$ contains another Howe subgroup of
$\OO(n)$ of the same dimension.

\ele

{\it Proof.}~ If $H$ does not contain another Howe subgroup of
$\OO(n)$ of the same dimension, then $H$ satisfies $\mr
C_{\OO(n)}(H) = \mr C_{\OO(n)}(H_0)$ by Lemma \rref{L-holind-2}.
Since $H_0 \subseteq \mr SH \subseteq H$, there holds $\mr
C_{\OO(n)}(H_0)\supseteq \mr C_{\OO(n)}(\mr SH)\supseteq \mr
C_{\OO(n)}(H)$, hence $C_{\OO(n)}(H_0) = \mr C_{\OO(n)}(\mr SH)$.
Intersecting with $\SO(n)$ and using $(\mr SH)_0 = H_0$ we obtain
$\mr C_{\SO(n)}(\mr SH) = \mr C_{\SO(n)}\big((\mr SH)_0\big)$.

Conversely, assume that $H$ contains another Howe subgroup $K$ of
$\OO(n)$ of the same dimension. Then $\mr SK\subseteq \mr SH$.
Since intersection with $\SO(n)$ does not change the dimension,
$\mr SK$ has the same dimension as $\mr SH$. It remains to show
that $\mr SK$ is a Howe subgroup of $\SO(n)$ and that it is
distinct from $\mr SH$. For that purpose, it suffices to show that
$K$ is $\mr S$-admissible. We may assume that $K$ is a direct predecessor of
$H$ and thus is obtained from $H$ by inverse field restriction of a factor
$\OOF2k$ or by merging a double factor $\OOF1k\times\OOF1l$. We check that
neither of these two operations produces a factor that causes $K$ or $\mr
C_{\OO(n)}(K)$ to belong to cases (A) or (B) of Lemma \rref{L-Mstable}. For
inverse field restriction of a factor $\OOF2k$ this is obvious, as it
produces a factor $\UUF1k$. Merging a double factor $\OOF1k\times\OOF1l$
produces the factor $\OOF1{k+l}$. This cannot make $K$ belong to case (A) nor
$\mr C_{\OO(n)}(K)$ to case (B) but it might make $K$ belong to case (B) or $\mr
C_{\OO(n)}(K)$ to case (A). If $K$ belongs to case (B), $k+l$ is odd, hence
exactly one of the two factors that are merged has odd multiplicity.
Moreover, among the remaining factors of $K$ there must be a single
$\OO(1)$-factor of odd multiplicity and no further $\OO$-factor of odd
multiplicity. Since all the remaining factors are also present in $H$, $H$
belongs to case (B) itself
(contradiction). If $\mr C_{\OO(n)}(K)$ belongs to case (A), there holds
$k+l=2$, hence $k=l=1$, and $K$ has no further $\OO$-factor of
odd rank. Then $\mr C_{\OO(n)}(H)$ belongs to case (B), which is a
contradiction, too.
 $\qed$
 \bigskip

Lemmas \rref{L-holind-1}--\rref{L-holind-3} and \rref{L-holind-4} imply

\btm\label{T-holind-SO}

A Howe subbundle $Q$ of a principal $\SO(n)$-bundle is
holonomy-induced if and only if the Howe subgroup of $\OO(n)$
generated by the structure group of $Q$ does not contain a factor
$\OOF2k$ nor a double factor $\OOF1k\times \OOF1l$.
 \qed

\etm

When $n$ is even, the conditions given in the theorem supersede conditions (A)
and (B) of Lemma \rref{L-Mstable}. Hence, in this case we arrive at the
following characterization. The holonomy-induced Howe subbundles of a
principal $\SO(n)$-bundle, $n$ even, are given by the reductions to the
subgroups $\mr SH$, where $H$ is a Howe subgroup of $\OO(n)$ which does not
belong to one of the following cases:
\medskip

--~ $H$ contains a factor $\OOF2k$,

--~ $H$ contains a double factor $\OOF1k\times\OOF1l$,

--~ $H$ contains a factor $\OOF m2$ with $m$ odd and no further $\OO$-factor of
odd rank,

--~ $H$ contains a double factor $\OOF k1 \times \OOF l1$ with $k$ and $l$ odd
and no further $\OO$-factor of odd rank.
\medskip

\begin{figure}

\begin{center}

\unitlength1.75cm

\begin{picture}(1,1)
\put(0,0.5){
 \marke{-0.3,0}{cr}{\mbox{\normalsize $\OO(2)$}}
 \lri{0,0}{tc}{\OOF12}
 \whole{1,0}{tc}{\UUF11}
 }
\end{picture}
 \hspace{5cm}
\begin{picture}(2,1)
\put(0,0.5){
 \marke{-0.3,0}{cr}{\mbox{\normalsize $\OO(3)$, $\SO(3)$}}
 \lri{0,0}{tc}{\OOF13}
 \lri{1,0}{tc}{\OOF11\stimes\UUF11}
 \whole{2,0}{tc}{\OOF31}
 }
\end{picture}

\begin{picture}(4,2)
\put(0,1){
 \marke{-0.3,0}{cr}{\mbox{\normalsize $\OO(4)$}}
 \lori{0,0}{tc}{\OOF14}
 \lri{0,0}{}{}

 \lrii{1,0.5}{bc}{\OOF12\stimes\UUF11}
 \luri{1,0.5}{}{}
 \lri{1,0}{tr}{\UUF12\!\!\!}
 \luri{1,0}{}{}

 \lri{2,0}{tc}{\UUF11\stimes\UUF11}
 \lori{2,-0.5}{tc}{\SpF11}

 \luri{3,0.5}{bc}{\OOF11\stimes\OOF31}
 \lri{3,0}{tl}{\!\!\UUF21}

 \whole{4,0}{tc}{\OOF41}
 }
\end{picture}

\begin{picture}(4,2)
\put(0,1){
 \marke{-0.3,0}{cr}{\mbox{\normalsize $\SO(4)$}}
 \lori{0,0}{tc}{\OOF14}
 \lri{0,0}{}{}

 \lri{1,0.5}{}{}
 \lri{1,0.5}{bc}{\UUF12{}^+}
 \lri{1,0}{tc}{\UUF12{}^-}
 \luri{1,0}{}{}

 \lri{2,0.5}{bc}{\SpF11{}^+}
 \lori{2,0}{bc}{\UUF11\stimes\UUF11}
 \lri{2,0}{}{}
 \lori{2,-0.5}{tc}{\SpF11{}^-}

 \luri{3,0.5}{bc}{\UUF21{}^+}
 \lri{3,0}{tc}{\UUF21{}^-}

 \whole{4,0}{tc}{\OOF41}
 }
\end{picture}

\begin{picture}(5,2)
\put(0,1){
 \marke{-0.3,0}{cr}{\mbox{\normalsize $\OO(5)$, $\SO(5)$}}
 \lori{0,0}{tc}{\OOF15}
 \lri{0,0}{}{}

 \lri{1,0.5}{bc}{\OOF13\stimes\UUF11}
 \luri{1,0.5}{}{}
 \lri{1,0}{tr}{\OOF11\stimes\UUF12\!\!\!}
 \luri{1,0}{}{}

 \lrii{2,0.5}{bc}{\OOF12\stimes\OOF31}
 \luri{2,0.5}{}{}
 \lri{2,0}{bl}{\!\!\!\!\!\!\!\!\!\!\!\OOF11\stimes\UUF11\stimes\UUF11}
 \luri{2,0}{}{}
 \lri{2,-0.5}{tc}{\OOF11\stimes\SpF11}

 \lrii{3,0}{bc}{\OOF31\stimes\UUF11}
 \loori{3,-0.5}{tc}{\OOF11\stimes\UUF21}

 \luri{4,0.5}{bc}{\OOF11\stimes\OOF41}

 \whole{5,0}{tc}{\OOF51}
 }
\end{picture}

\begin{picture}(6,2.5)
\put(0,1){
 \marke{-0.3,0}{cr}{\mbox{\normalsize $\OO(6)$}}
 \lori{0,0}{tc}{\OOF16}
 \lri{0,0}{}{}
 \luri{0,0}{}{}

 \lori{1,0.5}{bc}{\OOF14\stimes\UUF11}
 \lri{1,0.5}{}{}
 \luri{1,0.5}{}{}
 \lori{1,0}{tr}{\OOF12\stimes\UUF12\!\!\!}
 \lri{1,0}{}{}
 \lurrri{1,0}{}{}
 \luri{1,0}{}{}
 \lori{1,-0.5}{tc}{\UUF13}

 \lrii{2,1}{bc}{\OOF13\stimes\OOF31}
 \lorri{2,0.5}{br}{\OOF12\stimes\UUF11\stimes\UUF11\!\!\!\!\!\!\!\!}
 \lri{2,0.5}{}{}
 \luri{2,0.5}{}{}
 \lori{2,0}{bc}{\!\!\!\UUF11\stimes\UUF12}
 \luri{2,0}{}{}
 \lori{2,-0.5}{tc}{\OOF12\stimes\SpF11}
 \lri{2,-0.5}{}{}

 \luri{3,0.5}{bl}{\UUF11\stimes\UUF11\stimes\UUF11}
 \lori{3,0}{bc}{\!\!\!\OOF12\stimes\UUF21}
 \lri{3,0}{}{}
 \lori{3,-0.5}{tc}{\UUF11\stimes\SpF11}

 \lri{4,1}{bc}{\OOF11\stimes\OOF31\stimes\UUF11}
 \luri{4,1}{}{}
 \lri{4,0.5}{bc}{~~~\OOF12\stimes\OOF41}
 \luri{4,0.5}{}{}
 \lori{4,0}{bc}{\!\!\!\UUF11\stimes\UUF21}
 \luri{4,0}{}{}
 \looori{4,-0.5}{tc}{\OOF32}
 \lri{4,-0.5}{}{}

 \luuri{5,1}{bc}{\OOF31\stimes\OOF31}
 \luri{5,0.5}{bc}{~\,\UUF11\stimes\OOF41}
 \lri{5,0}{tc}{\OOF11\stimes\OOF51}
 \lori{5,-0.5}{tc}{\UU31}

 \whole{6,0}{tc}{\OOF61}
 }
\end{picture}

\begin{picture}(6,1.5)
\put(0,0.5){
 \marke{-0.3,0}{cr}{\mbox{\normalsize $\SO(6)$}}
 \lori{0,0}{tc}{\OOF16}
 \lri{0,0}{}{}

 \luri{1,0.5}{bc}{\OOF14\stimes\UUF11}
 \lri{1,0}{tc}{\UUF13}

 \lori{2,0}{tc}{\UUF11\stimes\UUF12}
 \lri{2,0}{}{}

 \luri{3,0.5}{bc}{\UUF11\stimes\UUF11\stimes\UUF11}
 \lri{3,0}{tc}{\UUF11\stimes\SpF11}

 \lori{4,0}{tc}{\UUF11\stimes\UUF21}
 \lri{4,0}{}{}

 \luri{5,0.5}{bc}{\OOF41\stimes\UUF11}
 \lri{5,0}{tc}{\UUF31}

 \whole{6,0}{tc}{\OOF61}
 }
\end{picture}

\end{center}

\caption{\label{F-HDgr-O} Hasse diagrams of the sets of conjugacy
classes of Howe subgroups which are generated by their identity
connected components for $\OO(n)$, $n=2,\dots,6$ and $\SO(n)$,
$n=3,\dots,6$. For the notation, see Figure \rref{F-HSG-O}.}

\end{figure}
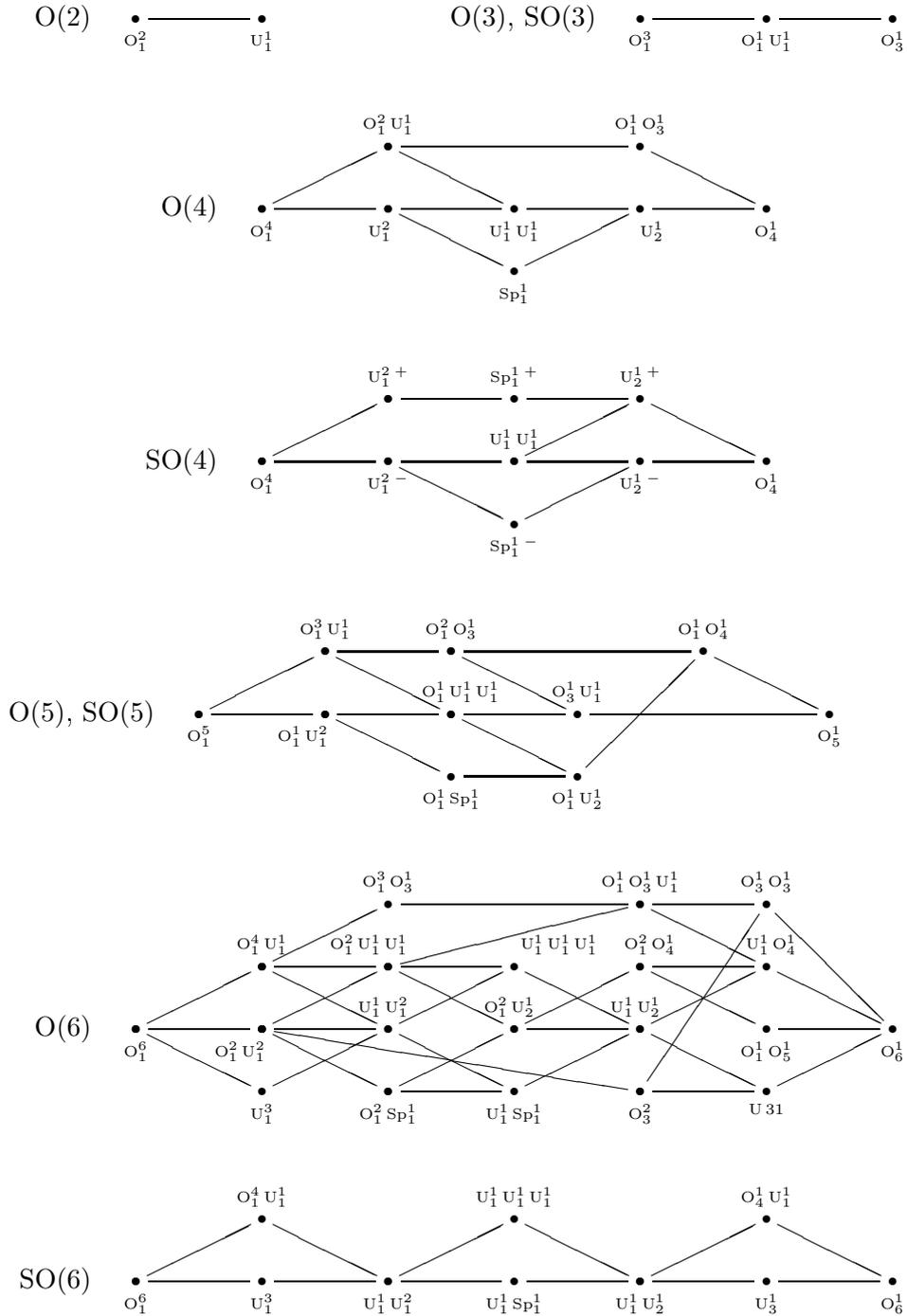

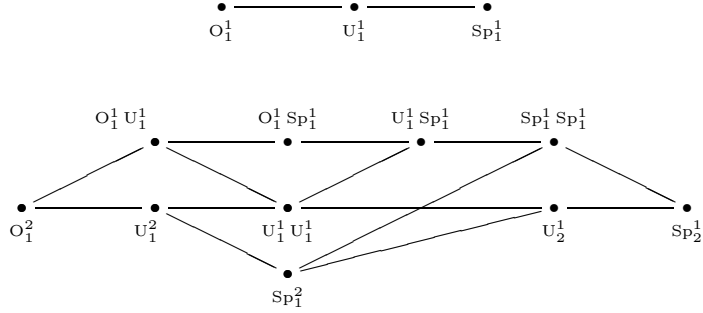
\begin{figure}

\begin{center}

\unitlength1.75cm

\begin{picture}(2,1)
\put(0,0.5){
 \lri{0,0}{tc}{\OOF11}
 \lri{1,0}{tc}{\UUF11}
 \whole{2,0}{tc}{\SpF11}
 }
\end{picture}

\begin{picture}(5,2)
\put(0,1){
 \lori{0,0}{tc}{\OOF12}
 \lri{0,0}{}{}

 \lri{1,0.5}{br}{\OOF11\stimes\UUF11}
 \luri{1,0.5}{}{}
 \lri{1,0}{tr}{\UUF12\!\!\!}
 \luri{1,0}{}{}

 \lri{2,0.5}{bc}{\OOF11\stimes\SpF11}
 \lori{2,0}{tc}{\UUF11\stimes\UUF11}
 \lrii{2,0}{}{}
 \lorri{2,-0.5}{tc}{\SpF12}
 \lorii{2,-0.5}{}{}

 \lri{3,0.5}{bc}{\UUF11\stimes\SpF11}

 \lri{4,0}{tc}{\UUF21}
 \luri{4,0.5}{bc}{\SpF11\stimes\SpF11}

 \whole{5,0}{tc}{\SpF21}
 }
\end{picture}

\end{center}

\caption{\label{F-HDgr-Sp} Hasse diagrams of the sets of conjugacy
classes of Howe subgroups which are generated by their identity
connected components for $G=\Sp(1)$ and $\Sp(2)$. For the
notation, see Figure \rref{F-HSG-O}. For identifications due to
$\Sp(1) \equiv \SU(2)$, see Figure \rref{F-HSG-Sp}.}

\end{figure}

For illustration, Figures \rref{F-HDgr-O} and \rref{F-HDgr-Sp} show the Hasse
diagrams of the sets of conjugacy classes of Howe subgroups satisfying
\eqref{G-holcond} for some low-dimensional classical compact Lie groups. These
are the structure groups of holonomy-induced Howe subbundles in the respective
situations. The diagrams can be extracted from the Hasse diagrams in Figures
\rref{F-HSG-O}, \rref{F-HSG-Sp} and \rref{F-HSG-SO} by removing vertices which
correspond to Howe subgroups containing a factor $\OOF2k$ or a double factor
$\OOF1k\times\OOF1l$.

\section{Factorization by the action of the structure group}
\label{S-factoriz} \label{factorization}


Let $P$ be a principal bundle with structure group $G=\OO(n)$,
$\SO(n)$ or $\Sp(n)$. In the last step of our program we have to
factorize the set of isomorphism classes of holonomy-induced Howe
subbundles of $P$ by the action of the structure group $G$ on
subbundles. Since above we have already restricted attention to
Howe subbundles of standard form, all we have to do now is to
analyze which of these Howe subbundles of standard form are
conjugate under the action of the structure group. Thus, below we
will assume that we are given a reduction $Q$ of $P$ to some Howe
subgroup $H$ of $G$ of standard form and determine which bundles
arise from $Q$ by transformation with $d\in G$ such that
$dHd^{-1}$ is of standard form again.


\subsection{Structure groups $\OO(n)$ and $\Sp(n)$}
\label{SS-factoriz-O}


First, consider $G=\OO(n)$ and $\Sp(n)$. Up to isomorphy, Howe
subbundles of $P$ of standard form are characterized by $r$-tuples
 \beq\label{G-HSB-data}
 \big(
(\KK_1,m_1,k_1,\alpha_1)
 ,\dots,
(\KK_r,m_r,k_r,\alpha_r)
 \big)
 \eeq
where

--~ $r$ is a positive integer,

--~ $\IF{\KK_1}{m_1}{k_1}\times\cdots\times\IF{\KK_r}{m_r}{k_r}$
is a Howe subgroup of $G$,

--~ $\alpha_i$ are characteristic classes of
$\I_{\KK_i}(m_i)$-bundles over $M$, satisfying the respective
reduction equations derived in Section \rref{S-HSB}. That is,
$\alpha_i = w,p,[e]$ for $\OO$-factors and $\alpha_i = c$ for
$\UU$- and $\Sp$-factors.

To derive how the structure group acts on these data we also have
to consider the connected components of Howe subbundles of
standard form. These are characterized analogously by the data
\eqref{G-HSB-data}, where now $\alpha_i$ are characteristic
classes of $\I_{\KK_i}(m_i)_0$-bundles over $M$. I.e., $\alpha_i =
w,p,e$ for $\SO$-factors and $\alpha_i = c$ for $\UU$- and
$\Sp$-factors.

We define the flip to be the following operation on factors. For
an $\SO$-factor, multiply the Euler class $e$ by $-1$. For a
$\UU$-factor, multiply the first Chern class $c_1$ by $-1$, i.e.,
replace $c$ by $\tilde c$. For $\OO$-factors and $\Sp$-factors,
the flip is not defined.

\btm\label{T-factoriz-O}

Let $P$ be a principal bundle with structure group $G=\OO(n)$ or
$\Sp(n)$ over a closed simply connected manifold of dimension $4$.
Two Howe subbundles of $P$ of standard form are conjugate under
the action of the structure group if and only if they can be
transformed into one another by a permutation of factors and by
flips.

\etm

{\it Proof.}~ First, let $G=\OO(n)$. Let there be given a Howe
subbundle of $P$ of standard form and let $H$ denote the
corresponding Howe subgroup. Let $d\in\OO(n)$ such that $dHd^{-1}$
is of standard form again. We have to find out how $d$ acts on the
data of a connected component $Q_0$ of $Q$ and derive from that
how it acts on the data of $Q$. Since $dHd^{-1}$ is of standard
form, we can decompose $d = d_1 d_2 d_3$, where $d_1$ is a pure
permutation of factors, $d_2\in\mr C_{\OO(n)}(H)$ and $d_3$ leaves
invariant each factor separately. The action of $d_1$ on $Q$ is by
the corresponding permutation of factors. The action of $d_2$ is
trivial. The group element $d_3$ can be further decomposed as $d_3
= d_{31} \oplus\cdots\oplus d_{3r}$ with $d_{3i}\in\mr
N_{\OO(\delta_im_ik_i)}\big(\IF{\KK_i}{m_i}{k_i}\big)$. Each
$d_{3i}$ defines an automorphism $\vp_i$ of the identity connected
component $\I_{\KK_i}(m_i)_0$ of $\I_{\KK_i}(m_i)$. According to
Lemma \rref{L-nmr-1}, up to a redefinition of $d_2$ we may assume
that $d_{3i} = a_i \oplus\cdots\oplus a_i$ ($k_i$ times) for some
$a_i\in\mr N_{\OO(\delta_im_i)}\big(\IF{\KK_i}{m_i}1\big)$. Then
$\vp_i$ is induced from the action of $a_i$ on the subgroup
$\big(\IF{\KK_i}{m_i}1\big)_0$ of $\OO(\delta_im_i)$ by
conjugation. We have to determine how $(\BB\vp_i)^\ast$ maps the
characteristic classes of $\I_{\KK_i}(m_i)_0$. If $\vp_i$ is
inner, $(\BB\vp_i)^\ast \alpha_i = \alpha_i$, because
$\I_{\KK_i}(m_i)_0$ is arcwise connected. In the following
situations, $\vp_i$ is outer, cf.\ Lemma \rref{L-nmr-2}:
 \smallskip

--~ $\KK_i=\RR$, $m_i$ even, $\det a_i = -1$. Here
$(\BB\vp_i)^\ast w(\SO(m_i)) = w(\SO(m_i))$, $(\BB\vp_i)^\ast
p(\SO(m_i)) = p(\SO(m_i))$ and $(\BB\vp_i)^\ast e(\SO(m_i)) = -
e(\SO(m_i))$.
 \smallskip

--~ $\KK_i = \CC$, $a_i = \II_{m_i,m_i}$. Here $(\BB\vp_i)^\ast
c(\UU(m_i)) = \tilde c(\UU(m_i))$.
 \smallskip

In either case, $(\BB\vp_i)^\ast$ induces a flip of the factor
under consideration. Conversely, since a flip of an $\SO$-factor
of odd rank is trivial, any flip of $Q_0$ can be generated in this
way. Finally, when passing to $Q$, flips of $\SO$-factors cancel
whereas flips of $\UU$-factors persist.

For $G=\Sp(n)$, the proof is analogous. Here, the automorphisms
$\vp_i$ are induced by the action of $d_{3i}\in\mr
N_{\Sp(m_ik_i)}\big(\IF{\KK_i}{m_i}{k_i}\big)$ on the subgroup
$\IF{\KK_i}{m_i}{k_i}$ of $\Sp(m_ik_i)$. A case by case inspection
of the outer automorphisms of $\OO(m)$, $\UU(m)$ and $\Sp(m)$
shows that here the $\vp_i$ are outer in the following situations:
 \smallskip

--~ $\KK_i=\RR$, $m_i$ even, $d_{3i} = a_{i} \oplus \cdots \oplus
a_{i}$ ($k_i$ times) with $a_i\in\OO(m_i)$ such that $\det a_i =
-1$;
 \smallskip

--~ $\KK_i=\CC$, $d_{3i} = \mr j \II_{m_i k_i}$;\, here $\mr j$
denotes the second quaternionic unit.
 \smallskip

The action of $(\mr B\vp_i)^\ast$ on the characteristic classes is
the same as for $G=\OO(n)$. Hence, the assertion.
 \qed
 \bigskip


\subsection{Structure group $\SO(n)$}
 \label{SS-factoriz-SO}


Let $P$ be a principal $\SO(n)$-bundle. As before, we will more
generally consider reductions to subgroups of type $\typeSHpm$,
see \eqref{G-def-SHpm} and the comment after this formula. Such
reductions will be referred to as reductions of type $\typeSHpm$.
Isomorphy classes of such reductions are characterized by the data
\eqref{G-HSB-data} for Howe subbundles of $P^{\OO(n)}$, together
with a compatible value of the invariant $\sigma$. Like in the
case of $G=\OO(n)$, to prove the result for the reductions of type
$\typeSHpm$ we have to consider their connected components, i.e.,
the reductions of $P$ to the identity connected components of Howe
subgroups of $\OO(n)$. These are characterized by the same data as
for $G=\OO(n)$ where now the reduction equations are those of the
$\SO(n)$-bundle $P$.

To formulate the result we need some more terminology.
$\SO$-factors of odd multiplicity, $\OO$-factors of odd
multiplicity and $\UU$-factors of odd rank and odd multiplicity
will be called signed factors. All other factors are called
unsigned. A flip is called signed or unsigned according to the
factor it is applied to. An $\SO$-factor or a $\UU$-factor is
called stable if it does not change under a flip, i.e., if for its
characteristic class(es) there holds $-e = e$ or $\tilde c = c$,
respectively. Otherwise, it is called unstable. An $\OO$-factors
is called stable (unstable) if it arises from a stable (unstable)
$\SO$-factor by extension. This definition makes sense, because
the $\SO$-factor in question is determined by the given
$\OO$-factor up to the sign of the Euler class. Note that to be
signed is an algebraic property, depending on the structure group
of the bundle reduction under consideration, whereas to be stable
is a topological property, depending on the bundle reduction
itself. Finally, if $\sigma$ can take the value $-1$, there exists
an unstable signed $\OO$-factor. In this case, we define the flip
of $\sigma$ as the passage to $-\sigma$. We assume this flip to be
signed. Thus, unsigned flips involve flips of factors only,
whereas signed flips involve flips of factors and of $\sigma$,
provided the latter is defined.

\btm\label{T-factoriz-SO}

Let $P$ be a principal bundle with structure group $G=\SO(n)$ over
a closed  simply connected manifold of dimension $4$.
Two bundle reductions of $P$ of type $\typeSHpm$ are conjugate under the
action of the structure group if and only if they can be transformed into one
another by a permutation of factors and by the following operations:
\\
-- If there exists a signed stable factor or an $\OO$-factor of odd rank, any
flip.
\\
-- Otherwise, unsigned flips and pairs of signed flips.

\etm

{\it Proof.}~ First, we consider bundle reductions of type $\typeSHp$. Let $H$
be a Howe subgroup of $\OO(n)$ and let $Q$ be a reduction of $P$ to the subgroup
$\mr SH^+$. Let $d\in\SO(n)$ such that $d(\mr SH^+)d^{-1}$
is of standard form, i.e., equal  to $\mr SK^+$ for some Howe subgroup $K$ of
$\OO(n)$ of standard form. We have to determine how $d$ acts on a connected
component $Q_0$ of $Q$ and derive from that how it acts on $Q^H$ and on
$\sigma$. We start with some preliminary considerations. The structure group of
$Q_0$ is $H_0$. If $d(\mr SH^+)d^{-1}$ is of standard form then so is its identity connected
component $d H_0 d^{-1}$. A similar argument as in the proof of
Theorem \rref{T-factoriz-O} shows that then $d$ can be decomposed as $d_1 d_2
d_3$, where $d_1$ is a pure permutation of the factors of $H_0$ (equivalently,
those of $H$), $d_2\in\mr C_{\OO(n)}(H_0)$ and
$$
d_3
 =
(a_1\oplus\stackrel{k_1}{\cdots}\oplus a_1)
\oplus \cdots \oplus
(a_r\oplus\stackrel{k_r}{\cdots}\oplus a_r)\,,
$$
with $a_i\in\mr N_{\OO(\delta_im_i)}\big((\IF{\KK_i}{m_i}1)_0\big)
\equiv \mr N_{\OO(\delta_im_i)}\big(\IF{\KK_i}{m_i}1\big)$. Since $d\in\SO(n)$,
 \beq\label{G-detcond}
(\det d_1) (\det d_2) (\det a_1)^{k_1} \cdots (\det a_r)^{k_r}
 =
1\,.
 \eeq
The only effect of $d_2$ is that it might compensate for a sign stemming from
$d_1$ or the $a_i$. We claim that for this purpose it is
sufficient that $d_2\in\mr C_{\OO(n)}(H)$. Indeed, since $\mr C_{\OO(n)}(H_0)$
is a Howe subgroup, it contains an element of negative determinant if and only
if it has an $\OO$-factor of odd multiplicity.
Then the double centralizer $K=\mr C_{\OO(n)}^2(H_0)$ contains an
$\OO$-factor of odd rank. Since $H_0\subseteq K \subseteq H$, $K$ has the same
dimension as $H$. It is therefore obtained from $H$ by inverse
field restriction of a factor $\OOF2k$ to the factor $\UUF1k$ or
by merging a double factor $\OOF1{k_1}\times\OOF1{k_2}$ to the
factor $\OOF1{k_1+k_2}$. In both cases, $H$ has at least the same
number of $\OO$-factors of odd rank as $K$, hence $\mr
C_{\OO(n)}(H)$ has an $\OO$-factor of odd multiplicity and
therefore contains an element of negative determinant.

It follows that we may restrict attention to elements $d$ of $\SO(n)$ for which
$d_2\in\mr
C_{\OO(n)}(H)$. For such elements, $d H d^{-1}$ is of standard form. If,
conversely, $dHd^{-1}$ is of standard form then so is $d(\mr SH^+)d^{-1}$. This
shows that the operations which apply to reductions of $P$ to $H_0$ are
given by those operations on reductions of $P^{\OO(n)}$ to $H_0$ which can be
implemented by $d\in\SO(n)$. We take the latter operations from the proof of
Theorem \rref{T-factoriz-O} and determine what signs they require. Like there,
let $\vp_i$ denote the automorphism of $\I_{\KK_i}(m_i)$ induced by $a_i$. We
check:

--~ The determinant of $d_1$ is given by the sign of the
corresponding permutation of the standard basis in $\RR^n$. If it
is negative then $H$ contains an $\OO$-factor of odd rank and
odd multiplicity.

--~ The flips of signed factors of $Q_0$ produce a sign in
\eqref{G-detcond}:~ indeed, since the multiplicity is odd, it suffices to
consider $\det a_i$. To induce a flip for the case of $\SO$-factors of even rank and
$\UU$-factors of odd rank, $\vp_i$
must be outer. Then Lemma \rref{L-nmr-2} yields $\det a_i = -1$.
The case of $\SO$-factors of odd rank is degenerate in that the
flip is trivial here. Nevertheless we may choose $a_i$ so that
$\det a_i=-1$ and assume that $\vp_i$ generates this (trivial) flip.

--~ The flips of unsigned factors of $Q_0$ do not produce a sign in
\eqref{G-detcond}:~ here either the multiplicity is even, in which
case the assertion is obvious, or the factor is a $\UU$-factor of
even rank, in which case the assertion follows from Lemma
\rref{L-nmr-2}.

Now we can prove Theorem \rref{T-factoriz-SO}.
If $Q^H$ has an $\OO$-factor of odd rank, $\mr C_{\OO(n)}(H)$ has an
$\OO$-factor of odd multiplicity. Then $\mr C_{\OO(n)}(H)$ contains an element
of negative determinant so that we may choose the sign of $d_2$ appropriately
to compensate for any sign produced by $d_1$ or the $a_i$ in
\eqref{G-detcond}. It follows that in this case, any permutation
and any flip can be applied to $Q_0$ and hence to $Q^H$ and to $\sigma$.
If $Q^H$ does not have an $\OO$-factor of odd rank, $\det d_1 = 1$. In addition,
$\mr C_{\OO(n)}(H)$ does not have an $\OO$-factor of odd multiplicity,
hence $\det d_2 = 1$. Hence, while permutations and
unsigned flips of $Q_0$ can be applied arbitrarily, signed flips have to be
paired. Since the properties of factors of $Q_0$ to be signed or unsigned and
stable or unstable carry over to the corresponding factors of $Q^H$,
signed/unsigned flips of $\UU$-factors of $Q_0$ translate into signed/unsigned
flips of factors of $Q^H$, flips of signed unstable $\SO$-factors
translate into flips of $\sigma$ and flips of unsigned $\SO$-factors cancel.
This proves the
assertion for the case where $Q^H$ does not possess a stable
signed $\OO$-factor. If, on the other hand, $Q^H$ has a stable signed
factor then $Q_0$ has a stable signed factor. By pairing any
signed flip with the flip of this factor we see that in effect any
flip can be applied to $Q_0$ and hence to $Q^H$ and to $\sigma$.

Finally, consider a bundle reduction $Q$ of type $\typeSHm$ of
$P$. Let $d\in\SO(n)$ such that $d(\mr SH^-)d^{-1}$ is of standard
form, i.e., equal to $\mr SK^-$ for some Howe subgroup $K$ of
$\OO(n)$ of standard form. Let $a_{(n)}$ be the element of
$\OO(n)$ chosen to define $\mr SH^-$ and put $\tilde d = a_{(n)} d
a_{(n)}^{-1}$. Let $\tilde Q$ be obtained by transforming $Q$ by
$a_{(n)}$ inside $Q^{\OO(n)}$ and let $\tilde P$ be obtained by
transforming $P$ by $a_{(n)}$ inside $P^{\OO(n)}$. Then $\tilde Q$
is a bundle reduction of type $\typeSHp$ of $\tilde P$ and $d(\mr
SH^-)d^{-1}$ is of standard form iff $\tilde d(\mr SH^+)\tilde
d^{-1}$ is of standard form. Thus, any operation applicable to
bundle reductions of type $\typeSHp$ applies to $\tilde Q$ and
hence to $Q$, by definition of the classifying data of $Q$.
 \qed
 \bigskip


\section{Summary:~ Classification of orbit types}
\label{S-summary}


We summarize the classification of orbit types obtained in this
paper. For the sake of completeness, we include the results for
$\mr U(n)$ and $\SU(n)$ derived in \cite{RSV:clfot}.

Let $M$ be a closed, simply connected\footnote
 {
The assumption that $M$ be simply connected can be dropped in case
$G=\UU(n)$ or $\SU(n)$. For the other groups, it can be weakend to
$H^1(M,\Z_2) = 0$, see Remark \rref{Bem-simplyconnected}. In any
case, the results hold as well for $\dim(M) =2,3$.
 }
manifold of dimension $4$
and let $P$ be a principal $G$-bundle over $M\, ,$ with $G$ being
a classical compact Lie group. Orbit types of the action of the
group of vertical automorphisms of $P$ on the affine space of
connections are 1:1 with isomorphism classes of holonomy-induced
Howe subbundles of $P$, factorized by the action of the structure
group.

In the following, $r, m_i, k_i$ denote positive integers and $w$,
$p$, $e$ and $c$ denote the Stiefel-Whitney, Pontryagin, Euler and
Chern classes, respectively. Let  $\tilde p(Q_i) = \sum_k (-1)^k
p_k(Q_i)$ and $\tilde c(Q_i) = \sum_k (-1)^k c_k(Q_i)$ and let
$\rho_2:H^\ast(M,\ZZ)\to H^\ast(M,\ZZ_2)$ be induced by reduction
mod $2$.

\paragraph{Isomorphism classes of Howe subbundles of $P$}~
 \medskip

\underline{$G=\UU(n)$:}~ The classes are labelled by
 \beq\label{G-HSB-data-U}
 \textstyle
 \big(
(m_1,k_1,c(Q_1))
 ,\dots,
(m_r,k_r,c(Q_r))
 \big) \, ,
 \eeq
where $Q_i$ are principal $\UU(m_i)$-bundles over $M$. These data are subject to
the relations
 \beqa\label{G-HSGcond-U}
n
 & = &
 \textstyle
\sum_{i=1}^r k_i \, m_i  \,,
\\ \label{G-redfml-U}
c(P) & = & c(Q_1)^{k_1}\dots c(Q_r)^{k_r}\,.
 \eeqa

\underline{$G=\SU(n)$:}~ The classes are labelled by the data
\eqref{G-HSB-data-U} and a characteristic class $\xi\in H^1(M,\ZZ_g)$ for
$\ZZ_g$-bundles, where $g$ is the greatest common divisor of the $k_i$. These
data are subject to the relations \eqref{G-HSGcond-U}, \eqref{G-redfml-U} and
$$
 \textstyle
\beta \xi = \sum_{i=1}^r \frac{k_i}{g} ~ c_1(Q_i)\,,
$$
where $\beta : H^1(M,\ZZ_g) \to H^2(M,\ZZ)$ is the connecting
(Bockstein) homomorphism associated with the sequence of
coefficient groups $0\to\ZZ\to\ZZ\to\ZZ_g\to0$.
 \medskip

\underline{$G = \Sp(n)$:}~ The classes are labeled by $r$-tuples
 \beq\label{G-HSB-data1}
 \textstyle
 \big(
(\KK_1,m_1,k_1,\alpha_1)
 ,\dots,
(\KK_r,m_r,k_r,\alpha_r)
 \big) \, ,
 \eeq
where $\KK_i=\RR$, $\CC$ or $\HH$ and $\alpha_i$ are
characteristic classes of principal bundles $Q_i$ over $M $ with
structure group $\OO(m_i)$ for $\KK_i = \RR$, $\UU(m_i)$ for
$\KK_i = \CC$ and $\Sp(m_i)$ for $\KK_i = \HH$. In detail,
$\alpha_i = \{w(Q_i),p(Q_i),[e](Q_i)\}$ for $\KK_i = \RR$ and
$\alpha_i = c(Q_i)$ for $\KK_i = \CC$ or $\HH$. Here $[e](Q_0)$ is
defined as the equivalence class $\{\pm e(Q_{i0})\}$, where
$Q_{i0}$ is a reduction of $Q_i$ to $\SO(m_i)$. These data are
subject to the relation \eqref{G-HSGcond-U} and
 \beq\label{G-redfml-c1}
 \textstyle
c(P)
 =
\prod\nolimits_{\KK_i=\RR} ~ \tilde p(Q_i)^{2 k_i}
 \cdot
\prod\nolimits_{\KK_i=\CC} ~ c(Q_i)^{k_i} \, \tilde c(Q_i)^{k_i}
 \cdot
\prod\nolimits_{\KK_i=\HH} ~ c(Q_i)^{k_i}\,.
 \eeq

\underline{$G = \OO(n)$:}~ The classes are labeled by the data
\eqref{G-HSB-data1}, subject to the relations
 \beqa\label{G-HSGcond-O}
n
 & = &
 \textstyle
\sum_{i=1}^r ~(\dim_{\RR} \KK_i) \, k_i \, m_i\,,
\\ \label{G-redfml-w1}
w(P)
 & = &
 \textstyle
 \prod\nolimits_{\KK_i=\RR} ~ w(Q_i)^{k_i}
 \cdot
\prod\nolimits_{\KK_i=\CC,\HH} ~ \rho_2\left(c(Q_i)\right)^{k_i}\,,
\\ \label{G-redfml-p1}
\tilde p(P)
 & = &
 \textstyle
 \prod\nolimits_{\KK_i=\RR} ~ \tilde p(Q_i)^{k_i}
 \cdot
\prod\nolimits_{\KK_i=\CC,\HH} ~ c(Q_i)^{k_i}\,\tilde
c(Q_i)^{k_i}\,,
\\ \label{G-redfml-O-e1}
[e](P)
 & = &
 \textstyle
\prod\nolimits_{\KK_i=\RR} ~ [e](Q_i)^{k_i}
 \cdot
\prod\nolimits_{\KK_i=\CC} ~ \left[c_{m_i}(Q_i)\right]^{k_i}
 \cdot
\prod\nolimits_{\KK_i=\HH} ~ \left[c_{2m_i}(Q_i)\right]^{k_i}\,.
 \eeqa
For $\KK_i = \RR$, the classes $\alpha_i$ are in addition subject
to the relations valid for the characteristic classes of
$\OO(m_i)$-bundles, see Corollary \rref{C-Obun} and the discussion
after Equation \eqref{G-w4}.
 \medskip

\underline{$G = \SO(n)$:}~ The classes are labelled by those of
the data \eqref{G-HSB-data1} which satisfy

1.~ if there is $i$ with $\KK_i=\RR$, $m_i = 2$ and $k_i$ odd
there is also $j\neq i$ with $\KK_j=\RR$ and $k_j$ odd,

2.~ if there are $i\neq j$ with $\KK_i = \KK_j = \RR$, $m_i = m_j
= 1$ and $k_i$, $k_j$ odd there is also $k\neq i,j$ with
$\KK_k=\RR$ and $k_k$ odd,

3.~ Conditions 1 and 2 with $m$ and $k$ exchanged.

Moreover, the following additional data occur. If there does not
exist $i$ with $\KK_i=\RR$ and $m_i$ odd or $k_i$ odd, or with
$\KK_i=\CC$ and $m_i$, $k_i$ odd, the Howe subgroup $H$ of
$\OO(n)$ defined by the data \eqref{G-HSB-data1} gives rise to two
distinct conjugacy classes of Howe subgroups of $\SO(n)$, see
Proposition \ref{P-cjg}. These two classes are distinguished by a
sign $\pm$. If furthermore there exists $i$ with $\KK_i=\RR$,
$k_i$ odd and $2e(Q_{i0})\neq 0$ for some reduction $Q_{i0}$ of
$Q_i$ to $\SO(m_i)$, the two reductions to the subgroup
$\SO(n)\cap H$ of the principal bundle with structure group $H$
defined by the data \eqref{G-HSB-data1} are nonisomorphic. They
are distinguished by a further sign $\sigma = \pm1$. All of these
data are subject to the relations
\eqref{G-HSGcond-O}--\eqref{G-redfml-p1} and
 \beq \label{G-redfml-SO-e1}
e(P) = \pm\sigma
 \cdot
\prod\nolimits_{\KK_i=\RR} ~ e^\circ(Q_i)^{k_i}
 \cdot
\prod\nolimits_{\KK_i=\CC} ~ c_{m_i}(Q_i)^{k_i}
 \cdot
\prod\nolimits_{\KK_i=\HH} ~ c_{2m_i}(Q_i)^{k_i}\,,
 \eeq
where $e^\circ(Q_i)$ are representatives for the equivalence
classes $[e](Q_i)$. As for $\OO(n)$, in case $\KK_i=\RR$, the
characteristic classes are subject to the relations listed in
Theorem \rref{T-SObun}.

 \medskip

\paragraph{Holonomy-induced Howe subbundles}

For $G=\UU(n)$ or $\SU(n)$, all Howe subbundles are holonomy-induced. For
$G = \OO(n)$, $\SO(n)$, or $\Sp(n)$, a Howe subbundle is holonomy-induced if
in the data \eqref{G-HSB-data1} there is neither an $i$ with $\KK_i=\RR$ and
$m_i=2$ nor an $i$ and a $j\neq i$ with $\KK_i=\KK_j=\RR$ and
$m_i=m_j=1$.

\paragraph{Factorization by the action of the structure group on bundle
reductions}

The structure group acts on the data \eqref{G-HSB-data1} and
\eqref{G-HSB-data-U} by permuting the members $(\KK_i,m_i,k_i,\alpha_i)$
and $(m_i,k_i,c(Q_i))$, respectively.
For $G=\Sp(n)$, $\OO(n)$ and $\OO(n)$, the structure group acts in addition on
the characteristic classes of the factors $Q_i$. For $G=\Sp(n)$ and $\OO(n)$
this action is given by $c(Q_i) \mapsto \tilde c(Q_i)$.
For $SO(n)$, we refer to Theorem \ref{T-factoriz-SO}.


\section{Examples}
\label{S-examples}


To illustrate the above results, we will determine the Howe
subbundles of principal bundles with structure groups $G=\OO(4)$,
$\SO(4)$, $\Sp(1)$ and $\Sp(2)$. We will start with $\Sp(1)$ and
$\Sp(2)$, because there is only one equation to be studied here.
\comment{In addition, we discuss bundle reductions occuring in the
$\SO(10)$-GUT, i.e., reductions of $G=\SO(10)$ to the subgroup
$\mr S\big(\UU(2)\times\UU(3)\big)$, the gauge group of the
standard model.}%
To be definite, we restrict attention to the base manifolds $M=\mr
S^2\times\mr S^2$ and $\CC\mr P^2$. Let us recall the ring
structure of $H^\ast(M,\ZZ)$ and $H^\ast(M,\ZZ_2)$. Let $\zeta$
denote the generator of $H^2(\mr S^2,\ZZ)$ and let $\gamma$ denote
the generator of $H^2(\CC\mr P^2,\ZZ)$. Then
$$
 \begin{array}{c|c|c|c|c}
 & \multicolumn{2}{l|}{M = \mr S^2\times\mr S^2} & \multicolumn{2}{l}{M = \CC\mr
 P^2}
\\
 & \text{group} & \text{generated by} & \text{group} & \text{generated by}
\\ \hline
H^2(M,\ZZ) & \ZZ\oplus\ZZ & \zeta\times 1, 1\times \zeta & \ZZ &
\gamma
\\
H^4(M,\ZZ) & \ZZ & \zeta\times\zeta & \ZZ & \gamma^2
\\
H^2(M,\ZZ_2) & \ZZ_2\oplus\ZZ_2 & \rho_2\zeta\times 1, 1\times
\rho_2\zeta & \ZZ_2 & \rho_2\gamma
\\
H^4(M,\ZZ_2) & \ZZ_2 & \rho_2\zeta\times\rho_2\zeta & \ZZ_2 &
\rho_2\gamma^2\,.
 \end{array}
$$
We use the following parameterization of characteristic classes of
factors $Q_i$. For $\UU$- and $\Sp$-factors,
$$
c_1(Q_i) = x_i \, \gamma
 \,,~~~~~~
c_2(Q_i) = z_i \, \gamma^2
$$
in case $\caseCPt$ and
$$
c_1(Q_i) = x_i \, \zeta\times 1 + y_i \, 1\times\zeta
 \,,~~~~~~
c_2(Q_i) = z_i \, \zeta\times\zeta
$$
in case $\caseStS$, where $x_i,y_i,z_i$ are integers. The only
$\OO$-factors we will come about have structure group $\OO(2)$ or
$\OO(3)$. In case of an $\OO(2)$-factor, $w_2(Q_i)$ and $p_1(Q_i)$
can be expressed in terms of $[e](Q_i)$, see Theorem
\rref{T-SObun} and Corollary \rref{C-Obun}. Writing, respectively,
$$
[e](Q_i) = f_i\,[\gamma]
 \,,~~~~~~
[e](Q_i) = [f_i\,\zeta\times 1 + g_i\,1\times\zeta]
$$
with integers $f_i,g_i\in\ZZ$, we obtain
$$
w_2(Q_i) = f_i\,\rho_2\gamma
 \,,~~~~~~
p_1(Q_i) = f_i^2\,\gamma^2
$$
for $M=\CPt$ and
$$
w_2(Q_i) = f_i \,\rho_2\zeta\times 1 + g_i \,1\times\rho_2\zeta
 \,,~~~~~~
p_1(Q_i) = 2f_ig_i \,\zeta\times\zeta
$$
for $M=\StS$. The integers $f_i,g_i$ are determined uniquely if we
require $f_i\geq 0$ or $(f_i,g_i)\geq (0,0)$ (lexicographic
ordering), respectively. In case of an $\OO(3)$-factor, we write
$$
w_2(Q_i) = s_i \,\rho_2\gamma
 \,,~~~~~~
p_1(Q_i) = a_i \,\gamma^2
$$
in case $M=\CPt$ and
$$
w_2(Q_i) = s_i \,\rho_2\zeta\times 1 + t_i \,1\times\rho_2\zeta
 \,,~~~~~~
p_1(Q_i) = a_i \,\zeta\times\zeta
$$
in case $M = \StS$, where $s_i,t_i,a_i\in\ZZ$. The parameters
$s_i,t_i$ are uniquely determined if we require $s_i, t_i = 0,1$.

Finally, let us note the following. With the exception of
$\SO(2)$, the center of $\OO(n)$, $\SO(n)$ or $\Sp(n)$
(corresponding to the Howe subgroup $\OOF1n$) is finite. Due to
the assumption that $M$ be simply connected, reductions to the
center are therefore necessarily trivial and hence occur exactly
when $P$ is trivial. They will not be mentioned below.


\subsection{$\Sp(1)$ and $\Sp(2)$}


The reduction equation is given by \eqref{G-redfml-c}. We
parameterize the characteristic class of $P$ by an integer $Z$,
i.e., $c_2(P) = Z \gamma^2$ for $\CPt$ and $c_2(P) = Z
\zeta\times\zeta$ for $\StS$. According to Theorem
\rref{T-factoriz-O}, reductions have to be identified iff they can
be transformed into one another by flips of $\UU$-factors and
permutations. To factorize by flips we require $x_i\geq 0$ or
$(x_i,y_i)\geq(0,0)$ (lexicographic ordering), respectively, for
any $\UU$-factor. To factorize by permutations we fix an order of
the factors $H_i$ for each conjugacy class of Howe subgroups $H$
and require that the parameters of identical factors (same field,
rank and multiplicity) increase w.r.t.\ lexicographic ordering. In
detail, for $i<j$ we require

--~ $f_i \leq f_j$ or $(f_i,g_i) \leq (f_j,g_j)$, respectively,
whenever $H_i = H_j = \OOF 2 k$,

--~ $(a_i,s_i)\leq (a_j,s_j)$ or $(a_i,s_i,t_i)\leq
(a_j,s_j,t_j)$, respectively, whenever $H_i = H_j = \OOF 3 k$;
here the restriction on $s_i$, $s_j$, $t_i$, $t_j$ to take the
values $0$ and $1$ only is understood,

--~ $(x_i,z_i)\leq(x_j,z_j)$ or $(x_i,y_i,z_i)\leq(x_j,y_j,z_j)$,
respectively, whenever $H_i = H_j = \UUF m k$,

--~ $z_i\leq z_j$ whenever $H_i = H_j = \SpF m k$.
\bigskip

Now consider $G=\Sp(1)$. The only nontrivial Howe subgroup is
$\UUF11$. It is holonomy-induced. Parameters are $x\in\ZZ$ for
$M=\CPt$ and $x,y\in\ZZ$ for $M=\StS$. The reduction equation
\eqref{G-redfml-c} reads $c(P) = c(Q)\tilde c(Q)$. In terms of
parameters, this amounts to
$$
Z = -x^2
 ~~~~ \text{ or } ~~~~
Z = -2xy\,,
$$
respectively. In case $\caseCPt$, a solution $x$ exists iff $-Z$
is a square. Due to $0\leq x$, the solution is unique. In case
$\caseStS$, solutions exist iff $Z$ is even. Under the condition
$(0,0)\leq (x,y)$ the solutions are parameterized as follows. If
$Z\neq 0$, $x=\ve_1,\cdots,\ve_s$, where $\ve_l$ are the
(nonnegative) divisors of $\frac{|Z|}{2}$. If $Z=0$, one has
$x=y=0$ and the two families of solutions $x>0$, $y=0$ and $x=0$,
$y>0$. Thus, over $\CPt$ the number of orbit types with stabilizer
isomorphic to $\UU(1)$ can be $0$ or $1$, whereas over $\StS$ it
can be any nonnegative integer or countably infinite.

\bre

By virtue of the isomorphism $\vp_{\HH,\CC}:\Sp(1)\to\SU(2)$,
$\UUF11$ is mapped to the subgroup of diagonal matrices in
$\SU(2)$. The above result is consistent with what is known about
reductions of $\SU(2)$-bundles to this subgroup.

\ere
\smallskip

Now consider $G=\Sp(2)$. Equations will be given in terms of
parameters only. First, we discuss holonomy-induced Howe
subbundles, see Figure \rref{F-HDgr-Sp}.
\medskip

$\UUF12$:~ In case $\caseCPt$, $Z = -2x^2$; in case $\caseStS$, $Z
= - 4xy$. The discussion is analogous to the case of the Howe
subgroup $\UUF11$ of $\Sp(1)$.
\medskip

$\OOF11\times\UUF11$:~ In case $\caseCPt$, $Z = -x^2$. In case
$\caseStS$, $Z = -2xy$. The parameters $x$ and $y$ belong to the
$\UU$-factor. In both cases, the discussion is analogous to the
case of the Howe subgroup $\UUF11$ of $\Sp(1)$.
\medskip

$\OOF11\times\SpF11$:~ In both cases, the equation is $Z = z$,
where $z$ belongs to the $\Sp$-factors. Hence, there exists a
unique reduction for any $P$.
\medskip

$\SpF12$:~ In both cases, the equation is $Z = 2z$. There exists a
reduction iff $Z$ is even and the reduction is unique.
\medskip

$\SpF11\times\SpF11$:~ In both cases, the equation is $Z = z_1 +
z_2$. Reductions exist for any $P$. They can be parameterized by
$z_1\in\ZZ$.
\medskip

$\UUF11\times\UUF11$:~ In case $\caseCPt$, $Z = -(x_1^2 + x_2^2)$.
There is no reduction for $Z>0$ and a unique reduction $x_1=x_2=0$
for $Z=0$. For $Z<0$, there is the classical result that for given
integer $n$ the number of solutions of the equation $a^2 + b^2 =
n$ satisfying $a\geq 0$ and $b > 0$ is given by $N(n) =
\sum_{d|n}\chi(d)$, where the sum runs over the divisors of $n$
and $\chi$ denotes the nontrivial Dirichlet character modulo $4$,
defined by $\chi(d) = 1$ if $d=1\mod 4$, $\chi(d) =-1$ if $d=3\mod
4$ and $\chi(d) = 0$ if $d$ is even \cite{Ireland}. First, it
follows that reductions exist if and only if those of the prime
factors $p$ of $-Z$ which satisfy $p = 3\mod 4$ appear with even
powers. That this condition is sufficient for the existence of reductions can
also be deduced directly from Fermat's 
theorem on sums of two squares and the Brahmagupta-Fibonacci
identity. Second, it follows that the number of reductions, i.e.,
solutions satisfying $0\leq x_1\leq x_2$, is $\frac{N(-Z)+1}{2}$
if $-Z$ is a square or double a square and $\frac{N(-Z)}{2}$
otherwise. For example, for small $n$ we have $N(-Z) = 0$ for $-Z
= 3,6$, $N(-Z) = 1$ for $-Z = 1,2,4,8,9$ and $N(-Z)=2$ for $-Z =
5,10$; yet the reduction is unique also in the last two cases. On
the other hand, it is easy to see that both $N(-Z)$ and the number
of reductions can take any nonnegative integer value.

In case $\caseStS$, $Z = -2(x_1y_1 + x_2 y_2)$. Reductions exist
iff $Z$ is even. They may be enumerated by choosing $(x_1,y_1)\geq
(0,0)$ arbitrarily and letting $(x_2,y_2)$ run through the
solutions of $x_2y_2 = - \frac Z 2 - x_1 y_1$ which satisfy
$(x_2,y_2)\geq (x_1,y_1)$ (for certain values of $(x_1,y_1)$ there
may be no such solutions).
\medskip

$\UUF11\times\SpF11$:~ In case $\caseCPt$, $Z = z-x^2$. In case
$\caseStS$, $Z = z - 2xy$. Here $x$ and $y$ belong to the
$\UU$-factor and $z$ belongs to the $\Sp$-factor. Reductions exist
for all $P$. They are parameterized by $x\geq 0$ in case $\caseCPt$
and $(x,y)\geq 0$ in case $\caseStS$.
\medskip

$\UUF21$:~ In case $\caseCPt$, $Z = 2 z - x^2$. Reductions always
exist. They are parameterized by $x=2l$ if $Z$ is even and $x=
2l+1$ if $Z$ is odd, where $l=0,1,2,\dots$ . In case $\caseStS$, $Z = 2 (z -
xy)$. Reductions exist iff $Z$ is even. They are parameterized by
$(x,y)\geq (0,0)$ then.
\medskip

Out of the Howe subgroups whose reductions are not
holonomy-induced, there is only one which does not consist
exclusively of $\OO(1)$-factors:
\medskip

$\OOF21$:~ Here the abstract reduction equation is $c(P) = \tilde
p(Q)^2$. Using the relation $p_1(Q) = [e](Q)^2$, see Theorem
\rref{T-SObun} and Corollary \rref{C-Obun}, we obtain $c_2(P) =
-2 [e](Q)^2$. In case $\caseCPt$, this equation becomes $Z = -2f^2$
with $f\geq 0$. In case $\caseStS$, $Z = -4fg$ with
$(f,g)\geq(0,0)$. The discussion of these equations is analogous
to the case of the Howe subgroup $\UUF11$ of $\Sp(1)$.


\subsection{$\OO(4)$}


We parameterize the characteristic classes of $P$ by
 \beq\label{G-ccP-CP2}
 w_2(P) =
S\,\rho_2\gamma
 \,,~~~~~~
p_1(P) = A \gamma^2
 \,,~~~~~~
[e](P) = F [\gamma^2]
 \eeq
for $M=\CPt$ and by
 \beq\label{G-ccP-S2S2}
 w_2(P) = S\,\rho_2\zeta\times 1 + T
\,1\times\rho_2\zeta
 \,,~~~~~~
p_1(P) = A \zeta\times\zeta
 \,,~~~~~~
[e](P) = F [\zeta\times\zeta]
 \eeq
for $M=\StS$, where $A,F,S,T$ are integers. $S,T,F$ are determined
uniquely under the conditions $S,T = 0,1$ and $F\geq 0$. These
conditions will be assumed to hold below. The characteristic
classes of $P$ are subject to a single relation which arises from
the relation $w_4(P) = \rho_2 [e](P)$ of Theorem \rref{T-SObun} by
replacing $w_4(P)$ by means of the fundamental relation, see
\eqref{G-w4}. Using \eqref{pontryagin1} we find that in terms of the parameters,
this relation reads, respectively,
 \beq\label{G-fundrel-par}
A - S^2 - 2F = 0 \mod 4
 \,,~~~~~~
A - 2ST - 2F = 0 \mod 4\,.
 \eeq
In particular, in case $M=\StS$, $A$ must be even. To obtain unique
representatives under the action of the structure group, see
Theorem \rref{T-factoriz-O}, the same conditions on the parameters
as for the structure groups $\Sp(1)$ and $\Sp(2)$ above have to be
imposed. The reduction equations are given by \eqref{G-redfml-w},
\eqref{G-redfml-p}, \eqref{G-redfml-O-e}. We discuss them in terms
of the parameters, starting with the holonomy-induced reductions.
The equation for $w_4(P)$ needs not be considered, because it
is automatically satisfied if the equation for $[e](P)$ holds.
\medskip

$\UUF12$:~ In case $\caseCPt$, $S=0\mod 2$, $A = 2 x^2$, $F = \pm x^2$.
Reductions exist iff $S=0 \mod 2$, $A=2F$ and $F$ is a square.
Under the condition $x\geq 0$, they are unique. In case
$\caseStS$, $S=T=0$, $A = 4 x y$ and $F = \pm 2xy$. Reductions
exist iff $S=T=0$ and $A=\pm 2F$. Under the condition $(x,y)\geq
(0,0)$, they are parameterized as follows. If $F\neq 0$,
$(x,y) = (\ve,\pm\frac{F}{2\ve})$, where $\ve$ runs through the (nonnegative)
divisors of $F$. If $F=0$, one has $x=y=0$ and the two families of
solutions $x>0$, $y=0$ and $x=0$, $y>0$.
\medskip

$\OOF12\times\UUF11 $:~ In case $\caseCPt$, $S= x \mod 2$, $A =
x^2$ and $F = 0$ where the parameter $x$ refers to the
$\UU$-factor. A reduction exists iff $F=0$ and $A$ is a square.
Due to $x\geq 0$, it is unique. Since the mod $4$ reduction of a
square is $0$ if the square is even and $1$ if the square is odd,
the equation for $S$ is automatically satisfied due to
\eqref{G-fundrel-par}.

In case $\caseStS$, $S=x\mod2$, $T=y\mod2$, $A
= 2 x y$ and $F = 0$. Reductions exist iff $F=0$ and ($S=1$ or $T=1$ or $A$ is
divisible by $8$). If $A=0$, \eqref{G-fundrel-par} implies $S=0$ or $T=0$.
Accordingly, reductions are parameterized by $x=0$, $y = 2l+T$ or $y=0$,
$x = 2l+S$, $l\geq 0$. If $A\neq 0$, reductions are parameterized as follows. If
$S=T=0$, $A$ must be divisible by $8$ and $x=2\ve_1,\dots,2\ve_s$, where
$\ve_i$ are the divisors of $\frac A 8$. If $S=1$, $T=0$, $x$ runs through the
odd divisors of $\frac A2$. Similarly for $y$ in case $S=0$, $T=1$. If $S=T=1$,
$x$ runs through the divisors of $\frac A2$.
\medskip

$\OOF11\times\OOF31 $:~ In case $\caseCPt$, $S=s \mod 2$, $A = a$,
$F=0$ where the parameters refer to the $\OO(3)$-factor. Together
with \eqref{G-fundrel-par}, these equations imply $a - s^2 = 0
\mod 4$, which is the fundamental relation for the $\OO(3)$-bundle
parameterized by $a$ and $s$. Hence, reductions exist iff $F=0$ and
are unique then. In case $\caseStS$, the above equations hold,
together with $T=t\mod 2$. The result is analogous.
\medskip

$\SpF11 $:~ Here, $A = 2z$, $F = \pm z$, $S=0$ or $S=T=0$,
respectively. Reductions exist iff $S=0$ or $S=T=0$, respectively,
and $A=\pm 2F$. They are unique.
\medskip

$\UUF21$:~ In case $\caseCPt$, $S = x \mod 2$, $A= x^2 - 2z$, $F=\pm z$.
Reductions exist iff $A - 2F$ or $A+2F$ is a square. They are unique in each of
these two cases. (That means in particular, if both $A - 2F$ or $A+2F$ are
squares and $F\neq 0$ there exist exactly $2$ solutions.) By the same
argument as in the discussion of the Howe subgroup $\OOF12\times\UUF11$,
\eqref{G-fundrel-par} implies that the equation for $S$ is automatically
satisfied.

In case $\caseStS$, $S=x\mod2$, $T=y\mod2$, $A=2(xy-z)$, $F=\pm z$. The last and
the third equation yield $xy = \frac A2\pm F$. Hence, in case $F\neq 0$,
potentially there are two families of reductions, corresponding to the two
signs. For each of these families, the discussion is analogus to the case of
reductions to the Howe subgroup $\OOF12\times\UUF11$ above, with $A$ replaced
by $A\pm2F$.
\medskip

$\UUF11\times\UUF11$:~ In case $\caseCPt$,
 \beq\label{G-U1U1-CP2}
S = (x_1 + x_2)\mod 2
 \,,~~~~~~
A= x_1^2+x_2^2
 \,,~~~~~~
F= x_1x_2\,.
 \eeq
We have omitted the equation $F= -x_1x_2$, because it does not have a
solution due to the requirement $0\leq x_1\leq x_2$.
Define $x_\pm := x_2 \pm x_1$. Then \eqref{G-U1U1-CP2} translates into
 \beq\label{G-U1U1-CP2-2}
S = x_+ \mod 2
 \,,~~~~~~
A + 2F = x_+^2
 \,,~~~~~~
A - 2F = x_-^2
 \eeq
and the requirement $0\leq x_1\leq x_2$ translates into the requirement $0\leq
x_-\leq x_+$. Systems \eqref{G-U1U1-CP2} and \eqref{G-U1U1-CP2-2} are
equivalent:~ the only
thing to be checked is that if $x_\pm$ is a solution of \eqref{G-U1U1-CP2-2}
then $x_1 = \frac{x_+ - x_-}{2}$ and $x_2 = \frac{x_+ + x_-}{2}$ are
integers. Since $x_+^2$ and $x_-^2$ differ by $4F$, this is obvious.
Thus, a reduction exists iff $A=S\mod 2$ and $A\pm 2F$ are both squares. If it
exists, it is unique.

In case $\caseStS$,
 \begin{align}\nonumber
S & = (x_1 + x_2)\mod 2
 \,,~~~~~~ &
T & = (y_1 + y_2)\mod 2
 \,,
\\ \label{G-U1U1-S2S2}
A & = 2(x_1y_1 + x_2y_2)
 \,,~~~~~~ &
F & = \pm (x_1y_2 + x_2y_1)
 \end{align}
Define $x_\pm := x_2 \pm x_1$ and $y_\pm := y_2 \pm y_1$.
Then \eqref{G-U1U1-S2S2} translates into
 \begin{align}\nonumber
 \textstyle
S
 & =
 \textstyle
x_+ = x_- \mod 2
 \,, &
 \textstyle
T
 & =
 \textstyle
y_+ = y_- \mod 2\,,
\\ \label{G-U1U1-S2S2-2}
 \textstyle
\frac A2 + F
 & =
 \textstyle
x_+y_+
 \,,~~~
\frac A2 - F = x_-y_-
 ~~~~~~\text{ or } &
 \textstyle
\frac A2 + F
 & =
 \textstyle
x_-y_-
 \,,~~~
\frac A2 - F = x_+y_+
 \end{align}
(recall that $A$ is now even due to the relation
\eqref{G-fundrel-par}; moreover we have added the obvious
equations $S=x_-\mod 2$ and $T=y_-\mod 2$). The requirement
$(0,0)\leq (x_1,y_1)\leq (x_2,y_2)$ translates into the
requirement $(0,0)\leq (x_-,y_-)\leq (x_+,y_+)$. It is
straightforward to check that systems \eqref{G-U1U1-S2S2} and
\eqref{G-U1U1-S2S2-2} are equivalent. We discuss the solutions of
the latter system.

--~ If $S=T=1$, the relation \eqref{G-fundrel-par} implies that
$\frac A2\pm F$ is odd. Hence, reductions always exist. They are
parameterized by arbitrary combinations of decompositions $\frac A2
+ F = a_+b_+$ and $\frac A2 - F = a_-b_-$ with
$(a_\pm,b_\pm)\geq(0,0)$. Then $(x_+,y_+) =
\max\{(a_+,b_+),(a_-,b_-)\}$ and $(x_-,y_-) =
\min\{(a_+,b_+),(a_-,b_-)\}$

--~ If $S=1$, $T=0$, the relation \eqref{G-fundrel-par} implies
that $\frac A2\pm F$ is even. Therefore, again, reductions always
exist. They are parameterized in the same way as in the case
$S=T=1$, with the additional condition that $a_\pm$ has to be odd.

--~ The case $S=0$ and $T=1$ is similar. Instead of $a_\pm$, $b_\pm$ has to
be odd.

--~ If $S=T=0$, the relation \eqref{G-fundrel-par} yields that
$\frac A2\pm F$ is even. For reductions to exist, however, both
$\frac A2 + F$ and $\frac A2 - F$ have to be divisible by $4$. In
this case, reductions are parameterized by arbitrary combinations
of decompositions $\frac{A + 2F}{8} = a_+b_+$ and $\frac{A -
2F}{8} = a_-b_-$ with $(a_\pm,b_\pm)\geq(0,0)$. Then $(x_+,y_+) =
\max\{(2a_+,2b_+),(2a_-,2b_-)\}$ and $(x_-,y_-) =
\min\{(2a_+,2b_+),(2a_-,2b_-)\}$
\bigskip

Next, we discuss the reductions which are not holonomy-induced.
\bigskip

$\OOF22$:~ In case $\caseCPt$, $S=0$, $A=2f^2$, $F=\pm f^2$; in
case $\caseStS$, $S=T=0$, $A=4fg$, $F=\pm 2fg$. The discussion is
analogous to that for the Howe subgroup $\UUF21$, with $x,y$
replaced by $f,g$\footnote{In fact, $[e](Q)$ here and $c_1(Q)$
there are related via the extension of the structure group to
$\OO(2)$. This holds similarly for the other reductions which are
not holonomy-induced.}. The present requirement $f\geq 0$ or
$(f,g)\geq (0,0)$, respectively, which ensures that the parameters
are uniquely determined corresponds to the requirement $x\geq 0$
or $(x,y)\geq (0,0)$, respectively, which singles out unique
representatives for the classes of bundle reductions under the
action of the structure group there.
\medskip

$\OOF21\times\OOF21$:~ In case $\caseCPt$, $S = (f_1 + f_2)\mod 2$, $A= f_1^2
+f_2^2$, $F= \pm f_1f_2$; in case $\caseStS$, $S = (f_1 + f_2)\mod 2$, $T = (g_1
+ g_2)\mod 2$, $A= 2(f_1g_1 + f_2g_2)$, $F= \pm (f_1g_2 + f_2g_1)$. The
discussion is analogous to that for the Howe subgroup $\UUF11\times\UUF11$, with $x_i,y_i$
replaced by $f_i,g_i$.
\medskip

$\OOF21\times\UUF11$:~ In case $\caseCPt$, $S = (f + x)\mod 2$, $A= f^2
+x^2$, $F= \pm fx$; in case $\caseStS$, $S = (f + x)\mod 2$, $T = (g
+ y)\mod 2$, $A= 2(fg + xy)$, $F= \pm (fy + xg)$. Variables $f,g$ 
refer to the $\OO$-factor, variables $x,y$ to the $\UU$-factor. The
discussion is analogous to that for the Howe subgroup $\UUF11\times\UUF11$, with
$x_1,y_1$ replaced by $f,g$ and $x_2,y_2$ replaced by $x,y$; the only difference
is that here permutations do not appear, hence the variables $f,g$ and $x,y$ are
independent from one another.
\medskip

$\OOF21\times\OOF12$:~ In case $\caseCPt$, $S=f\mod 2$, $A=f^2$, $F=0$;
in case $\caseStS$, $S=f\mod 2$, $T=g\mod 2$, $A=2fg$, $F=0$. The parameters
refer to the last factor. The discussion is analogous to that for the Howe
subgroup $\OOF12\times\UUF11$, with $x,y$ replaced by $f,g$.
\medskip

For $\OOF11\times\OOF11\times\UUF11$ and
$\OOF11\times\OOF11\times\OOF21$, see $\OOF12\times\UUF11$.


\subsection{$\SO(4)$}


The characteristic classes $w_2(P)$ and $p_1(P)$ are parameterized
as in \eqref{G-ccP-CP2} and \eqref{G-ccP-S2S2}. The Euler class
will be parameterized by $e(P) = F\gamma^2$ in case $\caseCPt$ and
by $e(P)= F\gamma\times\gamma$ in case $\caseStS$. Here $F$ can
take any integer value. Again, the characteristic classes of $P$
are subject to the single relation \eqref{G-fundrel-par}. The
reduction equations are given by \eqref{G-redfml-w},
\eqref{G-redfml-p} and \eqref{G-redfml-SO-e}. In comparison with
the case of structure group $\OO(4)$, the corresponding equations
for the parameters $A$, $S$ and $T$ are the same, whereas the
equation for $F$ is modified as follows:

--~ $F$ can take any integer value,

--~ The sign is either positive or negative, according to whether
the Howe subgroup $\mr SH^+$ or $\mr SH^-$ is considered. Thus,
solutions for positive sign belong to $\mr SH^+$ and solutions for
negative sign belong to $\mr SH^-$. If $\mr SH^+$ and $\mr SH^-$
are conjugate in $\SO(4)$, only the equation with the positive
sign has to be taken into account.

--~ The factor $\sigma$ is added.

First, we discuss holonomy-induced Howe subbundles. Here,
$\sigma=1$, because if $G=\SO(4)$ then $\sigma=-1$ requires an
$\OO(2)$-factor. The system of equations is therefore the same as
for the corresponding subgroup of $\OO(4)$. We comment on how the
discussion has to be modified and we derive the conditions on the
parameters of the reductions which have to be imposed in order to
obtain unique representatives w.r.t.\ the action of the structure
group $\SO(4)$. Generally, $F\in\ZZ$ and solutions for positive
sign belong to $\mr SH^+$ and solutions for negative sign belong
to $\mr SH^-$.
 \medskip

$\UUF12{}^\pm$:~ In case $\caseCPt$ the condition $\langle$$A=2F$
and $F$ is a square$\rangle$  has to be replaced by
$\langle$$A=\pm2F$ and $\pm F$ is a square$\rangle$. In
particular, in the case of $\UUF12{}^+$, necessarily $F\geq0$
whereas in the case of $\UUF12{}^-$, $F\leq 0$. Since the factor
$\UUF12{}^\pm$ is unsigned, any flip is allowed. Hence, we have to
require $x\geq 0$ in case $\caseCPt$ and $(x,y)\geq (0,0)$ in case
$\caseStS$. This is the same condition as in the case of structure
group $\OO(4)$.
 \medskip

$\UUF11\times\UUF11$:~ Only the positive sign has to be discussed.
In case $\caseCPt$ this was so before for another reason. In case
$\caseStS$ this amounts to setting $(x_+,y_+) = (a_+,b_+)$ and
$(x_-,y_-) = (a_-,b_-)$. Since the factors of $\UUF11\times\UUF11$
are both signed, flips can be applied in pairs only. We thus have
to require $(x_1,x_2) \geq (0,0)$ in case $\caseCPt$ and
$((x_1,y_1),(x_2,y_2))\geq ((0,0),(0,0))$ (lexicographic ordering)
in case $\caseStS$. We leave it to the reader to translate these
conditions to $(x_\pm,y_\pm)$.
 \medskip

$\SpF11{}^\pm$:~ The discussion of the reduction equations is
modified in the standard way. The action of the structure group
$\SO(4)$ is trivial.
 \medskip

$\UUF21{}^\pm$:~ Again, the discussion of the reduction equations
is modified in the standard way. Since the factor $\UUF21$ is
unsigned, it can be flipped. This leads to the same conditions on
the parameters as in the case of $\OO(4)$.
 \medskip

Next, we discuss Howe subbundles which are not holonomy-induced,
omitting Howe subgroups consisting enirely of $\OO(1)$-factors. To
define the invariant $\sigma$, as unique representatives for the
classes in $\mr PH^2(M,\ZZ)$ we choose the elements of
$H^2(M,\ZZ)$ with nonnegative (w.r.t.\ lexicographic ordering)
coefficients w.r.t. the generators $\gamma$ and $\gamma\times 1$,
$1\times\gamma$, respectively. For an integer $x$, define $\sgn(x)
= 1$ if $x\geq 0$ and $\sgn(x) = -1$ otherwise. For integers $x$,
$y$ define $\sgn(x,y) = 1$ if $(x,y)\geq (0,0)$ and $\sgn(x,y) =
-1$ otherwise.
 \medskip

$\OOF22{}^\pm$:~ Since the multiplicity is even, $\sigma = 1$.
Therefore, the reduction equations are the same as for $\OO(4)$
and the discussion is modified in the standard way. The action of
the structure group $\SO(4)$ is trivial.
 \medskip

$\OOF21\times\OOF21$:~ In the equation for $F$, $\sigma$ can be
omitted if one allows $f_i$ and $(f_i,g_i)$, respectively, to take
arbitrary values. This amounts to seeking reductions to the
identity connected component, which coincides with the Howe
subgroup $\UUF11\times\UUF11$ discussed above. For any reduction
$(x_1,y_1)$, $(x_2,y_2)$ found there, we define:~

--~ in case $\caseCPt$, $\sigma = \sgn(x_1) \sgn(x_2)$ and $f_i =
\sgn(x_i) x_i$, $i=1,2$,

--~ in case $\caseStS$, $\sigma = \sgn(x_1,y_1) \sgn(x_2,y_2)$ and
$(f_i,g_i) = \sgn(x_i,y_i) (x_i,y_i)$, $i=1,2$.

This way, due to the conditions imposed on $x_i$ and $(x_i,y_i)$,
respectively, each reduction to $\OOF21\times\OOF21$ arises from
exactly one reduction to $\UUF11\times\UUF11$. For the same
reason, the action of the structure group $\SO(4)$ is already
factored out. For completeness, let us just state what this action
amounts to. As operations there occur the interchange of factors
and the flip of $\sigma$. Since both factors of
$\OOF21\times\OOF21$ are signed, flips have to be applied in pairs
which means that $\sigma$ remains unchanged unless one of the
factors is stable and the other one is not. That is, $f_1=0$ and
$f_2\neq 0$ or $(f_1,g_1) = (0,0)$ and $(f_2,g_2) \neq (0,0)$,
respectively, or vice versa.


\section*{Acknowledgements}

The authors are very much indebted to M.\ \v{C}adek for helpful
hints concerning the classification of real vector bundles. They
are also grateful to J.\ Huebschmann for discussing with them
several tools from algebraic topology used in the paper.

A.\ Hertsch acknowledges financial support by the International
Max Planck Research School of the Max Planck Institute for
Mathematics in the Sciences (MPI-MIS) at Leipzig.



\begin{appendix}




\section{Universal bundles}
\label{universal bundles}


We collect some basic facts about universal bundles and
classifying spaces. For details, see e.g.\ \cite[\S 19]{steenrod} or \cite{Hus}.

For every Lie group $G$ and every positive integer $n$ there
exists a principal $G$-bundle $(\EE G)_n \ra (\BB G)_n$ such that
for every connected $(n-1)$-dimensional $CW$-complex $X$, the
assignment of the pull-back bundle $f^\ast (\EE G)_n$ to a map
$f:X\to (\BB G)_n$ induces a bijection from the set of homotopy
classes of maps $X\to (\BB G)_n$ onto the set of isomorphism
classes of principal $G$-bundles over $X$. The bundle $(\EE G)_n$
is referred to as an $n$-universal bundle for $G$, $(\BB G)_n$ is
referred to as an $n$-classifying space of $G$ and the map
$f:X\to(\BB G)_n$ associated with a principal $G$-bundle $P$ over
$X$ is referred to as an $n$-classifying map of $P$. Given a
principal $G$-bundle $E\ra B$, this bundle is $n$-universal for
$G$ if and only if $B$ is path connected and
 \beq\label{G-piE}
\pi_i(E) = 0 \,,~~~~~~ 1\leq i< n\,.
 \eeq
For the groups considered in this paper, $n$-universal bundles are
constructed as follows. For $\KK=\RR$, $\CC$, $\HH$ and positive
integers $k < l$ let $V^{\mathbb{K}}_{l,k}$ denote the Stiefel
manifold of orthonormal $k$-frames in $\KK^l$ and let
$G^{\mathbb{K}}_{l,k}$ denote the Grassmann manifold of
$k$-dimensional subspaces in $\KK^l$. By mapping an orthonormal
frame to the subspace it spans one obtains a principal bundle
$V^{\mathbb{K}}_{l,k}\ra G^{\mathbb{K}}_{l,k}$
with structure group $\I_\KK(k)$ (see Section \rref{S-HSG} for this
notation), known as the $\KK$-Stiefel bundle of $k$-frames in
dimension $l$. Since $\pi_i(V^{\mathbb{K}}_{l,k}) = 0$ for $i <
(\dim_\RR\KK)\,(l-k+1) - 1$, the Stiefel bundle in dimension
$l = m+k-1$ is $n$-universal for $\I_\KK(k)$, where $n=(\dim_\RR\KK)\,m - 1$.

The $n$-universal bundles of a Lie group $G$ can be arranged such
that there exist embeddings $(\EE G)_{n_i}\into (\EE G)_{n_{i+1}}$
for an increasing sequence $n_1<n_2<\dots$ . The inductive limit
of such a sequence provides a universal bundle $\EE G\ra \BB G$
for the group $G$. A general construction of the universal bundle
as the infinite join of $G$ is given in \cite{milnor}, see also \cite{Hus}. Since
$\pi_i(\EE G) = 0 $ for all $i$, the long exact homotopy sequence
of the universal bundle implies
 \beq\label{ub4}
\pi_i(G)=\pi_{i+1}(\BB G)
 \,,~~~~~~
i=0,1,2,\dots~.
 \eeq
The assignment $G\mapsto\BB G$ can be made a functor as follows.
Let $\phi: G \to H$ be a Lie group homomorphism and let $P$ be a
principal $H$-bundle over $X$. The fiber bundle
$P^{\phi}=P\times_HG$ associated with $P$ by virtue of the left
action of $G$ on $H$ by $(g,h) \mapsto \phi(g) h$ is a principal
$H$-bundle over $X$, where the action of $H$ is induced from the
action of $H$ on itself by right multiplication. Define
$\BB\phi:\BB G \to \BB H$ to be the classifying map of the
principal $H$-bundle $(\EE G)^\phi\to\BB G$. The assignment
$\phi\mapsto \BB\phi$ is indeed functorial,
$$
\BB(\psi\circ\phi) = \BB\psi\circ B\phi
 \,,~~~~~~
\BB\id_G = \id_{\BB G}\,.
$$
The classifying map of $P^\phi$ is $\BB\phi\circ f : X\to\BB H$,
where $f:X\to\BB G$ is the classifying map of $P$.

In the special case where $\phi$ is a Lie subgroup embedding,
$P^\phi$ is usually denoted by $P^H$ and is called the extension
of $P$ to the structure group $H$. Then $P$ is a reduction of
$P^H$ to the subgroup $G$ of $H$. Since the principal $G$-bundle
$\EE H \to \EE H/G$ is universal for $G$, the classifying map
$\BB\phi$ can be identified with the projection in the fiber
bundle $\EE H/G \to \BB H$ with fiber $H/G$. For example, the
classifying space $\BB\SO(n)$ is the $2$-fold covering of
$\BB\OO(n)$.


\section{Eilenberg-MacLane spaces}
\label{e-m}\label{A-em}


Let $\pi$ be an Abelian group. Up to homotopy equivalence, there
exists a unique space $X$ with $\pi_i(X) = \pi$ if $i=n$ and
$\pi_i(X) = 0$ otherwise. This space is known as the
Eilenberg-MacLane space $K(\pi,n)$. It can be realized as a
$CW$-complex. The following properties of Eilenberg-MacLane spaces are used in
the paper:
 \smallskip

1.~ According to the Hurewicz Theorem and the Universal
Coefficient Theorem, there is a canonical isomorphism
$$
H^n(K(\pi,n),\pi) = \Hom(\pi,\pi)\,.
$$
This way, the identity of $\pi$ is mapped to a distinguished class
$\iota_n\in H^n(K(\pi,n),\pi)$, called the fundamental class of
$K(\pi,n)$. For any $CW$-complex $X$, the assignment of $f^\ast\iota_n$ to a
map $f:X\to K(\pi,n)$ defines a bijection from the set of homotopy classes of
maps $X\to K(\pi,n)$ onto the cohomology group $H^n(X,\pi)$.
 \smallskip

2.~ From the long exact homotopy sequence of the path-loop fibration
$\Omega K(\pi,n)\into PK(\pi,n)\ra K(\pi,n)$ of the space $K(\pi,n)$ one
obtains $\Omega K(\pi,n) = K(\pi,n-1)$.
 \smallskip

3.~ From the long exact homotopy sequence of the product bundle
$K(\pi_1,n)\times K(\pi_2,n)\ra K(\pi_2,n)$ one obtains $K(\pi_1,n)\times
K(\pi_2,n) = K(\pi_1\oplus \pi_2,n)$.
 \smallskip

Furthermore, the notions of transgression and suspension
homomorphism are used. Let $F\stackrel{\iota}{\into}
E\stackrel{p}{\ra} B$ be a fibration and $\delta: H^*(F)\ra
H^*(E,F)$ be the connecting homomorphism of the long exact
sequence of the pair $(E,F)$. Let $\tilde{H}^*(B) = H^*(B,\ast)$
denote the reduced cohomology groups and $\tilde{p}:
(E,F)\ra(B,\ast)$ the induced map. Define subsets
$$
T^*(F) = \delta^{-1}\tilde{p}^*\tilde{H}^*(B) \subset H^\ast(F)
 \,,~~~~~~
S^*(B) = (\tilde{p}^*)^{-1}\delta H^*(F) \subset \tilde
H^\ast(B)\,.
$$
The transgression is the homomorphism $$\tau: T^*(F)\ra S^*(B)/\ker
(\tilde{p}^*)$$ defined by $\tau(u)=[u']$, where $\delta u =
\tilde{p}^*u'$. The suspension homomorphism is the homomorphism
$$\sigma:S^*(B)\ra T^*(F)/\im\, (\iota^*)$$ defined by
$\sigma(v)=[v']$, where $\tilde{p}^*v=\delta v'$. Due to $\ker
\tau = \im\,\iota^*$ and $\ker\sigma = \ker\tilde{p}^*$, $\tau$
and $\sigma$ induce inverse isomorphisms
\[
T^*(F)/\im\, \iota^* = S^*(B)/\ker \tilde{p}^*.
\]
Now consider the fibration $K(\Z_r,n-1)\stackrel{\iota}{\ra} PK(\Z_r,n)
\stackrel{p}{\ra} K(\Z_r,n)$. The Serre exact sequence for this fibration
contains a portion
 \begin{align*}
H^{n-1}(PK(\Z_r,n),\Z_r)
 \stackrel{\iota^*}{\ra}
H^{n-1}(K(\Z_r,n-1),\Z_r)
 \stackrel{\tau}{\ra}
H^n(K(\Z_r,n),\Z_r)
 \stackrel{p^*}{\ra}
H^n(PK(\Z_r,n),\Z_r) \, .
 \end{align*}
Since $H^{i}(PK(\Z_r,n),\Z_r)=0$ for all $i$,
$\im\,(\iota^*)=0=\im\, p^*$. It follows that $\tau$ itself is an
isomorphism here and $\sigma$ is its inverse. Let $\kappa_{n-1}$ and
$\kappa_n$ denote the fundamental class of $K(\Z_r,n-1)$ and
$K(\Z_r,n)$ respectively. Since $\kappa_{n-1}$ and $\kappa_n$ are the unique
generators of $H^{n-1}(K(\Z_r,n-1),\Z_r)$ and $H^n(K(\Z_r,n),\Z_r)$
respectively, we get
 \beq\label{sigma-EMcL}
\tau(\kappa_{n-1})=\kappa_n
 \,,~~~~~~
\sigma(\kappa_n) = \kappa_{n-1}.
 \eeq

The definitions of $\tau$ and $\sigma$ given here follow \cite{thomas1}.
There is an equivalent definition of the transgression as
a certain differential of the Serre spectral sequence. For this as
well as for further details we refer to \cite{mccleary}.


\section{Cohomology operations}
\label{cohomology operations}


A cohomology operation is a map $\Theta=\Theta_X:H^m(X,\pi_1)\ra H^n(X,\pi_2)$
defined for all spaces $X$, with a fixed choice of $m$
and $n$, $\pi_1$ and $\pi_2$ satisfying the naturality property that for all
maps $f:X\ra Y$ the following diagram commutes:
\begin{align*}
 \xymatrix{
H^m(Y,\pi_1)\ar[d]^-{f^*}\ar[r]^-{\Theta_Y}& H^n(Y,\pi_2)\ar[d]^-{f^*}\\
H^m(X,\pi_1)\ar[r]^-{\Theta_X}& H^n(X,\pi_2)
}
 \end{align*}
Let $\iota_m\in H^m(K(\pi_1,m),\pi_1)$ be the fundamental class, see Appendix
\rref{A-em}. For fixed $m,n,\pi_1$ and $\pi_2$, the assignment of
$\Theta(\iota_m)$ to $\Theta$ defines a bijection between the
set of all cohomology operations $\Theta:H^m(X,\pi_1)\ra H^n(X,\pi_2)$
and the cohomology group $H^n(K(\pi_1,m),\pi_2)$. See e.g.\ \cite[Prop.4I.1]{at}
for a proof. The following cohomology operations are used in the text:
 \smallskip

1.~ Let $R$ be a ring. The transformation $H^m(X,R)\ra H^{mp}(X,R)$,
$\alpha\mapsto \alpha^p$ is a cohomology operation since
$f^*(\alpha^p)=(f^*(\alpha))^p$. This example shows that
cohomology operations need not be homomorphisms.
 \smallskip

2.~ The Bockstein homomorphisms are cohomology operations. We use in particular
the Bockstein homorphisms $\beta_m: H^n(X,\Z_m)\ra H^{n+1}(X,\Z_m)$ associated
with the coefficient sequence $0\ra\Z_m\stackrel{m}{\ra} \Z_{m^2}\ra \Z_m\ra 0$
and $\beta:H^n(X,\Z_m)\ra H^{n+1}(X,\Z)$ associated with the coefficient
sequence $0\ra\Z\stackrel{m}{\ra}\Z\stackrel{\rho_m}{\ra} \Z_m\ra 0$. They are
related via
 \beq\label{G-bock}
\beta_m = \rho_m\circ\beta\,.
 \eeq
We use this relation for the case $m=2$.
 \smallskip

3.~ The Steenrod square $\Sq^i : H^n(X,\Z_2) \to H^{n + i}(X,\Z_2) $, see e.g.\ \cite{at}, is a cohomology operation uniquely defined by
the following axioms:
 \begin{align}
\label{steenrod1}
\Sq^0 &= \mathrm{id}\, ,
\\
\label{steenrod2}
\Sq^i x &= x^2  \, , \quad x \in H^i(X,\Z_2)\, ,
\\
\label{steenrod3}
\Sq^i x &= 0  \, , \quad x \in H^j(X,\Z_2)\, , \, j < i\, ,
\\
\label{steenrod4}
\Sq^i(x+y) &= \Sq^i(x)+\Sq^i(y)\, ,
\\
\label{steenrod5}
\Sq^i(xy) &= \sum_{j=0}^i\Sq^j(x)\Sq^{i-j}(y)\, .
 \end{align}
One has $\Sq^1 = \beta_2$, the Bockstein homomorphism associated with the
sequence of coefficient groups $0\ra\Z_2\stackrel{2}{\ra} \Z_{4}\ra \Z_2\ra 0$.
Hence \eqref{G-bock} implies
 \beq\label{steenrod-bockstein}
\Sq^1 = \beta_2 = \rho_2\circ\beta\,.
 \eeq

4.~ The Pontryagin square  $\frak{P}: H^{2k}(X,\Z_2) \to H^{4k}(X,\Z_4)$, see \cite{thomas3}, is a cohomology operation uniquely defined by the
following axioms:
 \begin{align}
 \label{pontryagin01}
\frak{P}\rho u & = u^2 \, ,
\\
\label{pontryagin02}
\rho\frak{P} v & = v^2 \, ,
\\
\label{pontryagin2}
\frak{P}(v_1+v_2) & = \frak{P}v_1+\frak{P}v_2  + \theta_*(v_1  v_2)\,,
\end{align}
for all $u \in H^{4k}(X,\Z_4)\, , $ $v,v_1,v_2 \in H^{2k}(X,\Z_2)\, .$
Writing $u = \rho_4 x$ with $x \in H^{2k}(X,\Z)$ in \eqref{pontryagin01}
we obtain
 \begin{align}\label{pontryagin1}
\frak{P}\rho_2x = \rho_4x^2 \, .
 \end{align}



\section{Characteristic classes}
\label{characteristic classes} \label{S-cc}


We collect the basic facts about characteristic classes for principal bundles
with structure group $G=\OO(n)$, $\SO(n)$ and $\Sp(n)$. For our purposes, it is
suitable to view characteristic classes of a $G$-bundle $P$ as being
obtained from generators $\alpha$ of the cohomology of the classifying space
$\BB G$ as
$$
\alpha(P) = f^\ast \alpha\,,
$$
where $f:X\to\BB G$ is the classifying map of $P$. This way, any relation
between generators translates into a relation for characteristic classes of
bundles.

\begin{thm}\label{stiefel-chern}

$H^*(\BB\OO(n),\Z_2)$ is the polynomial ring
$\Z_2[w_1,\dots, w_n]$, with $w_i\in H^i(\BB\OO(n),\Z_2)$.
Similarly, in the complex case $H^*(\BB\UU(n),\Z)$ is the
polynomial ring $\Z[c_1,\dots, c_n]$, with $c_i\in
H^{2i}(\BB\UU(n),\Z)$.

\end{thm}

See \cite[Thm.\ 3.2]{vb} for a proof. The generators
$w_i\in H^i(\BB\OO(n),\Z_2)$ are called
Stiefel-Whitney classes, the generators $c_i\in
H^{2i}(\BB\UU(n),\Z)$ are called Chern classes.

\begin{thm}\label{App-thm2} ~

$1.$~ All torsion elements in $H^*(\BB\OO(n),\Z)$ are of order $2$, and
$H^*(\BB\OO(n),\Z)$ modulo torsion is the polynomial ring
$\Z[p_1,\dots,p_k]$ for $n=2k$ or $2k+1$.
 \smallskip

$2.$~ All torsion elements in $H^*(\BB\SO(n),\Z)$ are of order $2$, and
$H^*(\BB\SO(n),\Z)$ modulo torsion is the polynomial
ring $\Z[p_1,\dots,p_k]$ for $n=2k+1$ and $\Z[p_1,\dots,p_{k-1},e]$ for $n=2k$, with
$e^2=p_k$ in the latter case.

\end{thm}

Proofs of this theorem can be found in \cite{vb} and \cite{at}.
The generators $p_i$ are called Pontryagin classes. The generator
$e$ is the Euler class of the universal real $n$-dimensional
vector bundle. This class is also present in case $n=2k+1$ where
it is a nontrivial $2$-torsion element, uniquely determined by the
equation
 \beq\label{G-ESW}
\rho_2 e = w_{n}\,,
 \eeq
see \cite[Prop.\ 3.13]{vb}. Application of the long exact sequence associated
with the coefficient sequence
$0\ra\Z\stackrel{2}{\ra}\Z\stackrel{\rho_2}{\ra}\Z_2\ra 0$ shows that the
torsion subgroup of $H^*(\BB\OO(n),\Z)$ is mapped injectively to
$H^*(\BB\OO(n),\Z_2)\, ,$ i.e. $H^*(\BB\OO(n),\Z)$ is completely
determined by the classes $p_i$ and $w_i$. The same argument
shows that $H^*(\BB\SO(n),\Z)$ is determined by $p_i, w_i$ and
$e$. The injectivity of $\rho_2$ on torsion elements also follows directly from
a Lemma of Borel, see \cite{borel}, Lemma 24.2.

There exist many relations between the above generators and,
hence, between the characteristic classes they define. First,
there are relations among the real characteristic classe. As
before, let $\beta : H^n(X,\ZZ_2)\to H^{n+1}(X,\ZZ)$ denote the
Bockstein homomorphism associated with the coefficient sequence $0
\to \Z \to \Z \to \Z_2 \to 0$ and let $\rho_m : H^n(X,\ZZ) \to
H^n(X,\ZZ_m)$ denote the homomorphism induced by reduction mod
$m$.

\begin{thm}\label{wu-formel-thm}

The following relations hold:
\begin{eqnarray}
\label{wu-formel}
\Sq^j w_m
 & = &
\sum\nolimits_{l=0}^j ~ {m-j+l-1 \choose l} \, w_{j-l}\,
w_{m+l}\,,
\\[0.2cm] \label{rel-StW-odd}
\frak{P}w_{2i+1}
 & = &
\rho_4 \circ \beta \circ \Sq^{2i} w_{2i+1} +
\theta_*\left(w_1\Sq^{2i}w_{2i+1} \right)\, ,
\\[0.2cm]  \label{rel-StW-even}
\frak{P}w_{2i}
 & = &
\rho_4 p_i
 +
\theta_*
 \left(
w_1\Sq^{2i-1}w_{2i}
 +
\sum\nolimits_{j=0}^{i-1} ~ w_{2j} w_{4i-2j}
 \right)\,,
 \end{eqnarray}

\end{thm}

See \cite{thomas3} and \cite{Wu} for a proof. Equation \eqref{wu-formel}
is known as the Wu formula.

\bpr\label{P-realrelCC}

There holds $\rho_2p_i = w_{2i}^2\,$.

\epr

{\it Proof:}~ Consider the coefficient sequence
 $
0 \to \ZZ_2 \stackrel{\theta}{\to} \ZZ_4 \stackrel{\rho}{\to} \ZZ_2 \to 0
 $.
By applying $\rho$ to \eqref{rel-StW-even} and using $\rho \circ \rho_4 =
\rho_2$ and \eqref{pontryagin02} we obtain
 $
0 = \rho\frak{P}w_{2i} - \rho_2p_i = w_{2i}^2 - \rho_2p_i\,.
 $
 \qed
 \bigskip

Second, there are relations between real, complex and quaternionic
characteristic classes. For clarity, in the following we label the
cohomology generators by the groups they belong to. Denote $\tilde
p = \sum_k (-1)^k p_k $ and $\tilde c = \sum_k (-1)^k c_k \, .$
Let
$$
\arraycolsep1pt
 \begin{array}{rccclcrccclcrcccl}
\vp_{\CC,\RR} & : & \UU(m) & \to & \OO(2m)
 & \qquad &
\vp_{\HH,\CC} & : & \Sp(n) & \to & \UU(2n)
 & \qquad &
\vp_{\HH,\RR} = \vp_{\CC,\RR} \circ \vp_{\HH,\CC} & : & \Sp(n) &
\to & \OO(4n)
\\
j_{\RR,\CC} & : & \OO(n) & \to & \UU(n)
 & &
j_{\CC,\HH} & : & \UU(n) & \to & \Sp(n)
 & &
j_{\RR,\HH} & : & \OO(n) & \to & \Sp(n)
 \end{array}
$$
be the embeddings defined by field restriction and field
extension, respectively. The $\vp_{\cdot,\cdot}$ can be chosen as
in Section \ref{S-HSG}. Recall that the Chern class for
$\Sp(n)$-bundles is defined by the element $c(\Sp(n)) :=
\big(\BB\vp_{\HH,\C}\big)^\ast c(\UU(2n)) \in
H^\ast(\BB\Sp(n),\ZZ)$.

\bpr\label{P-relBvp}

The following relations hold:
\begin{align*}
\big(\BB\vp_{\CC,\RR}\big)^\ast w(\SO(2n))
 & = \rho_2 \, c(\UU(n))\,,
 &
(\BB j_{\RR,\CC})^\ast c(\UU(n)) & = \tilde p(\OO(n))\,,
\\
\big(\BB\vp_{\HH,\RR}\big)^\ast w(\SO(4n))
 & = \rho_2 \, c(\Sp(n))\,,
 &
(\BB j_{\CC,\HH})^\ast c(\Sp(n))
 & =
c(\UU(n))\,\,\tilde c(\UU(n))\,,
\\
\big(\BB\vp_{\CC,\RR}\big)^\ast \tilde p(\SO(2n))
 & = c(\UU(n))\, \tilde c(\UU(n))\,,~~~~~~~~~
 &
(\BB j_{\RR,\HH})^\ast c(\Sp(n)) & = \tilde p(\OO(n))^2\,,
\\
\big(\BB\vp_{\HH,\RR}\big)^\ast \tilde p(\SO(4n))
 & = c(\Sp(n))\, \tilde c(\Sp(n))\,, \\
\big(\BB\vp_{\CC,\RR}\big)^\ast e(\SO(2n))
 & = c_n(\UU(n))\,, \\
\big(\BB\vp_{\HH,\RR}\big)^\ast e(\SO(4n))
 & = c_{2n}(\Sp(n))\, .
\end{align*}

\epr

{\it Proof.}~ The relations for $\vp_{\CC,\RR}$ are well known,
see \cite{Hus}, \S 17 and \S 20, Proposition 8.4. Since
$\vp_{\HH,\RR} = \vp_{\CC,\RR}\circ\vp_{\HH,\CC}$, the relations
for $\vp_{\HH,\RR}$ follow from those for $\vp_{\CC,\RR}$ by
replacing $c(\UU(n))$ by $c(\Sp(n))$. The relation for
$j_{\RR,\CC}$ is an equivalent definition of Pontryagin classes.
To prove the relation for $j_{\CC,\HH}$, observe that by
definition of $c(\Sp(n))$, we have
$$
(\BB j_{\CC,\HH})^\ast c(\Sp(n))
 =
(\BB j_{\CC,\HH})^\ast\,(\BB\vp_{\HH,\CC})^\ast\, c(\UU(2n)) \, .
$$
We compute
$$
\vp_{\HH,\CC}\circ j_{\CC,\HH} (a)
 =
\left[\begin{array}{cc} \ol a & 0 \\ 0 & a \end{array}\right]
 \,,~~~~~~
a\in\UU(n)\,.
$$
Since $c(\UU(n))\mapsto \tilde c(\UU(n))$ under the map $\BB\UU(n)\to\BB\UU(n)$
induced by conjugation of matrices, $(\BB
j_{\CC,\HH})^\ast\,(\BB\vp_{\HH,\CC})^\ast\, c(\UU(2n)) = c(\UU(n))\,\,\tilde 
c(\UU(n))$. To prove the relation for $j_{\RR,\HH}$, we compute
$$
(\BB j_{\RR,\HH})^\ast c(\Sp(n))
 =
(\BB j_{\RR,\CC})^\ast\,(\BB j_{\CC,\HH})^\ast c(\Sp(n))
 =
(\BB j_{\RR,\CC})^\ast \big\{ c(\UU(n)) \, \tilde c(\UU(n))\big\} \, .
$$
Due to $(\BB j_{\RR,\CC})^\ast \tilde c(\UU(n))  = \tilde
p(\OO(n))\, ,$ the assertion follows.
 \qed


\section{The Moore-Postnikov tower}
\label{moore postnikov tower}


Below we work in the pointed category, hence all maps are assumed to preserve base points.

A map $f:X\ra Y$ is called an $n$-equivalence if $f_*:\pi_i(X)\ra\pi_i(Y)$ is an isomorphism for $i<n$ and an
epimorphism for $i=n$. If $f_*$ is an isomorphism for all $i$ then
$f$ is called a weak homotopy equivalence.
If $f:X\ra Y$ is an $n$-equivalence and $K$ is a $CW$-complex
then composition with $f$ defines a bijection from the set of homotopy classes of maps $K\to X$ onto the set of homotopy classes of maps $K\to Y$, see e.g.\ \cite[Cor.11.13]{bredon}.

\begin{prop}\label{Lemma-SerreS}

Let $f : X \ra Y$ be an $(n+1)$-equivalence
and let $F \into X \ra Y$ be the
corresponding fibration obtained by homotopy equivalent
deformation of $X$. Assume that $Y$ is simply connected. Then the
map
$$
f^*:H^k (X,\Z) \ra H^k (Y,\Z) \, ,
$$
is an isomorphism for all $k \leq n\, ,$ and a monomorphism for $k=n+1$.
\end{prop}
\emph{Proof}: From the exact homotopy sequence of the fibration $F
\into X \ra Y$ we conclude that $\pi_k (F) = 0\, ,$ for $k \leq
n\, .$ Using the Hurewicz theorem we get $H_k(F)=\{0\}$ for all
$k\leq n$. Inserting this in the universal coefficient theorem
$$
H^k (X,\Z) \cong F\big(H_k (X,\Z)\big) \oplus
\mathrm{Tor}\big(H_{k-1}(X,\Z)\big) \, ,
$$
we obtain $H^k(F,\Z)=\{0\}$ for all $k\leq n$. Since $Y$ is simply
connected we obtain (using again the Hurewicz theorem and the
universal coefficient theorem) $H^1(Y,\Z) = 0$. Moreover, since
$Y$ is simply connected, we have a trivial action of $\pi_1(Y)$ on
$H^*(F,\Z)\, .$ These facts imply that we have a Serre sequence up
to $(n+2)$:
\begin{align*}
\xymatrix{ \cdots\ar[r] & H^{n-1}(F,\Z)\ar[r] &
H^n(Y,\Z)\ar[r]^-{f^*} & H^n(X,\Z)\ar[r] & H^n(F,\Z)\\
 \ar[r] & H^{n+1}(Y,\Z)\ar[r]^-{f^*}  & H^{n+1}(X,\Z)\ar[r]
 & H^{n+1}(F,\Z)\ar[r]
 & \cdots
}
\end{align*}
Since $H^k(F,\Z)=\{0\}$ for all $k\leq n$, the assertion follows.
 \bewe
 \bigskip

The Moore-Postnikov-tower of a fibration $F\into E\stackrel{p}{\ra} B$ is a commutative diagram
\begin{align*}
\xymatrix@R=0.5cm{
& &\ar[d]\\
& &E_n\ar[d]^{p_n}\\
& &\vdots\ar[d]\\
& &E_1\ar[d]^{p_1}\\
E\ar[rr]^{p} \ar[rru]^{q_1}  \ar[rruuu]^{q_n}& & B
}
\end{align*}
such that

1.~ the maps $q_n: E\ra E_n$ are $n$-equivalences;

2.~ the maps $P_n=(p_1\circ p_2\circ\dots\circ p_n):E_n\ra B$ induce
isomorphisms on $\pi_i$ for $i>n$ and an injection for $i=n$;

3.~ the map $p_{n+1}:E_{n+1}\ra E_n$ is a fibration with fiber
$K(\pi_n(F),n)$.
 \bigskip

In Theorem \rref{chs3} we use the convenient fact that every fibration $p:E\ra B$ between connected CW-complexes has a Moore-Postnikov tower, see e.g.\ \cite{at}. For a detailed exposition of Moore-Postnikov theory see \cite{thomas1}.

\end{appendix}

\end{document}